\documentclass{article}

\usepackage{amsmath}
\usepackage[psamsfonts]{amssymb}
\usepackage{latexsym}
\usepackage{amsthm}
\usepackage{cmmib57}
\usepackage{amscd}
\usepackage{longtable}

\def\aivsize{0}
\if1\aivsize
\fi

\newcommand{\goth}[1]{\mathfrak{#1}}
\newcommand{\scr}[1]{\mathcal{#1}}
\def\idty{{\leavevmode{\rm 1\ifmmode\mkern -4.7mu\else\kern -.3em\fi
      I}}}
\renewcommand{\Bbb}[1]{\if1#1\idty\else\mathbb{#1}\fi}

\newcommand{\supp}{\operatorname{supp}}
\newcommand{\singsupp}{\operatorname{singsupp}}
\newcommand{\WF}{\operatorname{WF}}

\newcommand{\CCR}{\operatorname{CCR}}

\newcommand{\ci}[2]{\scr{E}(#1,#2)}
\newcommand{\cni}[2]{\scr{D}(#1,#2)}
\newcommand{\restr}{{\hspace{-1pt}\mbox{\large $\upharpoonright$}\hspace{0pt}}}
\newcommand{\Lh}{\operatorname{span}}
\newcommand{\Ran}{\operatorname{Ran}}
\newcommand{\Dom}{\operatorname{Dom}}
\newcommand{\Id}{\operatorname{Id}}
\renewcommand{\Re}{\operatorname{Re}}
\renewcommand{\Im}{\operatorname{Im}}

\newcommand{\Lin}{\operatorname{L}}

\newcommand{\BU}{\goth{A}}
\newcommand{\JetC}{E^l}

\newcommand{\JetRa}[1]{E^{#1}_\Bbb{R}}
\newcommand{\JetCa}[1]{E^{#1}}
\newcommand{\OPS}{K}

\newcommand{\Gr}{\operatorname{Gr}}

\newtheorem{thm}{Theorem}[section]
  
\newtheorem{defi}[thm]{Definition}
\newtheorem{prop}[thm]{Proposition}
\newtheorem{lem}[thm]{Lemma}
\newtheorem{kor}[thm]{Corollary}
\newenvironment{pf}{\begin{proof}}{\end{proof}} 

\begin{document}

\title{Quantum fields on timelike curves}
\author{Michael Keyl\thanks{electronic mail: \texttt{M.Keyl@TU-BS.DE}}\\
\small
TU-Braunschweig, Inst. Math. Phys, Mendelssohnstra{\ss}{}e 3, D-38106
Braunschweig}
\date{\today}
\maketitle

\begin{abstract}
  A quantum field $\Phi(x)$ exists at an event $x \in M$ of space-time
  $(M,g)$ in general only as a quadratic form $\Phi(x)$. Only after
  smearing $\Phi(x)$ with a smooth test function $f$ we get an operator
  $\Phi(f)$. In this paper the question is considered whether it is
  possible as well to smear $\Phi(x)$ with a singular test function $T$ 
  (i.e. test distributions) supported by a smooth timelike curve
  $\gamma$. It is shown that this is always possible if $\Phi(x)$ satisfies
  the micro local spectrum condition and $T$ belongs to a special
  class of distributions which retain some regularity in timelike
  directions (i.e. along $\gamma$). In the free field case these results
  are used to define some kind of time-translation along $\gamma$ which
  generalizes global space-time translations of Minkowski space.
\end{abstract}

\section{Introduction}
\label{sec:intro}

The most fundamental object of algebraic quantum field theory is a net
$\scr{O} \mapsto \scr{A}(\scr{O})$ which associates to each open and bounded
region $\scr{O}$ of space-time $(M,g)$ a *-algebra $\scr{A}(\scr{O})$
(in most cases a C* or von Neumann algebra) in such a way that the
self adjoint elements of $\scr{A}(\scr{O})$ describe local observables
measurable in $\scr{O}$ (see \cite{HAAG92} and the references therein
for details). Many measurements in general relativity are,
however, observer dependent and it seems to be reasonable to associate
observables not only to space-time regions but also to
\emph{worldlines}. To demonstrate this with a simple example, consider
the free scalar field $\Phi(x)$ in Minkowski space $(\Bbb{R}^4,\eta)$ and an
observer with worldline  $\gamma(t) = (t,\mathbf{x}(t))$. (Note that this
is in general not the proper time parameterization.) Quantities of
the form  
\if1\aivsize
\begin{equation} \label{eq:54}
  C(\gamma,f) = \Phi(T)^*\Phi(T) \ \mbox{with} \ \Phi(T) = \int_{\Bbb{R}^4}
  T(t,\mathbf{x}) \Phi(t,\mathbf{x}) dt d\mathbf{x} \mbox{and} \
  T(t,\mathbf{x}) = \delta(\mathbf{x} - \mathbf{x}(t)) f(t)
\end{equation}
\else
\begin{gather} \label{eq:54}
  C(\gamma,f) = \Phi(T)^*\Phi(T)\\
  \mbox{with} \ \Phi(T) = \int_{\Bbb{R}^4} T(t,\mathbf{x}) \Phi(t,\mathbf{x})
  dt d\mathbf{x} \ \mbox{and} \ T(t,\mathbf{x}) = \delta(\mathbf{x} -
  \mathbf{x}(t)) f(t)  \notag
\end{gather}
\fi
where $f:\Bbb{R} \to \Bbb{C}$ denotes a Schwartz function, describe
simple, point-like particle detectors moving along $\gamma$ (see 
\cite{FREHAAG87} for a more detailed discussion of such models). From 
the works of Fulling \cite{fulling73}, Unruh \cite{UNRUH76}, Bisognano
and Wichmann \cite{bisognano75} and others we know that a uniformly
accelerated observer $\gamma_a$ sees the free field vacuum as a thermal
state at finite temperature while it is of course a ground state with
respect to an inertial observer $\gamma_i$. With appropriately chosen $f$
the observables $C(\gamma_a,f)$ and $C(\gamma_i,f)$ can therefore be used
to distinguish the worldlines $\gamma_a$ and $\gamma_i$. More generally we can
say that at least some information about the geometry of worldlines
is reflected in the structure of the observable families $\{C(\gamma,f)
\, | \, f \in \scr{S}(\Bbb{R}) \}$. 

Hence for the study of observer dependent effects in quantum field
theory it is reasonable to consider quantum fields ``smeared'' by
distributional test ``functions'' $T(x)$ with $\supp T \subset \Ran \gamma$ and
to study their relation to the geometry of $\gamma$. A quantum field,
however, is a very singular object, i.e. $\Phi(x)$ exists in general only
as quadratic form. It is therefore not clear whether objects like
$\Phi(T)$ in Equation (\ref{eq:54}) exists as an operator and how they
should be defined in a mathematically precise way. For a free field in
a globally hyperbolic space-time an analysis of this kind is partially
carried out by Wollenberg \cite{WOLLENBERG93,wollenberg98}, showing
that the algebras  
\begin{displaymath}
  \scr{A}(\gamma) = \bigcup_{\scr{O} \supset \Ran \gamma} \scr{A}(\scr{O}).
\end{displaymath}
which can be associated to any (i.e. not necessarily timelike) curve,
contain some information about the causal character of $\gamma$.

In this paper we are using methods from micro local analysis,
in particular wave front sets of distributions, to consider a much
bigger class of fields. The concept of wave front sets, originally
introduced by Duistermaat and H{\"o}rmander
\cite{hoermander71,duistermaat72} in the context of hyperbolic partial
differential equations, was recently applied with great success in
quantum field theory: In \cite{RADZIKOWSKI96} Radzikowski has shown
that Hadamard states can be characterized in a very elegant way in
terms of the wave front set of its two-point function. Based on this
result Brunetti, Fredenhagen and K{\"o}hler gave in
\cite{BRFRKOE96,brunetti97} a micro local generalization of the
spectrum condition. One reason why micro local methods are so useful
in quantum field theory depends on the fact that wave fronts provides
conditions under which the product of two distributions is
defined. The existence of products of distributions is, however,
closely related to the problem raised in the last paragraph. To
explain this remark in greater detail consider a coordinate system $u
:M \supset M_u \ni p \mapsto u(p) := (t, x^1 \ldots, x^{n-1}) \in \Bbb{R}^n$ of space-time,
such that $t \mapsto u^{-1}(t,0,\ldots,0) \in M_u \subset M$ coincides with the
(parametrized) world-line $\gamma$, and a distribution of the form 
\begin{equation} \label{eq:62}
  T(f) = \int_\Bbb{R} \sum_{|\alpha| \leq l} a_\alpha(t) \frac{\partial^{|\alpha|}f(t,0)}{\partial x^\alpha} dt,
  \quad a_\alpha \in \scr{D}(\Bbb{R}) \ \forall |\alpha| \leq l
\end{equation}
supported by $\gamma$. The corresponding generalization of the expression
$\Phi(T)$ from Equation (\ref{eq:54}) is now given by
\begin{equation} \label{eq:44}
  \Phi(T) = \int_\Bbb{R} \sum_{|\alpha| \leq l} a_\alpha(t) \frac{\partial^{|\alpha|}\Phi(t,0)}{\partial x^\alpha} dt, 
\end{equation}
i.e. $\Phi(T)$ is, as before, the quantum field ``smeared'' by the
singular test function $T$. If we disregard for a moment the fact that
$\Phi$ is operator valued and not just a (numerical) distribution we can
give the formal expression in Equation (\ref{eq:44}) a mathematical
precise meaning by defining $\Phi(T) := (\Phi T)(\mathbf{1})$, where $\Phi T$
denotes the product of the distributions $\Phi$ and $T$ (which we assume
to exist, of course) and $\mathbf{1} \in \scr{E}(M)$ is the function with
$\mathbf{1} \equiv 1$ (note that the support of the distribution $\Phi T$ is
compact if the support of $T$ is compact, hence $(\Phi T)(\mathbf{1})$
exists). Applying this idea to Wightman distributions and combining it
with the reconstruction theorem of quantum field we can show (Theorem
\ref{prop:muloc}) that for a quantum field satisfying the micro local
spectrum condition all the operators $\Phi(T)$ exist with a common dense
domain if the distribution $T$ is regular or of the form given in
Equation (\ref{eq:62}).

The second main result presented in this paper concerns a new look at
representations of space-time translations. It is well known and in
fact one of the major problems of quantum field theory in curved
space-times that there is in a generic space-time no replacement for
global space-time translations of Minkowski space. The possibility
however to restrict a quantum field $\Phi$ to a timelike curve $\gamma$ leads
naturally to the more general question whether there is a way to
translate $\Phi$ \emph{along} $\gamma$. We will show that at least in the free
field case, this is indeed possible. To outline the corresponding
construction, consider the C*-algebra $\scr{A}^1(\gamma)$ generated by unitary
operators $\exp\bigl(i \Phi(T)\bigr)$ with test distributions $T$ which are
supported by $\gamma$ and appropriately regular (more precisely $T$ is
given as in Equation (\ref{eq:62}) with $l=1$; see
Sect. \ref{sec:distri-curves} and \ref{sec:time-translations} for 
details). Using a result of Demoen et al \cite{demoen77} we can show
that there exists a family $\alpha_t$ of completely positive maps on
$\scr{A}(\gamma)$ such that the translated Weyl operators
$\alpha_t\bigl[\exp\bigl(i \Phi(T)\bigr)\bigr]$ coincide up to a numerical
factor with $\exp\bigl(i \Phi(T_t)\bigr)$ where $T_t$ is the distribution
pushed forward in the following way\footnote{This construction seems
  to depend on the coordinate system $(M_u,u)$ given above. We will
  see however that only a reference frame is needed to define $T_t$
  and this is a physically satisfactory dependency.}
\begin{displaymath} 
  T_t(f) = \int_\Bbb{R} \sum_{|\alpha| \leq l} a_\alpha(t'-t) \frac{\partial^{|\alpha|}f(t',0)}{\partial x^\alpha} dt.
\end{displaymath}
It is obvious that this is the most direct generalization of the
space-time translation automorphisms of the Minkowski space theory. It
is therefore reasonable to claim that the $\alpha_t$ reflect the history of
the observer with worldline $\gamma$.

The paper is organized as follows: In Section \ref{sec:qf} we will
give a short summary of some well known material about quantum fields
which will be used throughout the paper. Section \ref{sec:QFsingTF}
contains some general considerations about smearing quantum fields
with distributions which are applied in Section \ref{sec:qf:qf-on-wl}
and \ref{sec:loc-alg} to a special class of distributions  which is
introduced in Section \ref{sec:distri-curves} (cf. Equation
(\ref{eq:62})). In Section \ref{sec:frskf-gen} and
\ref{sec:minkowski-space} the free scalar field is treated as an 
example and in Section \ref{sec:time-translations} the structures
developed so far are used to define the time-translations along
worldlines outlined in the last paragraph. In the last Section we will
discuss some ideas how the results of this paper can be used to study
observer dependent aspects of quantum field theory. In the Appendix we
have postponed some more technical proofs (Appendices
\ref{sec:pf-ext-thm} and \ref{sec:retadvKG}) and given some background
material on micro local analysis, differential operators and
jet-bundles, and the global Hadamard condition (Appendices
\ref{sec:wfs}, \ref{sec:jetbun} and \ref{sec:hadamard}).

\section{Quantum fields}
\label{sec:qf}

We will start with a brief review of some well known material on
quantum fields. Hence consider a (strongly causal) space-time
$(M,g)$ and an \emph{hermitian quantum field} $\Phi(f)$, i.e. a map
$\scr{D}(M) \ni f \mapsto \Phi(f) \in \Lin(D_0,\scr{H})$ from the set
$\scr{D}(M)$ of smooth, compactly supported, complex valued  functions
on $M$ into the space $\Lin(D_0,\goth{H})$ of (unbounded) operators on
a Hilbert space $\scr{H}$ with dense domain $D_0 \subset \scr{H}$. As usual
$\Phi$ should satisfy the following aditional conditions  1. $f \mapsto \langle u,
\Phi(f) v\rangle$ is a distribution on $M$ for all $u, v \in D_0$, 2. The domain
$D_0$ is invariant, i.e. $\Phi(f) D_0 \subset D_0$ for all $f \in \scr{D}(M)$
3. There is a vector $\Omega \in D_0$ cyclic for the  *-algebra generated by
all $\Phi(f)$, 4. $\Phi(\bar f) = \Phi(f)^*\restr D_0$ for all $f\in\scr{D}(M)$
and 5. $[\Phi(f), \Phi(h)] u = 0$ for all $u \in D_0$ and all $f,h \in
\scr{D}(M)$ with spacelike separated supports. Cyclicity of the vacuum
$\Omega$ implies immediately that the span of expressions of the form
$\Phi(f_1) \cdots \Phi(f_n) \Omega$ defines a domain $\tilde{D}_0 \subset D_0$, which is, as
well as $D_0$, dense and invariant. We will assume therefore without
loss of generality that $\tilde{D}_0 = D_0$ holds throughout this paper.

Consider now the \emph{Borchers-Uhlmann algebra}, i.e. the topological 
tensor algebra   
\begin{displaymath}
  \BU = 1 \oplus \scr{D}(M) \oplus \scr{D}(M^2) \oplus \dots
  \oplus \scr{D}(M^n) \oplus \cdots,
\end{displaymath}
which is together with the map $f^*(x_1, \dots, x_n)  =
\overline{f(x_n, \dots, x_1)}$ a topological *-algebra. Each quantum
field defines a state\footnote{A state is in this context a positive,
  continuous, linear functional. Note that in contrast to C*-algebras
  continuity is on $\BU$ not implied by positivity.} or
\emph{Whightman functional} $\scr{W} : \BU \to \Bbb{C}$ on
$\BU$ by  
\begin{equation}
  \label{eq:state}
  \scr{W} = 1 \oplus \scr{W}^{(1)} \oplus \scr{W}^{(2)} \oplus \cdots
\end{equation}
where the \emph{n-point distributions} $\scr{W}^{(n)}$ are given by
\begin{equation}
  \label{eq:state-2}
  \scr{W}^{(n)}(f_1 \otimes \dots \otimes f_n) = \langle \Omega,
  \Phi(f_1)\dots \Phi(f_n) \Omega   \rangle.
\end{equation}
Each state $\scr{W}$ defines on the other hand a unique quantum field such
that (\ref{eq:state}) and (\ref{eq:state-2}) hold. The cyclic
representation $(\scr{H}, \Phi, \Omega)$ of $\BU$ related to $\Phi$ 
by $\Phi(f_1 \otimes \dots \otimes f_n) = \Phi(f_1) \dots \Phi(f_n)$ is
the (unbounded operator version of the) well known GNS representation
corresponding to $\scr{W}$. 

From a physical point of view of greater importance as the fields
itself are local von Neumann algebras to which the $\Phi(f)$ are
affiliated. The most simple way to define such algebras is given
if the common domain $D_0$ of the $\Phi(f)$ is a domain of essential
self adjointness for all $\Phi(f)$ with real valued test function
$f$. In this case we can associate to each open, relatively compact
region $\scr{O} \subset M$ of space-time the von Neumann algebra
\begin{equation}
  \label{eq:vNNet1}
  \scr{R}(\scr{O}) := \left\{ e^{i\Phi(f)} \, \big| \, \overline{f} =
      f, \ \supp f \subset \scr{O} \right\}''.
\end{equation}
The $\scr{R}(\scr{O})$ form obviously an \emph{isotone} family, i.e. if 
$\scr{O}_1 \subset \scr{O}_2$ holds, $\scr{R}(\scr{O}_1) \subset
\scr{R}(\scr{O}_2)$ holds as well. This means that the family
$\bigl(\scr{R}(\scr{O})\bigr)_{\scr{O} \subset \scr{B}(M)}$ forms a
\emph{net} of von Neumann algebras. Here $\scr{B}(M)$ denotes the set
of all admissible regions, i.e. $\scr{B}(M) := \{ \scr{O} \subset M \, 
| \, \scr{O}$ open $\overline{\scr{O}}$ compact $\}$. We will assume
in addition that $\bigl(\scr{R}(\scr{O})\bigr)_{\scr{O} \subset
  \scr{B}(M)}$ is a \emph{causal} net, i.e. the two algebras
$\scr{R}(\scr{O}_1)$ and $\scr{R}(\scr{O}_2)$ commute if the
corresponding regions are \emph{spacelike separated}. Note that this
property is not implied by the corresponding assumption on the fields
(see \cite{RESI1}, Sec. VIII.5). However under some additional
assumptions (e.g. quasianalyticity of the vacuum vector and some kind
of Reeh Schlieder theorem \cite[Prop. 13.2.3]{BW}) causality of the
net $\bigl(\scr{R}(\scr{O})\bigr)_{\scr{O} \subset \scr{B}(M)}$ can be 
derived from causality of the fields (see \cite[Sec. 13.2.2]{BW} for
details). Physically the $\scr{R}(\scr{O})$ are interpreted in terms
of \emph{bounded, local observables} of the field. More precisely each 
self adjoint element of $\scr{R}(\scr{O})$ describes a bounded, local
observable of the fields measurable in the space-time region $\scr{O}
\subset M$. 

If self adjointness of the $\Phi(f)$, as described above, is not given, 
the definition in Eq. (\ref{eq:vNNet1}) is not applicable. In this case 
we should use the \emph{weak commutant}, which is given for an
arbitrary set $\scr{P} \subset \Lin(D_0,\scr{H})$ of (unbounded)
operators by 
\begin{displaymath}
  \scr{P}'_w := \{ A \in B(\scr{H}) \, | \, \langle B^*u, Av \rangle = 
  \langle A^*u, Bv\rangle \ \forall B \in \scr{P} \ \forall u,v \in D_0
  \}.
\end{displaymath}
Now we can define von Neumann algebras $\scr{R}(\scr{O})$
alternatively by
\begin{equation}
  \label{eq:vNNet2}
  \scr{R}(\scr{O}) := \bigl(\{ \Phi(f) \, | \, \supp f \subset \scr{O}
  \}'_w\bigr)'. 
\end{equation}
If the $\Phi(f)$ are essentially self adjoint for real valued $f$
Eqs. (\ref{eq:vNNet1}) and (\ref{eq:vNNet2}) are equivalent
\cite[Sec. 13.2.1, 13.2.2]{BW}, which justifies the usage of the same
symbol. The advantage of (\ref{eq:vNNet2}) is that it works
without additional assumptions. In passing we will note here that it
is at least under special assumtions on the geometry of  $(M,g)$
(e.g. existence of a transitively acting isometry group; see
\cite[Ch. 14]{BW}) also possible to \emph{start} with a net 
$\bigl(\scr{R}(\scr{O})\bigr)_{\scr{O} \subset \scr{B}(M)}$ and  to
\emph{construct} the fields. However we will not use these results
here. Instead we will allways assume that the fields exist in the
described way. 

We have not yet talked about symmetries and translational invariance
of the vacuum, which plays a central role in Minkowski space quantum
field theory. The reason is that these concepts are, due to the lack
of a nontrivial isometry group, almost useless in a generic space-time.
However there are some promising ideas to replace at leat the spectrum
condition by some assumptions on the \emph{wave front set} of the n-point
distributions of the field $\Phi(f)$. A good choice for our purposes is
the \emph{micro local spectrum condition} introduced by 
Brunetti, Fredenhagen and K{\"o}hler \cite{BRFRKOE96}. 

\begin{defi} \label{def:muSC-def}
  Let us consider the set  $\Gr_n$ of finite, unordered graphs
  with  vertices $\{1,\ldots,n\}$ whose edges always occur in both admissible 
  directions. Assume further that no graph $\scr{G} \in \Gr_n$ has an
  empty set of edges and that no edge has the same vertex as source
  and target\footnote{\label{note:1} These two conditions are not
    present in the original definition. However they only single out
    degenerate cases: A graph $\scr{G}$ with no edges makes obviously not
    much sense in the current context and edges with $s(e) = t(e)$ do
    not contribute  to the sum in Equation (\ref{eq:14}), because
    $k(e,x_i)$ \emph{and} $k(e^{-1},x_i) = -k(e,x_i)$ occurs there due
    to items \ref{item:6} and \ref{item:7}.}. An \emph{immersion} of
  $\scr{G} \in \Gr_n$ is a triple $(x,\gamma,k)$ of maps such that: 
  \begin{enumerate}
  \item 
    $x$ maps vertices of $\scr{G}$ to points of $M$.
  \item 
    $\gamma$ maps edges of $\scr{G}$ to picewise smooth curves in $M$ with source
    $s\bigl(\gamma(e)\bigr) = x\bigl(s(e)\bigr)$ and target
    $t\bigl(\gamma(e)\bigr) = x\bigl(t(e)\bigr)$. 
  \item 
    $k$ maps edges to covariantly constant, causal covector fields
    $k(e)$ anlong $\gamma(e)$.
  \item \label{item:6}
    The edge $e^{-1}$ with the opposite direction of $e$ is mapped by
    $\gamma$ to the curve $\gamma(e^{-1})$ inverse to $\gamma(e)$.
  \item \label{item:8}
    $k_e$ is future pointing iff the source $s(e)$ of $e$ is smaller
    than its target $t(e)$ and 
  \item \label{item:7}
    $k(e^{-1}) = -k(e)$.
  \end{enumerate}
  A quantum field $\Phi$ satisfies the \emph{microlocal spectrum
    condition} ($\mu$SC) if the wave front sets of its $n-$point
  functions satisfy  
  \if1\aivsize
  \begin{multline}  \label{eq:14}
    \WF(\scr{W}^{(n)}) \subset \{ (x_1, k_1; \dots; x_n, k_n) \in
    T^*M^n \setminus \{0\}  \, | \, \exists \scr{G} \in \Gr_n \
    \exists \ \mbox{immersion} \ (x,\gamma,k) \ \mbox{of} \ \scr{G}
    \\ \mbox{such that} \ x(i) = x_i \, \forall i = 1, \dots, m \
    \mbox{and} \ k_i = \sum_{s(e)=i} k(e;x_i) \}.
  \end{multline}
  \else
  \begin{multline}  \label{eq:14}
    \WF(\scr{W}^{(n)}) \subset \{ (x_1, k_1; \dots; x_n, k_n) \in
    T^*M^n \setminus \{0\}  \, | \, \exists \scr{G} \in \Gr_n \\
    \exists \ \mbox{immersion} \ (x,\gamma,k) \ \mbox{of} \ \scr{G}
    \mbox{such that} \ x(i) = x_i \, \forall i = 1, \dots, m \\
    \mbox{and} \ k_i = \sum_{s(e)=i} k(e;x_i) \}.
  \end{multline}
  \fi
\end{defi}

It is shown in \cite{BRFRKOE96} that at least Wick ordered products of
free fields (including free fields itself) belong to this
class. Applications concerning interacting fields, especially
renormalizability, can be found in \cite{brunetti97}.

\section{Quantum fields with singular test functions}
\label{sec:QFsingTF}

It is well known that quantum fields do not exist as operator valued fields
but only as \emph{operator valued distributions.} This means that it
is in general impossible to define something like $\Phi(x) = \Phi(\delta_x)$ as
an operator (here $\delta_x$ denotes the delta-distribution at $x \in
M$). However this does not imply that it is impossible to evaluate a
quantum field on any distribution. A natural way to define $\Phi(T)$ for
a distribution $T \in \scr{E}'(M)$ is to consider limits of the form
$\lim_{k\to\infty} \Phi(T_k) u$ for an element $u \in D$ and a sequence $\Bbb{N} \ni
k \mapsto T_k \in \scr{D}(M)$ of smooth functions converging weakly to
$T$. However this idea has the drawback that it is not clear whether
limits of this kind (if they exist) depend on the chosen sequence. It
is therefore more reasonable to consider convergence in the space
$\scr{D}'_\Gamma(M)$ of distributions with wave front set contained in a
closed cone $\Gamma \subset T^*M$ (See Appendix \ref{sec:wfs} for a review of
some material about wave front sets). If we have an additional closed
cone $\Sigma$ which contains the wave front sets of all distributions
 $f \mapsto \langle u, \Phi(f) v\rangle$ and satisfies
\begin{equation}  \label{eq:19}
   \Gamma \oplus \Sigma := \{ (x,\xi_1+\xi_2) \in T^*M \, | \, (x,\xi_1) \in \Gamma, \ (x,\xi_2) \in \Sigma \}
   \subset T^*M \setminus  \{0\},   
\end{equation}  
(i.e. there is no element of the form $(x,0)$ in $\Gamma \oplus \Sigma$) it follows  
from Theorem \ref{thm:distri-extend} that limits $\lim_{k \to \infty} \Phi(T_k)
u$ (if they exist) depend on $T$ but not on the sequence $\Bbb{N} \ni k
\mapsto T_k \in \scr{D}(M)$.  Hence we can define:

\begin{defi} 
  \label{eq:qf-extend-0}
  Consider a compactly supported distribution $T \in \scr{E}'(M)$
  such that ${\rm WF}(T) \subset \Gamma$ holds with a closed cone $\Gamma$ satisfying 
  (\ref{eq:19}) and a sequence $\Bbb{N} \ni k \mapsto T_k \in \scr{D}(M)$
  converging in $\scr{D}'_\Gamma(M)$ to $T$. We define $\Phi(T): D_0 \to
  \scr{H}$ as the unique (if it exists) operator satisfying 
  $\|\,\cdot\,\|-\lim_{k \to \infty} \Phi(T_k) u = \Phi(T) u$ for all $u$. 
\end{defi}

Consider now a linear subspace $\goth{D}$ of $\scr{E}'(M)$ such that
$\scr{D}(M) \subset \goth{D}$ and such that $\Phi(T)$ exists for each
$T \in \goth{D} \setminus \scr{D}(M)$. We can define in analogy to the
Borchers-Uhlmann algebra $\BU$ a tensor algebra $\BU(\goth{D})$
generated by ``test-distributions'' 
\begin{equation}
  \label{eq:bu-goth-d}
  \BU(\goth{D}) := \Bbb{C} \oplus \goth{D} \oplus (\goth{D} \otimes
  \goth{D}) \oplus \dots \oplus \goth{D}^{\otimes n} \cdots.
\end{equation}
Note that all tensor products and direct sums in this expression are
\emph{purely algebraic} (to avoid topological difficulties) , i.e.  
\begin{displaymath}
  \goth{D}^{\otimes n} := \Lh \{ T_1 \otimes \dots \otimes T_n \, | \, T_j \in 
  \goth{D} \} \subset \scr{D}'(M^n)
\end{displaymath}
and the direct sum means \emph{finite direct sums}.  

The purpose of this *-algebra is to carry an extension $\goth{W}$ of
the Wightman functional $\scr{W}$ which is simply given by
\begin{equation}
\label{eq:gothw-def}
 \goth{W}^{(n)}(T_1 \otimes \dots \otimes T_n) = \langle \Omega,
 \Phi(T_1) \dots \Phi(T_n) \Omega \rangle.
\end{equation}
However this functional is defined only if the operators $\Phi(T)$ can be 
extended to an invariant, dense domain $D \subset \scr{H}$. If the existence 
of such a domain is not a priori known it is more convenient to define
$\goth{W}^{(n)}$ directly as a continuous extension of $\scr{W}^{(n)}$
to a distribution space $\scr{D}_\Gamma'(M^n)$, where $\Gamma \subset T^*M$ is a
closed cone with $\Gamma \oplus \WF(\scr{W}^{(n)}) \subset T^*M^n \setminus \{0\}$ (cf. Theorem
\ref{thm:distri-extend} and Theorem \ref{thm:distri-restr}). This
idea motivates the following definition.

\begin{defi}  \label{def:extend-qf}
  To each monomial $\mathbf{T} = T^{(1)} \otimes \cdots \otimes T^{(n)} \in \goth{D}^{\otimes
    n}$ we can associate a closed cone $\Gamma(\mathbf{T}) \subset T^*M^n \setminus \{0\}$
  which is recursively defined by the following properties
 \begin{enumerate}
  \item 
    $n=1$ implies $\WF(\mathbf{T}) = \Gamma(T)$,
  \item 
    if $\mathbf{T} \in \goth{D}^{\otimes n}$ and $\mathbf{S} \in \goth{D}^{\otimes
      m}$ the following recursion relation holds (cf. Equation
    (\ref{eq:18})) 
    \if1\aivsize
    \begin{equation} \label{eq:24}
         \Gamma(\mathbf{T} \otimes \mathbf{S}) = \Gamma(\mathbf{T}) \odot \Gamma(\mathbf{S}) = 
         \bigl(\Gamma(\mathbf{T}) \times \Gamma(\mathbf{S})\bigr) \cup \bigl([M^n \times \{0\}]
         \times \Gamma(\mathbf{S}\bigr) \cup \bigl(\Gamma(\mathbf{T}) \times [M^m \times \{0\}]\bigr) 
       \end{equation}
    \else
    \begin{multline} \label{eq:24}
         \Gamma(\mathbf{T} \otimes \mathbf{S}) = \Gamma(\mathbf{T}) \odot \Gamma(\mathbf{S}) = \\
         \bigl(\Gamma(\mathbf{T}) \times \Gamma(\mathbf{S})\bigr) \cup \bigl([M^n \times \{0\}]
         \times \Gamma(\mathbf{S}\bigr) \cup \bigl(\Gamma(\mathbf{T}) \times [M^m \times \{0\}]\bigr) 
    \end{multline}
    \fi
  \end{enumerate}
  A quantum field $f \mapsto \Phi(f)$ with $n$--point functions $\scr{W}^{(n)}$ 
  is called \emph{extendible} to a distribution space $\goth{D} \subset
  \scr{E}'(M)$ with $\scr{D}(M) \subset \goth{D}$ if $\Gamma(\mathbf{T}) \oplus
  \WF(\scr{W}^{(n)}) \subset T^*M \setminus \{0\}$ is satisfied for all $n \in
  \Bbb{N}_0$ and all $\mathbf{T} \in \goth{D}^{\otimes n}$.
\end{defi}

The family of cones $\Gamma(\mathbf{T})$ is chosen in such a way that
1. $\WF(\mathbf{T}) \subset \Gamma(\mathbf{T})$ holds and 2. the sequence
\if1\aivsize
$\Bbb{N} \ni j \mapsto \mathbf{T}_j \otimes \mathbf{S}_j \in \scr{D}(M^{n+m})$
\else
\begin{displaymath}
  \Bbb{N} \ni j \mapsto \mathbf{T}_j \otimes \mathbf{S}_j \in \scr{D}(M^{n+m})
\end{displaymath} 
\fi
converges in $\scr{D}_{\Gamma(\mathbf{T} \otimes  \mathbf{S})}'(M^{n+m})$ to
$\mathbf{T} \otimes \mathbf{S}$ if $j \mapsto \mathbf{T}_j$ and $j \mapsto \mathbf{S}_j$
converge in $\scr{D}_{\Gamma(\mathbf{T})}'(M^n)$ respectively
$\scr{D}_{\Gamma(\mathbf{S})}'(M^m)$ to $\mathbf{T}$ respectively
$\mathbf{S}$ (cf. Proposition \ref{prop:wf-tensprod-2}). Hence we can
define a functional $\goth{W}$ on the algebra $\goth{A}(\goth{D})$ by
\begin{equation} \label{eq:25}
  \goth{W}^{(n)}(T^{(1)} \otimes \cdots \otimes T^{(n)}) := \lim_{l \to \infty}
  \scr{W}^{(n)}(T^{(1)}_l \otimes \cdots \otimes T^{(n)}_l), 
\end{equation}
where the sequences
\if1\aivsize
 $\Bbb{N} \ni l \mapsto T^{(j)}_l \in \scr{D}(M)$, $j=1,\ldots,n$
\else
\begin{displaymath}
  \Bbb{N} \ni l \mapsto T^{(j)}_l \in \scr{D}(M), \quad j=1,\ldots,n
\end{displaymath}
\fi
converge in $\scr{D}_{\WF(T^{(j)})}'(M)$ to $T^{(j)}$. By Theorem
\ref{thm:distri-extend} this limit exists and depends only on
$\mathbf{T} = T^{(1)} \otimes \cdots \otimes T^{(n)}$. If we consider in particular a
regular $\mathbf{T}$, i.e. $\mathbf{T} \in \scr{D}(M^n)$ we can choose
the constant sequence $l \mapsto \mathbf{T}$ in Equation (\ref{eq:25}) which 
converges in $\scr{D}_\Gamma'(M^n)$ to $\mathbf{T}$ for any $\Gamma$. Hence we
get $\scr{W}^{(n)}(\mathbf{T}) = \goth{W}^{(n)}(\mathbf{T})$ in this
case, and this means that $\goth{W}$ is really an \emph{extension} of 
$\scr{W}$. Summarizing this discussion we get the following:

\begin{prop} \label{prop:scrW-extend}
  If the quantum field $f \mapsto \Phi(f)$ is extendible to a distribution space
  $\goth{D} \subset \scr{E}'(M)$ its $n$--point functions $\scr{W}^{(n)}$
  can be extended to $\goth{D}^{\otimes n}$ in exactly one way such that
  \begin{equation} \label{eq:26}
    \goth{W}^{(n)}(\mathbf{T}) = \lim_{l \to \infty} \scr{W}^{(n)}(\mathbf{T}_l),
  \end{equation}
  holds for any sequence $\Bbb{N} \ni l \mapsto \mathbf{T}_l \in
  \scr{D}(M^n)$ converging in $\scr{D}_{\Gamma(\mathbf{T})}'(M^n)$ to
  $\mathbf{T}$. 
\end{prop}

Let us reconsider now our original definition of $\goth{W}$ in
Equation (\ref{eq:gothw-def}). It is natural to ask whether it
coincides with the expression given in Proposition
\ref{prop:scrW-extend}. This includes in particulal the question
whether extendibility of $\Phi(f)$ in the sense of Definition
\ref{def:extend-qf} implies the existence of $\Phi(T)$ for all $T \in
\goth{D}$ (cf. Definition \ref{eq:qf-extend-0}) and of an invariant
dense domain $D \subset \scr{H}$. The following theorem states that this is
indeed the case.

\begin{thm}
  \label{prop:extend-qf}
  A quantum field $f \mapsto \Phi(f)$ which is extendible to the
  distribution space $\goth{D} \subset \scr{E}'(M)$ has the following
  properties 
  \begin{enumerate}
  \item \label{item:10}
    $\Phi(T)$ exists for all $T \in \goth{D}$ in the sense of
    Def. \ref{eq:qf-extend-0}. 
  \item \label{item:11}
    There is a dense, invariant (i.e. $\Phi(T) D \subset D$ for all $T \in
    \goth{D}$) domain $D$ with $\Omega \in D$ and $D \subset
    D\bigl(\overline{\Phi(T)}\bigr)$ for all $T \in \goth{D}$.  
  \item \label{item:12}
    For each sequence $\Bbb{N} \ni l \mapsto  \mathbf{T}_l \in \scr{D}(M^n)$
    converging in $\scr{D}_{\Gamma(\mathbf{T})}'(T)$ to a monomial
    $\mathbf{T} = (T^{(1)} \otimes \ldots \otimes T^{(n)}) \in \goth{D}^{\otimes n}$ 
    the limit
    \if1\aivsize
    $\lim_{l \to \infty} \Phi(\mathbf{T}_l) \Omega = \Phi(T^{(1)}_l) \cdots \Phi(T^{(n)}_l) \Omega$
    \else
    \begin{displaymath}
      \lim_{l \to \infty} \Phi(\mathbf{T}_l) \Omega = \Phi(T^{(1)}_l) \cdots \Phi(T^{(n)}_l) \Omega
    \end{displaymath}
    \fi
    exists and coincides with $\Phi(\mathbf{T}) \Omega = \Phi(T^{(1)}) \cdots
    \Phi(T^{(n)}) \Omega$. 
  \end{enumerate}  
\end{thm}

The proof of this theorem is somewhat lengthy and technical. Therefore
we have postponed it to Appendix \ref{sec:pf-ext-thm}. Let us consider
now local von Neumann algebras. In analogy to Eq. (\ref{eq:vNNet2}) we
can define  
\begin{equation} \label{eq:58}
  \widetilde{\scr{R}}(\scr{O}) := \bigl(\{ \Phi(T) \, | \, T \in
  \goth{D}(\scr{O}) \}'_w\bigr)' \ \mbox{with} \ \goth{D}(\scr{O}) = \{
  T \in \goth{D} \, | \, \supp T \subset \scr{O} \} 
\end{equation}
Obviously we have $\scr{R}(\scr{O}) \subset
\widetilde{\scr{R}}(\scr{O})$.  The next proposition says that even
equality holds.

\begin{prop}
  \label{prop:scrR-distrib}
  For a quantum field $\Phi(f)$ extendible to
  $\goth{D} \subset \scr{D}'(M)$  we have 
\begin{displaymath}
  \scr{R}(\scr{O}) = \bigl(\{ \Phi(T) \, | \, T \in
  \goth{D}(\scr{O}) \}'_w\bigr)'. 
\end{displaymath}
where $\scr{R}(\scr{O})$ denotes the local von Neumann algebra defined 
according to Equation (\ref{eq:vNNet2}) and $\goth{D}(\scr{O})$ is
given in Equation (\ref{eq:58}).
\end{prop}

\begin{pf}
  We have to show that 
    \begin{equation}\label{eq:6}
    \langle A^*u, \Phi(f)v \rangle = \langle \Phi(f)^*u, Av\rangle
    \quad \forall u,v \in D_0 \ \forall f \in \scr{D}(\scr{O}) 
  \end{equation}
  is equivalent to
  \begin{equation} \label{eq:7}
       \langle A^*u, \Phi(T)v \rangle = \langle \Phi(T)^*u, Av\rangle
    \quad \forall u,v \in D \ \forall T \in \goth{D}(\scr{O})
  \end{equation}
  The implication (\ref{eq:7}) $\Rightarrow$ (\ref{eq:6}) is trivial
  because we have $\scr{D}(\scr{O}) \subset \goth{D}(\scr{O})$ and $D_0 \subset
  D$. To prove the other direction note that there are, according to
  Theorem \ref{prop:extend-qf} item \ref{item:12} (cf. also
  \ref{lem:qf-extend-3}), sequences $l \mapsto \mathbf{T}_l = T^{(1)}_l \otimes \cdots \otimes
  T^{(n)}_l \in \scr{D}(M^n)$ and $l \mapsto \mathbf{S}_l = S^{(1)}_l \otimes \cdots \otimes
  S^{(m)}_l \in \scr{D}(M^m)$ converging in
  $\scr{D}_{\Gamma(\mathbf{T})}(M^n)$ respectively
  $\scr{D}_{\Gamma(\mathbf{S})}(M^m)$ to $\mathbf{T}$ or $\mathbf{S}$,
  satisfying $\lim_{l \to \infty} \Phi(\mathbf{T}_l) \Omega = u$ and $\lim_{l \to \infty}
  \Phi(\mathbf{S}_l) \Omega = v$. In addition we have $\lim_{l\to\infty} \Phi(T_l)x =
  \Phi(T)x$ and $\lim_{l\to\infty}\Phi(\bar{T}_l)x = \Phi(\bar T)x = \Phi(T)^*x$ for
  each $x\in D$. If $A$ satisfies Equation (\ref{eq:6}) we get 
  \begin{displaymath}
    \langle A^* \Phi(\mathbf{T}_l) \Omega, \Phi(T_l) \Phi(\mathbf{S}_l) \Omega \rangle = \langle \Phi(T_l)^*
    \Phi(\mathbf{T}_l) \Omega, A \Phi(\mathbf{S}_l) \Omega \rangle 
  \end{displaymath}
  because $\Phi(\mathbf{T}_l) \Omega \in D_0$ and $\Phi(\mathbf{S}_l)\Omega \in
  D_0$. Taking the limit $l \to\infty$ Equation (\ref{eq:7}) follows.  
\end{pf}

\section{Distribution supported by smooth curves}
\label{sec:distri-curves}

The purpose of this paper is the study of quantum fields  which are
concentrated on timelike curves, or, using the terminology of the last 
section, to extend quantum fields to distributions which are supported 
by such curves. To proceed in this direction, it is useful to discuss
first some properties of this special kind of distributions. Hence let 
us consider a (not necessarily timelike) smooth curve $\gamma : (a,b) \to M$ 
and a compactly supported distribution $T$ with $\supp T \subset \Ran(\gamma)$. In
an appropriate coordinate system $T$ can be 
represented obviously by a distribution $\tilde T \in
\scr{E}'(\Bbb{R}^n)$ with support on $\Bbb{R} \times \{0\} \subset \Bbb{R} \times
\Bbb{R}^{n-1}$. Therefore the following theorem tells us something
about the basic structure of $T$:

\begin{thm} \label{thm:curve-distri-thm-1}
  Consider a compactly supported distribution $T \in
  \scr{E}'(\Bbb{R}^n)$ of order $k$ with support contained in 
  $\Bbb{R} \times \{0\} \subset \Bbb{R} \times \Bbb{R}^{n-1}$. Then we have for a smooth 
  test function $\Bbb{R} \times \Bbb{R}^{n-1} \ni (t,x) \mapsto f(t,x) \in \Bbb{C}$
  \begin{equation} \label{eq:27}
    T(f) = \sum_{|\alpha| \leq k} T_\alpha(f_\alpha) \ \mbox{with} \ f_\alpha(t) =
    \frac{\partial^{|\alpha|}f(t,0)}{\partial x^{\alpha}},
  \end{equation}
  where the $T_\alpha$ are compactly supported distributions on $\Bbb{R}$
  of order $k -|\alpha|$.
\end{thm}

\begin{pf}
  See \cite[Theorem 2.3.5]{HOER1}.
\end{pf}

This result gives us a lower bound on the wave front set of $T$.

\begin{prop} \label{prop:distri-curve-prop-1}
  Consider a distribution $T \in \scr{E}'(M)$ of finite order and with
  support contained in the  image of the curve $\gamma$. Then we have
  \begin{displaymath}
    \WF(T) \supset \{ (\gamma(t),\theta) \in T^*M \, | \, \gamma(t) \in \supp T, \ \theta\cdot\gamma'(t) = 0 \}. 
  \end{displaymath}
\end{prop}

\begin{pf}
  Without loss of generality we can assume that $T \in
  \scr{E}'(\Bbb{R}^n)$ holds with $\supp T \subset \Bbb{R} \times \{0\} \subset \Bbb{R} \times
  \Bbb{R}^{n-1}$. Hence to calculate the wave front set of $T$ we have 
  to consider its Fourier transform\footnote{According to the
    definition of the wave front set we have to calculate the
    Fourier transform of $fT$ for appropriate functions in
    $\scr{D}(\Bbb{R}^n)$. However this would not affect the proof
    substantially, in other words we can assume without loss of
    generality that $f \equiv 1$ holds.\label{note:2}} 
  \begin{displaymath}
    \Bbb{R} \times \Bbb{R}^{n-1} \ni (\rho,\xi) \mapsto \hat T(\rho,\xi) =
    T(e^{-i \langle \rho,\xi; \,\cdot\,\rangle}).
  \end{displaymath}
  Using Theorem \ref{thm:curve-distri-thm-1} this leads to
  \begin{equation} \label{eq:29}
    \hat T(\rho,\xi) = \sum_{|\alpha| \leq k} (-i)^{|\alpha|} \xi^\alpha \hat T_\alpha(\rho)  
  \end{equation}
  where the $T_\alpha$ are, as in Theorem \ref{thm:curve-distri-thm-1},
  distributions on $\Bbb{R}$. This equation shows that the function
  $\Bbb{R}^+ \ni \lambda \mapsto \hat T(0, \lambda \xi)$ grows polynomially for each $\xi \in
  \Bbb{R}^{n-1}$. Hence $(t,0;0,\xi) \in (\Bbb{R} \times \Bbb{R}^{n-1}) \times
  (\Bbb{R} \times \Bbb{R}^{n-1})$ can not be a regular directed point
  whenever $(t,0) \in \supp T$.
\end{pf}

The discussion of the last section shows that the class of quantum
fields which are extendible to a special distribution space is bigger
if the wave front set of these distributions is as small as possible. 
Hence we will concentrate in the following our discussion to those $T$,
where $\WF(T)$ exactly coincides with the lower bound derived in the
last proposition. Hence let us define:

\begin{defi} \label{def:1}
  A distribution $T \in \scr{E}'(M)$ of finite order with $\supp T$ 
  $\subset \Ran (\gamma)$ for a smooth curve $\gamma: (a,b) \to M$ is called
  \emph{as regular as possible} if 
  \begin{equation} \label{eq:28}
       \WF(T) = \{ (\gamma(t),\theta) \in T^*M \, | \, \gamma(t) \in \supp T, \ \theta\cdot\gamma'(t) =
       0 \} 
  \end{equation}
  holds. We will denote the space of all distributions of this kind
  with $\goth{D}^\infty(\gamma)$ or simply $\goth{D}(\gamma)$. The subspace of
  all order $l$ distributiond in $\goth{D}(\gamma)$ is denoted by
  $\goth{D}^l(\gamma)$. 
\end{defi}

For the rest of this section we will develop a special
``parametrization'' of $\goth{D}^l(\gamma)$ in terms of jet-bundles (see
Appendix \ref{sec:jetbun} for a short review of this concept). The
first step in this direction is the following proposition.

\begin{prop} \label{prop:gothDl}
  The spaces $\goth{D}^l(\gamma)$ just defined can be characterized
  alternatively by: $T \in \goth{D}^l(\gamma)$ $\iff$ $T=P^\dagger T_\gamma$, i.e. $T(f) =
  T_\gamma(Pf)$, where $T_\gamma$ is the distribution given by
  \begin{displaymath} 
    T_\gamma(f) = \int_a^b f(\gamma(t)) dt,  
  \end{displaymath}
  $P$ is a $l^{\rm th}$ order differential operator defined
  around $\gamma$ and $P^\dagger$ denotes its formal adjoint. 
\end{prop}

\begin{pf}
  As in the proof of Proposition \ref{prop:distri-curve-prop-1} we can 
  assume without loss of generality that $T \in \scr{E}'(\Bbb{R}^n)$
  holds, with $\supp T \subset \Bbb{R} \times \{0\}$. In this case we have $T 
  = PT_\gamma$ with a differential operator $P$ iff all the distributions
  $T_\alpha$ in Equation (\ref{eq:27}) are regular, i.e. $T$ has the form
  (using the notations of Theorem \ref{thm:curve-distri-thm-1})
  \begin{displaymath}
    Tf = \sum_{|\alpha| \leq l} \int_\Bbb{R} a_\alpha(t) \frac{\partial^{|\alpha|}f(t,0)}{\partial x^\alpha} dt,
  \end{displaymath}
  where the $a_\alpha$ are smooth, compactly supported functions on
  $\Bbb{R}$. It is easy to see (cf. Prop. \ref{prop:wf-delta} and
  \ref{prop:wf-diffop}) that each distribution of this kind has 
  wave front set as in Equation (\ref{eq:28}).

  To prove the other implication let us assume that $T$ has the
  general form of Theorem \ref{thm:curve-distri-thm-1} and that its  
  wave front set satisfies Equation (\ref{eq:28}). Hence the Fourier
  transform of $T$ is given by Equation (\ref{eq:29}). Since
  $\WF(T)$ does not contain an element of the form $(t,0;\rho,0)$ we get
  for $\hat T$ (cf. Footnote \ref{note:2})  
  \begin{displaymath}
    |\hat T(\rho,0)| = |\hat T_0(\rho)| \leq \frac{C_N}{(|1+|\rho|)^N} \ \forall \rho \in 
    \Bbb{R} \ \forall N \in \Bbb{N},
  \end{displaymath}
  where $\hat T_0$ denotes $\hat T_\alpha$ with $\alpha = 0$ and $C_N$ is a
  constant. This implies obviously that $T_0$ is regular. For an
  higher order multiindex $\alpha$ we can modify this argument by
  considering the Fourier transform of the product $p_\alpha T$ where $p_\alpha$
  denotes the monomial $\Bbb{R} \times \Bbb{R}^{n-1} \ni (t,x) \mapsto p_\alpha(t,x) =
  i^{|\alpha|} x^\alpha$. Since $\WF(p_\alpha T) \subset \WF(T)$ (cf. Proposition
  \ref{prop:wf-diffop}) we get similar to the case $\alpha = 0$ the inequality 
  \begin{displaymath}
     |\widehat{p_\alpha T}(\rho,0)| = |\frac{\partial^{|\alpha|}\hat T(\rho,0)}{\partial \xi^\alpha}| = |\hat 
    T_\alpha(\rho)| \leq \frac{C_{N,\alpha}}{(|1+|\rho|)^N} \ \forall \rho \in \Bbb{R} \ \forall N \in
    \Bbb{N}.
  \end{displaymath}
  Hence all the distributions $T_\alpha$ are regular, i.e. $T_\alpha \in
  \scr{D}(\Bbb{R})$ as stated.
\end{pf}
 
The differential operator $P$ of Proposition \ref{prop:gothDl} can be
expressed according to Proposition \ref{prop:jetbun-1} by a section
$\psi$ of the vector bundle $J^l(U,\Bbb{C})^*$, where $U$ is an open
neighbourhood of $\gamma$. To determine the distribution $T$ it is even
sufficient to know the value of $\psi$ only on the curve $\gamma$. Hence the
distribution space $\goth{D}^l(\gamma)$ can be parametrized by $l$--jets as 
described in the following theorem.

\begin{thm} \label{thm:jet2distri}
  Consider the space $\scr{D}\bigl(\Ran(\gamma),\JetC(\gamma)\bigr)$ of smooth,
  compactly supported sections of the vectorbundle $\JetC(\gamma) :=
  J^l(M,\Bbb{C})^*\restr \gamma$ which in turn denotes the restriction of
  the dual of the $l$--jet bundle $J^l(M,\Bbb{C})$ to $\Ran(\gamma)$.
  Then there is a surjective linear map 
  \begin{displaymath}
    \scr{D}\bigl(M,\JetC(\gamma)\bigr) \ni \psi \mapsto T_\psi \in \goth{D}^l(\gamma) \
    \mbox{with} \ T_\psi(f) = \int_a^b \psi(t)j^l_{\gamma(t)}f(t)dt.
  \end{displaymath}
  Hence we have
  \begin{equation}
    \label{eq:gothDkgamma-def}
    \goth{D}^l(\gamma) = \{ T_\psi \in \scr{D}'(M) \, | \, \psi \in
    \scr{D}\bigl(M,\JetC(\gamma)\bigr) \}  \ \mbox{and} \ \goth{D}(\gamma) = \bigcup_{l
      \in  \Bbb{N}} \goth{D}^l(\gamma).  
  \end{equation}
\end{thm}

\begin{pf}
  The statement is an immediate concequence of Proposition
  \ref{prop:gothDl} and \ref{prop:jetbun-1}.
\end{pf}

The map $\psi \mapsto T_\psi$ just defined is, as stated, surjective but
\emph{not} injective, in other words we can not associate a unique
section $\psi \in \scr{D}\bigl(M,\JetC(\gamma)\bigr)$ to a distribution $T \in
\goth{D}^l(\gamma)$. To understand the reason consider again $T \in
\scr{E}'(\Bbb{R}^n)$ and $\gamma(t) = (t,0) \in \Bbb{R}\times \Bbb{R}^{n-1}$. A
$j^{\rm th}$ order differential operator along $\gamma$ is given by
\begin{displaymath}
  (Pf)(t) = \sum_{|\alpha| + j \leq l} a_{\alpha,j}(t) \frac{\partial^{|\alpha| + j}f(t,0)}{\partial x^\alpha \partial t^j}
\end{displaymath}
the corresponding distribution is therefore
\begin{displaymath}
  T(f) = \int_\Bbb{R} \sum_{|\alpha| + j \leq l} a_{\alpha,j}(t) \frac{\partial^{|\alpha| +
      j}f(t,0)}{\partial x^\alpha \partial t^j} dt.
\end{displaymath}
By partial integration this is equivalent to 
\begin{displaymath}
  T(f) = \int_\Bbb{R} \sum_{|\alpha| + j \leq l} (-1)^j \frac{\partial^ja_{\alpha,j}(t)}{\partial t^j}
  \frac{\partial^{|\alpha|}f(t,0)}{\partial x^\alpha} dt.
\end{displaymath}
Hence the operator 
\begin{displaymath}
  (\tilde Pf)(t) = \sum_{|\alpha| + j \leq l} (-1)^j \frac{\partial^ja_{\alpha,j}(t)}{\partial t^j}
  \frac{\partial^{|\alpha|}f(t,0)}{\partial x^\alpha}
\end{displaymath}
leads to the same distribution. This observation motivates the next
proposition.

\begin{prop} \label{prop:trans-jet}
  Consider a coordinate chart $(M_u,u)$ defined around $\gamma$
  (i.e. $\Ran(\gamma) \subset M_u$) and an open neighbourhood $V$ of
  $0 \in \Bbb{R}^{n-1}$ with the following properties: $u(M_u) = (a,b) \times
  V$ and $u\bigl(\gamma(t)\bigr) = (t,0)$. For each distribution $T \in
  \goth{D}^l(\gamma)$ there is exactly one $\psi \in
  \scr{D}\bigl(\Ran \gamma,\JetC(\gamma)\bigr)$ with $T = T_\psi$ and with local
  representative of the form ($f_u := f \circ u$ with $f 
  \in \scr{D}\bigl((a,b) \times V\bigr)$):
  \begin{equation} \label{eq:30}
    (P_\psi f_u)(t) = (\psi \circ j^lf_u)(t) =  \sum_{|\alpha| \leq l} a_\alpha(t)
    \frac{\partial^{|\alpha|}f(t,0)}{\partial x^\alpha}  
  \end{equation}
  with $(t,x) \in (a,b) \times V$, $f_u = f \circ u$ and $f \in \scr{D}\bigl(\Ran \gamma
  \times V\bigr)$. 
\end{prop}

\begin{pf}
  We have just seen that an operator of this kind exist for all $T \in
  \goth{D}^l(\gamma)$. To prove the uniqueness identify $T$ with its local 
  representative, i.e. assume $T \in \scr{E}'\bigl((a,b)\times V\bigr)$ and
  consider for each $t \in (a,b)$ the restriction $T\restr (\{t\} \times V)$ of
  $T$ to the submanifolds $\{t\} \times V$, which exists due to Theorem
  \ref{thm:distri-restr}. $T\restr (\{t\} \times V)$ is for all $t \in (a,b)$ a
  distribution in $\scr{E}'(V)$ with support equal to $\{0\}$. Hence
  $T\restr (\{t\} \times V)$ is a finite linear combination of derivatives of
  the delta-distribution. In other words
  \begin{displaymath}
    \bigl(T\restr (\{t\} \times V)\bigr)(h) = \sum_{|\alpha| \leq l} b_\alpha
    \frac{\partial^{|\alpha|}h(0)}{\partial x^\alpha} 
  \end{displaymath}
  with test function $h \in \scr{D}(V)$. However if $T$ is given as in
  Equation (\ref{eq:30}) we get
  \begin{displaymath}
    \bigl(T\restr (\{t\} \times V)\bigr)(h) = \sum_{|\alpha| \leq l} a_\alpha(t)
    \frac{\partial^{|\alpha|}h(0)}{\partial x^\alpha}. 
  \end{displaymath}
  This leads to the condition $a_\alpha = b_\alpha$ which fixes the
  differential operator $P_\psi$ and thereofre the section $\psi$.
\end{pf}

It is useful to note at this point that the condition on $\psi \in
\scr{D}\bigl(\Ran \gamma,\JetC(\gamma)\bigr)$ which makes the map $\psi \mapsto T_\psi$
unique, is coordinate dependent. More precisely the class of sections
defined implicitly in Proposition \ref{prop:trans-jet} depends on the
$l$--jet $j^lu\restr \gamma$ of the coordinate map along $\gamma$. For us, this
is not a problem, because we will allways have a canonical choice for
$j^lu\restr \gamma$.

\begin{kor} \label{kor:1}
  Consider a timelike curve $\gamma$ and an orthonormal frame $e_\nu$,
  $\nu=0,\ldots,3$ along $\gamma$ with $e_0(t) = \gamma'(t)$. Assume in addition
  without loss of generality that $0 \in \Dom \gamma$. Then we can choose an
  open neighbourhood $\tilde U$ of $0 \in \Bbb{R}^4$ such that the map
  \begin{equation} \label{eq:13}
    \tilde U \ni (t,\mathbf{x}) \mapsto \exp_{\gamma(t)}\left(\sum_{i=1}^3 x^i
      e_i(t)\right) \in U \subset M
  \end{equation}
  defines a coordinate system around $\Ran \gamma \cap U$. For each $T \in
  \goth{D}^l(\gamma)$ there is exactly one $\psi \in \scr{D}\bigl(\Ran \gamma,
  E^l(\gamma)\bigr)$ such that Proposition \ref{prop:trans-jet} holds with
  the coordinate system just introduced.
\end{kor}

\begin{pf}
  The statement is a simple concequence of elementary properties of
  the exponential map and of Proposition \ref{prop:trans-jet}.
\end{pf}

\section{Quantum fields along worldlines}
\label{sec:qf:qf-on-wl}

Now we are able to consider quantum fields which are concentrated on
timelike curves. This implies especially that $\gamma$ denotes a
\emph{timelike} curve for the rest of the paper. The space of ``test
distributions'' $\goth{D}$ to which quantum fields should be extended
(cf. Definition \ref{def:extend-qf}) is naturally defined as the union
of $\scr{D}(M)$ and all possible $\goth{D}(\gamma)$, i.e. 
\begin{equation}
  \label{eq:gothD-def}
  \goth{D}(M) := \scr{D}(M) \cup \bigcup_{\gamma \ \text{smooth,
      timelike}}  \goth{D}(\gamma).  
\end{equation}
Note that in the union on the left hand side of this equation
\emph{all} possible smooth, timelike curves occur, including those
which are only reparametrizations of one another. However it is not
necessary to consider only distinguished parametrizations (e.g. only
proper time parametrizations) because the spaces $\goth{D}^l(\gamma)$
are independent of the parametrization of $\gamma$.

It is reasonable to assume that physically realistic models can be
extended to this special space of test distributions because we can
show that this is true for all quantum fields satisfying $\mu$SC (see
Sec. \ref{sec:qf}):

\begin{thm}
\label{prop:muloc}
  Each quantum field satisfing $\mu$SC can be extended in the sense of
  Prop. \ref{prop:extend-qf} to the test distribution space $\goth{D}(M)$
  defined in Eq. (\ref{eq:gothD-def}).   
\end{thm}

\begin{pf}
  According to Definition \ref{def:extend-qf} we have to check that
  there is no element of the form $(x_1,\ldots,x_n;0,\ldots,0)$ in
  $\WF(\scr{W}^{(n)}) \oplus \Gamma(\mathbf{T})$ where $\mathbf{T} := T^{(1)} \otimes
  \ldots \otimes T^{(n)}$ is an element of $\goth{D}(M)^{\otimes n}$.
 
  To compute $\Gamma(\mathbf{T})$ let us consider first the case
  $n=2$. By Equation (\ref{eq:24}) we have 
  \if1\aivsize
  \begin{multline*}
    \Gamma(T^{(1)}\otimes T^{(2)}) = \bigl( \WF(T^{(1)}) \times \WF(T^{(2)})\bigr) \cup
    \bigl( [\supp(T^{(1)}) \times \{0\}] \times \WF(T^{(2)})\bigr) \cup \\
    \bigl(\WF(T^{(1)}) \times [\supp(T^{(2)}) \times \{0\}]\bigr).
  \end{multline*}
  \else
  \begin{multline*}
    \Gamma(T^{(1)}\otimes T^{(2)}) = \\ \bigl( \WF(T^{(1)}) \times \WF(T^{(2)})\bigr) \cup
    \bigl( [\supp(T^{(1)}) \times \{0\}] \times \WF(T^{(2)})\bigr) \cup \\
    \bigl(\WF(T^{(1)}) \times [\supp(T^{(2)}) \times \{0\}]\bigr).
  \end{multline*}
  \fi
  The elements of $\goth{D}(M)$ are either regular (i.e. smooth function)
  or concentrated on smooth timelike curves $\gamma_1, \gamma_2$ (i.e. $T^{(j)} \in 
  \goth{D}^l(\gamma)$). Hence we get, due to Proposition \ref{prop:wf-delta} 
  and \ref{prop:wf-diffop} in combination with Proposition
  \ref{prop:gothDl}: 
  \if1\aivsize
  \begin{displaymath}
    \Gamma(T^{(1)} \otimes T^{(2)}) = \bigl(N(\gamma_1) \times N(\gamma_2)\bigr) \cup \bigl( [M \times
    \{0\}] \times N(\gamma_2) \bigr) \cup \bigl( N(\gamma_1) \times [M \times \{0\}]\bigr),
 \end{displaymath}
 \else
 \begin{multline*}
    \Gamma(T^{(1)} \otimes T^{(2)}) = \\ \bigl(N(\gamma_1) \times N(\gamma_2)\bigr) \cup \bigl( [M \times
    \{0\}] \times N(\gamma_2) \bigr) \cup \bigl( N(\gamma_1) \times [M \times \{0\}]\bigr),
 \end{multline*}
 \fi
 where $N(\gamma) \subset T^*M$ denotes the normal bundle of the curve $\gamma$, i.e.
 \begin{displaymath}
   N(\gamma) = \{ k \in T^*M\restr  \gamma \, | \, k(\gamma') = 0 \}. 
 \end{displaymath}
 Assume now that $\mathbf{T}$ is an $n$--fold tensor product. By
 applying the arguments just discussed recursively we see (although
 an explicit calculation is somewhat involved) that
 $(x_1, \ldots, x_n; k_1, \ldots, ,k_n) \in \Gamma(\mathbf{T})$ implies $k_j = 0$ or
 $k_j$ spacelike for all $j=1,\ldots,n$. 
 
 The wave front set of $\scr{W}^{(n)}$ is, according  to $\mu SC$,
 given by Defintion \ref{def:muSC-def}. Hence consider a finite graph
 $\scr{G} \in \Gr_n$ and an immersion $(x,\gamma,k)$ of $\scr{G}$. Since the set of
 nodes of $\scr{G}$ is finite there is at least one node $j\in \{1,\ldots,n\}$ such
 that 1. the set of edges starting in $j$ is not empty and 2. for each
 edge $e$ with $s(e)=j$  we have $t(e) > s(e)$. (In Definition
 \ref{def:muSC-def} we have excluded graphs without edges and the case
 $s(e) =  t(e)$; see footnote \ref{note:1}.) Together with item
 \ref{item:8} of Definition \ref{def:muSC-def} and Equation
 (\ref{eq:14}) this implies that there is for each $(x_1,\ldots,x_n;
 k_1,\ldots,k_n) \in \WF(\scr{W}^{(n)})$ at least one $j=1,\ldots,n$ such that
 $k_j \in T^*M$ is causal. This shows together with the property of
 $\WF(\mathbf{T})$ derived in the last paragraph that there is no element
 of the form $(x_1,\ldots,x_n,0,\ldots,0) \in \WF(\scr{W}^{(n)}) \oplus
 \WF(\mathbf{T})$ and this completes the proof.
\end{pf}

Let us consider now the extension $\goth{W}^{(n)}$ of the $n$-point
function $\scr{W}^{(n)}$ defined in Proposition
\ref{prop:scrW-extend}. The parametrization of $\goth{D}^l(\gamma)$ in
terms of jet-bundles allows us to interpret the $\goth{W}^{(n)}$ as a
family of vector bundle valued distributions. To do this let us
introduce the \emph{external tensor product}\footnote{Note that the
  usual tensor product symbol $\otimes$ is occupied in the context of vector
  bundles already for another construction (the ordinary tensor
  product).} of $k$ respectively $j$
copies of the bundles $\JetC(\gamma)$ and $M \times \Bbb{C}$:
\if1\aivsize
\begin{multline*}
  \JetC(\gamma)^{\boxtimes k} \boxtimes (M \times \Bbb{C})^{\boxtimes j} = 
  \bigcup_{\substack{(t_1,\ldots,t_k) \in (a,b)^k\\ (x_1,\ldots,x_j) \in M^j}}
  \JetC_{\gamma(t_1)}(\gamma) \otimes \cdots \otimes \JetC_{\gamma(t_k)}(\gamma) \otimes (\{x_1\} \times \Bbb{C}) \otimes \cdots \otimes
  (\{x_j\} \times \Bbb{C}), 
\end{multline*}
\else
\begin{multline*}
  \JetC(\gamma)^{\boxtimes k} \boxtimes (M \times \Bbb{C})^{\boxtimes j} = \\
  \bigcup_{\substack{(t_1,\ldots,t_k) \in (a,b)^k\\ (x_1,\ldots,x_j) \in M^j}}
  \JetC_{\gamma(t_1)}(\gamma) \otimes \cdots \otimes \JetC_{\gamma(t_k)}(\gamma) \otimes (\{x_1\} \times \Bbb{C}) \otimes \cdots \otimes
  (\{x_j\} \times \Bbb{C}), 
\end{multline*}
\fi
where $\JetC_{\gamma(t_i)}(\gamma)$ denotes the fiber of $\JetC(\gamma)$ over
$\gamma(t_i)$. Using in addition the map $\scr{D}\bigl(\Ran \gamma,\JetC(\gamma)\bigr) 
\ni \psi \mapsto T_\psi \in \goth{D}^l(\gamma)$ defined in Theorem \ref{thm:jet2distri} we
get the following:

\begin{prop} \label{prop:gothW2distri}
  The linear map
  \if1\aivsize
  \begin{multline} \label{eq:31}
    \scr{D}\bigl((\Ran \gamma)^k \times M^j,\JetC(\gamma)^{\boxtimes k} \boxtimes (M \times
    \Bbb{C})^{\boxtimes j}\bigr) \ni \psi_1 \boxtimes \cdots \boxtimes \psi_k
    \boxtimes f_1 \boxtimes \cdots \boxtimes f_j \\ \mapsto
    \goth{W}^{k+j}(T_{\psi_1} \otimes \cdots \otimes T_{\psi_k} \otimes f_1 \otimes \cdots \otimes f_j)
  \end{multline}
  \else
  \begin{multline} \label{eq:31}
    \scr{D}\bigl((\Ran \gamma)^k \times M^j,\JetC(\gamma)^{\boxtimes k} \boxtimes (M \times
    \Bbb{C})^{\boxtimes j}\bigr) \\ \ni \psi_1 \boxtimes \cdots \boxtimes \psi_k
    \boxtimes f_1 \boxtimes \cdots \boxtimes f_j  \mapsto
    \goth{W}^{k+j}(T_{\psi_1} \otimes \\ \cdots \otimes T_{\psi_k} \otimes f_1 \otimes \cdots \otimes f_j)
  \end{multline}
  \fi
  is a vector bundle valued distribution.
\end{prop}

\begin{pf}
  The statement follows from the fact that the functional defined in
  Equation (\ref{eq:31}) coincides in the case $l=0$ with the
  restriction of $\scr{W}^{(k+j)}$ to $(\Ran \gamma)^k \times M^j$. If $l > 0$ we
  have to decompose the functional in terms of a natural frame of the
  vector bundle $\JetC(\gamma)^{\boxtimes k} \boxtimes (M \times \Bbb{C})^{\boxtimes
    j}$ (cf. Equation (\ref{eq:32})). Each component we get in this
  way is a differential of the restriction considered in the $l=0$
  case, hence a distribution.
\end{pf}

Of particular importance for us is the case $j=0$, because we get 
a family of distributions
\begin{equation}
  \label{eq:scrWsubgammal-def}
  \scr{W}_\gamma^{(l;n)}(\psi_1 \boxtimes \dots \boxtimes \psi_n) =
  \goth{W}^{(n)}(T_{\psi_1} \otimes \dots \otimes T_{\psi_n})
\end{equation}
which defines a state on the *-algebra
\begin{equation} \label{eq:59}
  \BU^l(\gamma) := \Bbb{C} \oplus
  \scr{D}\bigl(\Ran \gamma,\JetC(\gamma)\bigr) \oplus
  \scr{D}\bigl((\Ran \gamma)^2,\JetC(\gamma) \boxtimes \JetC(\gamma)\bigr) \oplus \cdots ,
\end{equation}
where the *-operation is given by
\begin{displaymath}
  \left(\theta_1 \boxtimes \dots \boxtimes \theta_n\right)^* :=
  \overline{\theta_n} \boxtimes \dots \boxtimes \overline{\theta_n} \ \mbox{and}
  \ \overline{\theta}(j^l_{\gamma(t)}f) := \overline{\theta(j^l_{\gamma(t)}\overline{f})}
\end{displaymath}
Constructing the generalized GNS-representation (cf. Section
\ref{sec:qf}) we get a $\JetC(\gamma)$-valued quantum field $\Phi_{\gamma,l}$. The
following Theorem shows how it is related to the original field $\Phi$.  

\begin{thm}
  \label{thm:Phik-def}
  Consider the Hilbert space 
  \begin{displaymath} 
    \scr{H}^l(\gamma) := \left(\Lh \{ \Phi(T_1) \cdots \Phi(T_n) \Omega \, | \, T_1 \otimes 
      \cdots \otimes T_n \in \goth{D}^l(\gamma)^{\otimes n}, \ n \in \Bbb{N}\}\right)^{\|\,\cdot\,\|-{\rm 
        cl}}
  \end{displaymath}
  and the operators $\Phi_\gamma^l(\psi) := \Phi(T_\psi)\restr
  \scr{H}^l(\gamma)$. Then the map
\begin{displaymath}
  \scr{D}\bigl(\Ran \gamma,\JetC(\gamma)\bigr) \ni \psi \mapsto \Phi_\gamma^l(\psi) := \Phi(T_\psi)
\end{displaymath}
 defines an $\JetC(\gamma)$-valued quantum field with $n$-point function
 (\ref{eq:scrWsubgammal-def}). 
\end{thm}

\begin{pf}
  This statement is an easy concequence of the definitions and of
  Proposition \ref{prop:gothW2distri}.
\end{pf}

Note that the quantum field $\Phi_\gamma^l$ carries less information as
the map $\goth{D}^l(\gamma) \ni T \mapsto \Phi(T)$. However $\Phi_\gamma^l$ is much easier to
study, because it is described completely in terms of the Wightman
functional $\scr{W}_\gamma^l$. The full field operators $\Phi(T)$ allways
need the knowledge of $\goth{W}$, which is a functional on
$\goth{D}(M)$ rather than $\goth{D}^l(\gamma)$. Hence a lot of geometry not
directly related to $\gamma$ comes in. To avoid this complication is the
major reason for us to introduce the fields $\Phi_\gamma^l$.

\section{Local algebras}
\label{sec:loc-alg}

Let us consider now local von Neumann algebras. In analogy to
Eq. (\ref{eq:vNNet2}) we can define for each smooth timelike curve
$\gamma: (a,b) \to M$ and for each $k \in \Bbb{N} \cup \{\infty\}$ a
von Neumann algebra $\scr{R}^l(\gamma)$ by 
\begin{equation}
\label{eq:scrRsubk-gamma-def}
  \scr{R}^l(\gamma) := \bigl(\{ \Phi(T) \, | \, T \in
  \goth{D}^l(\gamma)\}'_w\bigr)' 
\end{equation}
Since $J^l(M,\Bbb{C})^*$ is for $l<k$ a subbundle of $J^k(M,\Bbb{C})$
we have $\scr{D}\bigl((\Ran \gamma),\JetC(\gamma)\bigr)$ $\subset
\scr{D}\bigl((\Ran \gamma),\JetCa{k}(\gamma)\bigr)$ and therefore
$\goth{D}^l(\gamma) \subset \goth{D}^k(\gamma)$. Hence the definition
of the  $\scr{R}^l(\gamma)$ implies immeditely that
$\scr{R}^l(\gamma)$ is a subalgebra of $\scr{R}^k(\gamma)$ if $l<k$
holds. In other words we get the inifinte sequence
\begin{equation}
  \label{eq:scrRsubk-seq}
  \scr{R}^0(\gamma) \subset \scr{R}^1(\gamma) \subset \dots \subset
  \scr{R}^l(\gamma) \subset \dots \subset \scr{R}^\infty(\gamma)
  \subset \scr{R}(\gamma)
\end{equation} 
where we have introduced the additional algebra
\begin{equation}
  \label{eq:scrR-gamma-def}
  \scr{R}(\gamma) = \bigcap_{\scr{O} \in \scr{B}(\gamma)}
  \scr{R}(\scr{O}), \quad \scr{B}(\gamma) := \left\{ \scr{O} \in
    \scr{B}(M)  \, | \, \gamma\bigl((a,b)\bigr) \subset \scr{O}
  \right\}.
\end{equation}
The relations $\scr{R}^l \subset \scr{R}(\gamma)$ for all $l \in
\Bbb{N} \cup \{\infty\}$ stated in Eq. (\ref{eq:scrRsubk-seq}) follow
directly from Prop. \ref{prop:scrR-distrib}. 

To interpret the algebras $\scr{R}^l(\gamma)$ and $\scr{R}(\gamma)$
let us review first some simple material about timelike curves and
worldlines (for a detailed exposition of observers and reference frame
in general relativity see the book of Sachs and Wu
\cite{SACHSWU}). First it is useful to distinguish
between \emph{parametrized} curves, i.e. smooth maps $\gamma : (a,b)
\to M$ and paths (curves without a distinguished parametrization)
i.e. an equivalence class of parametrized curves (where two curves are
defined to be equivalent if there is a smooth, strictly monotone
reparametrization). Each parametrized curve $\gamma$ determines a
unique path which we will denote by $\gamma$ as well, as long
as confusion can be omitted. A timelike path describes physically the
\emph{worldline} of an observer, while the choice of a special
parametrization fixes the \emph{clock} used by the observer. Among all
possible parametrizations of a given path the \emph{proper time
  parametrizations} are distinguished by the condition
$g(\gamma'(t),\gamma'(t)) = -1$ for all $t \in (a,b)$. Physically
proper time is measured by so called \emph{standard clocks} which are 
experimentally approximated up to a very high degree of accuracy by 
atomic clocks. 

Let us come back now to the algebras $\scr{R}^l(\gamma)$ and
$\scr{R}(\gamma)$. Since $\scr{R}(\scr{O})$ contains observables
measurable in the region $\scr{O}$ the definition of $\scr{R}(\gamma)$ 
in Eq. (\ref{eq:scrR-gamma-def}) implies immediately that
$A=A^*\in\scr{R}(\gamma)$ is an observable which is measurable in any
region containing $\gamma$, in other words measurable \emph{along}
$\gamma$. This means we can interpret self adjoint elements of
$\scr{R}(\gamma)$ as bounded observables the observer $\gamma$ can
measure in the time-interval $(a,b)$. In the special case of $A=A^*
\in \scr{R}^l(\gamma)$ with finite $l$ only observables depending
at most on $l^{th}$ derivatives of the field are considered. The
algebras $\scr{R}^\infty(\gamma)$ lie somewhat intermediate between
$\scr{R}(\gamma)$ and $\scr{R}^l(\gamma)$ with $l \in \Bbb{N}$. It is reasonable
to conjecture that $\scr{R}(\gamma) =\scr{R}^\infty(\gamma)$
holds for physically realistic models (cf. Section
\ref{sec:minkowski-space}). 

The set of $T \in \goth{D}^l(\gamma)$ is not changed by
reparametrizations and hence the algebras $\scr{R}^l(\gamma)$ are
identical for worldlines belonging to the same path. This is very 
plausible from a physical point of view, because the set of
observables measurable during certain period should not depend on the
clock with which time is measured. From a more formal point of view
this means that we should consider the set of smooth timelike path as
the index set of the $\scr{R}^l(\gamma)$ (and not parametrized
curves). If we introduce in addition the ordering relation $\gamma_1
\subset \gamma_2 :\iff \Ran \gamma_1 \subset \Ran \gamma_2$ we get again an isotone family of
von Neumann algebras which is causal in the same way as the family
$\scr{R}(\scr{O})$. Hence the $\scr{R}^l(\gamma)$ form as well as the
$\scr{R}(\scr{O})$ a causal net of von Neumann algebras.

Let us consider now the state $\scr{R}^l(\gamma) \ni A \mapsto \langle \Omega, A
\Omega \rangle$. Its GNS representation is $(\scr{H}^l(\gamma),\eta_\gamma^l,\Omega)$ with
\begin{equation}
  \label{eq:2}
  \scr{R}^l(\gamma) \ni A \mapsto \eta_\gamma^l(A) = P_\gamma^l A P_\gamma^l  \in  \scr{B}(\scr{H}^l(\gamma)), 
\end{equation}
where $P_\gamma^l$ denotes the projection onto $\scr{H}^l(\gamma)$ (cf. Theorem
\ref{thm:Phik-def}). This representation is related to the fields
$\Phi_\gamma^l$ by  
\begin{equation} \label{eq:36}
  \scr{M}_\gamma^l(\mu) := \eta_\gamma^l\bigl(\scr{R}^l(\mu)\bigr)'' = \bigl( \{
  \Phi_\gamma^l(\psi) \, | \, \psi \in \scr{D}\bigl(\Ran \gamma,\JetC(\mu)\bigr) \}'_w\bigr)',
\end{equation}
with $\mu \subset \gamma$. A state $\rho$ on $\scr{R}^l(\gamma)$ is normal with respect to 
$\eta_\gamma^l$ if the observer $\gamma$ can prepare $\rho$ out of the vacuum using
only operations from $\scr{R}^l(\gamma)$. We will see that 
this is a serious restriction as long as $l$ is finite. Hence
considering the algebras $\scr{M}_\gamma^l(\mu)$ instead of $\scr{R}^l(\mu)$
means to consider limited measuring possibilities of the
observer. However the $\scr{M}_\gamma^l(\mu)$ are easier to study than the 
$\scr{R}^l(\mu)$, due to their direct relation to the fields $\Phi_\gamma^l$
and the corresponding Wightman functionals $\scr{W}_\gamma^l$, cf. the
discussion at the end of Section \ref{sec:qf:qf-on-wl}. 

\section{The free scalar field}
\label{sec:frskf-gen}

Let us consider now the free scalar field on a globally hyperbolic
$(M,g)$ space-time as a particular example. We will start with a short
review of this model (see \cite{WALDBOOK} and the references therein
for details). 

As a consequence of global hyperbolicity  there exist unique advanced
and retarded fundamental solutions $G^\pm$ of the Klein-Gordon
equation (see Appendix \ref{sec:retadvKG}). This means there are
continuous operators\footnote{We are considering in this section only
  real valued fields. However the generalization to the complex case
  is in most places straight forward.} $G^\pm : \scr{D}(M,\Bbb{R}) \to
\scr{E}(M,\Bbb{R})$ with  $G^\pm (\Box - m^2) f = (\Box - m^2) G^\pm f = f$ and
$\supp G^\pm(f) \subset J^\pm(\supp f)$ for all $f \in \scr{D}(M,\Bbb{R})$. Here
$\Box$ denotes d'Alembertian on $(M,g)$ and $J^\pm$ denotes the causal
past/future. $G^\pm$ gives rise to an antisymmetric bilinear form $G
:\scr{D}(M,\Bbb{R}) \times \scr{D}(M,\Bbb{R})$ by     
\begin{equation}
  \label{eq:sympl-form-def}
  G(f,h) = \int_M Gf(x)h(x) \lambda_g(x),
\end{equation}
where $Gf = G^+f - G^-f$. Since $G$ is degenerate it is not a
symplectic form on $\scr{D}(M,\Bbb{R})$. In other words, it is
necessary to consider the quotient space\footnote{If confusion
  can be avoided we will identify in the following functions
  $f \in \scr{D}(M,\Bbb{R})$ and the corresponding equivalence classes
  $f \in \scr{P}$.} $\scr{P} := \scr{D}(M,\Bbb{R})/\ker G$. This leads to
the real symplectic vector space $(\scr{P},G)$ and to the
corresponding CCR-algebra $\CCR(\scr{P},G)$. Alternatively we can
consider an arbitrary (smooth) Cauchy surface $\Sigma \subset M$ and the space of
compactly supported, smooth initial data, i.e. $\scr{D}(\Sigma,\Bbb{R}^2)$
which is a symplectic space too, if we equip it with the symplectic
form 
\begin{displaymath}
  \tilde G(f_1,k_1;f_2,k_2) = \int_\Sigma f_1(x) k_2(x) \lambda_\Sigma(x) - \int_\Sigma
  f_2(x) k_1(x) \lambda_\Sigma(x),
\end{displaymath}
where $\lambda_\Sigma$ denotes the natural (i.e. induced by the metric) volume
form on $\Sigma$. A symplectic isomorphism between $\scr{P}$ and
$\scr{D}(\Sigma,\Bbb{R}^2)$ is given by the map
\begin{equation} \label{eq:42}
  \scr{P} \ni f \mapsto \iota_\Sigma(f) := \Bigl((Gf)\restr \Sigma, \bigl[\partial_t (Gf)\bigr]
  \restr \Sigma\Bigr) \in \scr{D}(\Sigma,\Bbb{R}^2).    
\end{equation}

A quasi-free, regular state $\omega$ on this C*-algebra is given by a ``one
particle structure'' i.e. a real linear function $\OPS: \scr{P} \to
\scr{K}$ into a separable Hilbert space $\scr{K}$, which is
symplectic: $\Im \langle \OPS(f), \OPS(h) \rangle = G(f,h)$ and has the property
that $\Ran \OPS + i \Ran \OPS$ is dense in $\scr{K}$. The state $\omega$ is
then defined by its generating functional $\omega(W(f)) = \exp(-\frac{1}{4}
\|\OPS(f)\|^2)$. Consider now the GNS representation $(\scr{H},\pi,\Omega)$ of
$\omega$ and the unitary operators $\pi\bigl(W(f)\bigr)$ (for $f \in
\scr{P}$). It is well known that we can identify $\scr{H}$ with the
symmetric Fock space $\scr{F}_+(\scr{K})$,  the cyclic vector $\Omega$ with
the corresponding Fock vacuum and the Weyl operators
$\pi\bigl(W(f)\bigr)$ with the exponentials
$\exp\bigl[i\Phi_S\bigl(\OPS(f)\bigr)\bigr]$ of Segal operators
$\Phi_S\bigl(K(f)\bigr)$. The free scalar quantum field $\Phi(f)$ on $(M,g)$
in the (quasi free) state $\omega$ is given by complex linear extension
of the map $\scr{D}(M,\Bbb{R}) \ni f \mapsto \Phi(f)$. It has the two-point
function 
\begin{equation} \label{eq:35}
  W^{(2)}(f\otimes h) =  \langle \OPS(f), \OPS(h) \rangle =  \Re \langle \OPS(f), \OPS(h) \rangle + i
  G(f,h).    
\end{equation}

For real valued $f$ the operators are essentially self adjoint. Hence
the von Neumann algebras $\scr{R}(\scr{O})$ can be defined according
to Equation (\ref{eq:vNNet1}).
However for this particular model there is a more direct way to define 
the $\scr{R}(\scr{O})$ without explicit use of the fields. If we
introduce first a causal net of C*-subalgebras of $\CCR(\scr{P},G)$ by
\if1\aivsize
\begin{equation} \label{eq:60}
  \scr{B}(M) \ni \scr{O} \mapsto \scr{A}(\scr{O}) := C^* \bigl( \{ W(f) \in
  \CCR(\scr{P},G) \, | \, \supp f \subset \scr{O} \} \bigr) \subset
  \CCR(\scr{P},G), 
\end{equation}
\else
\begin{multline} \label{eq:60}
  \scr{B}(M) \ni \scr{O} \mapsto \scr{A}(\scr{O}) := \\
  C^* \bigl( \{ W(f) \in \CCR(\scr{P},G) \, | \, \supp f \subset \scr{O} \}
  \bigr) \subset \CCR(\scr{P},G), 
\end{multline}
\fi
we get 
\begin{equation} \label{eq:45}
  \scr{R}(\scr{O}) = \{ \pi\bigl(W(f)\bigr) \, | \, \supp f \subset \scr{O} \}''. 
 = \pi\bigl(\scr{A}(\scr{O})\bigr)''.
\end{equation}

To complete the description of the model we have to restrict the class 
of admissible states $\omega$, because there is, due the lack of a
translation group as in Minkowski space, no distinguished vacuum
state. As a replacement we have to assume that $\omega$ is, as a
physically reasonable state, a ``Hadamard state''. This means that
$\omega$ has to satisfy the \emph{global Hadamard condition}, first
introduced in mathematically rigorous way by Kay and Wald in
\cite{KAY+WALD91}. Physically this condition says that the two point
function $\scr{W}^{(2)}$ has the same short distance behaviour as
the Minkowski vacuum. The exact definition of a Hadamard state is
however somewhat involved. Therefore we have postponed it to the
Appendix \ref{sec:hadamard}. An equivalent, but simpler
characterisation of Hadamard states, introduced by Radzikowski
\cite{RADZIKOWSKI96} can be given in terms of the wave front set of
$\scr{W}^{(2)}$: A quasi free, regular state $\omega$ is a Hadamard state
iff $\WF(\scr{W}^{(2)})$ is given by  
\if1\aivsize
\begin{equation}
\label{eq:hadamard}
  \WF(\scr{W}^{(2)}) = \{ (x_1,k_1;x_2,k_2) \in T^* (M \times M)
  \setminus \mathbf{0} \, | \, (x_1,k_1) \sim (x_2,-k_2), \ k_1 \
  \text{future pointing} \ \},
\end{equation}
\else
\begin{multline}
\label{eq:hadamard}
  \WF(\scr{W}^{(2)}) = \{ (x_1,k_1;x_2,k_2) \in T^* (M \times M)
  \setminus \mathbf{0} \, | \, \\ (x_1,k_1) \sim (x_2,-k_2), \ k_1 \
  \text{future pointing} \ \},
\end{multline}
\fi
where $(x_1,k_1) \sim (x_1,-k_1)$ means that $(x_1,k_1)$ and
$(x_2,-k_2)$ are on the same null geodesic strip (cf. Appendix
\ref{sec:retadvKG}).

Eq. (\ref{eq:hadamard}) and Definition \ref{def:muSC-def} imply
immediately that a Hadamard state satisfies the microlocal spectrum
condition. Hence we can apply  Theorem \ref{prop:muloc} which implies
that the algebras $\scr{R}^l(\gamma)$ exist and we can use all the
machinery introduced in Section \ref{sec:qf:qf-on-wl}. However in this
particular case there is a more direct way to define these von Neumann
algebras because we can associate to each smooth worldine $\gamma$ and each
$l \in \Bbb{N} \cup \{\infty\}$ a C*-subalgebra $\scr{A}^l(\gamma)$ of
$\CCR(\scr{P},G)$ such that $\scr{R}^l(\gamma) =
\pi\bigl(\scr{A}^l(\gamma)\bigr)''$ holds. The explicit construction is
described by the following proposition.  

\begin{prop}
  Consider the unique, weakly continuous extension of $G^\pm$ to  
  $\scr{E}'(M,\Bbb{R})$, i.e. $G^\pm(T)(f) = T\bigl(G^\mp(f)\bigr)$   
  (cf. Appendix \ref{sec:retadvKG}) and the subspace
  $\goth{D}(M,\Bbb{R})$ consisting of \emph{real valued} elements of
  $\goth{D}(M)$. 
  \begin{enumerate}
  \item \label{item:14} 
    $GT = G^+T - G^-T$ is for each $T \in \goth{D}(M,\Bbb{R})$ a smooth
    function with support contained in $J^+(\supp T) \cup J^-(\supp T)$.
  \item \label{item:15}
    The distribution $G \in \scr{D}'(M \times M,\Bbb{R})$ is extendible to
    $\goth{D}(M,\Bbb{R}) \otimes \goth{D}(M,\Bbb{R})$ and has the form
    \begin{displaymath}
      G(T \otimes S) = S\bigl(G(T)\bigr) = - T\bigl(G(S)\bigr) = 
      \frac{1}{2i}(\goth{W}^{(2)}(T \otimes S) - \goth{W}^{(2)}(S \otimes T)),
    \end{displaymath}
    where $\goth{W}^{(2)}$ is the (extension to $\goth{D}(M)$ of the)
    two-point function associated to a  Hadamard state. 
  \item \label{item:16}
    For each $T \in \goth{D}(M,\Bbb{R})$ there is a $f \in
    \scr{D}(M,\Bbb{R})$ such that $GT = Gf$. Hence the symplectic
    space $(\goth{D}(M,\Bbb{R})/ \ker G, G)$ can be identified with
    $(\scr{P},G)$ in a natural way. 
  \end{enumerate}
\end{prop}

\begin{pf}
  Item \ref{item:14}. From Corollary \ref{kor:kg-inhom-1} we know that
  $\WF(G^\pm T) \subset \WF(T)$, hence $\WF(GT) \subset \WF(T)$ holds as well. But
  $GT$ is at the same time a solution of the homogeneous Klein-Gordon
  equation. Together with Theorem \ref{thm:prop-sing-thm} this implies
  that each $\theta \in \WF(GT)$ is a null covector. Concequently the
  characterization of $T \in \goth{D}^l(\Ran \gamma,\Bbb{R})$ in Definition
  \ref{def:1} leads to $\WF(GT) = \emptyset$, which is equivalent to
  regularity of $GT$. 

  Item \ref{item:15}. From Equation \eqref{eq:35} we know that $G$ is
  simply the imaginary part of the two-point function. Together with
  Proposition \ref{prop:1} this implies $\WF(G) \subset
  \WF(\scr{W}^{(2)})$. Hence we see by Lemma \ref{kor:distri-extend} 
  and the characterization of Hadamard states in Equation
  \eqref{eq:hadamard} that $G(T \otimes S)$ exists and can be defined by
  $G(T \otimes S) = \lim_{k,l\to\infty} G(T_k \otimes S_l)$, where $\Bbb{N} \ni k
  \mapsto \ni T_k \in \scr{D}(M, \Bbb{R}), \Bbb{N} \ni l \mapsto S_l \in \scr{D}(M,
  \Bbb{R})$ are sequences converging in $\scr{D}'_{\WF(T)}(M)$
  respectively $\scr{D}'_{\WF(S)}(M)$ to $T$ and $S$. On the other
  hand weak continuity of the map $\scr{E}'(M,\Bbb{R}) \ni T \mapsto G(T) \in 
  \scr{D}'(M,\Bbb{R})$ implies that  
  \begin{displaymath}
    \lim_{k \mapsto \infty} \int_M G(T_k)(x)S_l(x) \lambda(x) = \int_M G(T)(x) S_l(x) \lambda(x)
  \end{displaymath}
  holds where $G(T)$ denotes the \emph{smooth} solution of the
  Klein-Gordon equation consider in item \ref{item:14}. 
  Since $S$ is compactly supported, the latter integral equals
  $S\bigl(G(T)\bigr)$ which was to show.

  Item \ref{item:16}. By \ref{item:14} we know that $GT$ is a smooth
  solution of the Klein-Gordon equation with support contained in
  $J^+(\supp T) \cup J^-(\supp T)$. Hence $GT$ has smooth, compactly
  supported initial data $\bigl((GT)\restr \Sigma, [\partial_t (GT)]\restr
  \Sigma\bigr)$ on each Cauchy surface $\Sigma$. This implies together with
  bijectivity of the map $\iota_\Sigma$ from Equation \eqref{eq:42} that $GT =
  Gf$ holds for each $f \in \scr{D}(M,\Bbb{R})$ with $\iota_\Sigma(f) =
  \bigl((GT)\restr \Sigma, [\partial_t (GT)]\restr \Sigma\bigr)$. 
\end{pf}

The last proposition shows that we can define a Weyl element $W(T)
\in \CCR(\scr{P},G)$ for each distribution $T \in
\goth{D}^l(M,\Bbb{R})$. It is therefore natural to associate the
C*-Algebra  
\begin{equation} \label{eq:61}
  \scr{A}^l(\gamma) := C^* \bigl( \{ W(T) \, | \, T \in \goth{D}^l(\gamma) \} \bigr)
  \subset \CCR(\scr{P},G)
\end{equation}
to each $\gamma$ and $l$. Hence the von Neumann algebras $\scr{R}^l(\gamma)$ are
given by $\scr{R}^l(\gamma) = \pi\bigl(\scr{A}^l(\gamma)\bigr)''$, in
analogy to Equation \eqref{eq:45}. To get the algebras
$\scr{M}_\gamma^l(\mu)$ defined in Equation \eqref{eq:36} we only have to
consider $\pi_\gamma^l\bigl(\scr{A}_l(\mu)\bigr)'' = \scr{M}_\gamma^l(\mu)$, where
$\pi_\gamma^l$ denotes the GNS representation of the state $\omega_\gamma^l = \omega \restr
\scr{A}^l(\gamma)$.   
 The fields $\Phi(T)$ and $\Phi_\gamma^l(T)$ can be derived from
the representations $\pi$ and $\pi_\gamma^l$ by the conditions
$\exp\bigl(i\Phi(T)\bigr) = \pi\bigl(W(T)\bigr)$ and
$\exp\bigl(i\Phi_\gamma^l(T)\bigr) = \pi_\gamma^l\bigl(W(T)\bigr)$.

If the subspace $\scr{P}^l(\gamma)$ generated by equivalence classes of
distributions $T \in \goth{D}^l(\gamma)$ is a symplectic subspace of
$\scr{P}$ (i.e. if the restriction of $G$ to $\scr{P}^l(\gamma) \times
\scr{P}^l(\gamma)$ is non-degenerate) we can look at $\omega_\gamma^l$ as a
quasi-free state on the CCR-algebra $\CCR(\scr{P}^l(\gamma),G)$ with a two
point function which can be derived from the Hadamard form given in
Appendix \ref{sec:hadamard}. E.g. for $l=0$ this leads to
\if1\aivsize
\begin{multline*}
  \scr{W}_\gamma^{(0;2)}(f \otimes h) = \\ 
  \lim_{\epsilon\to0} \int_{\Bbb{R}\times\Bbb{R}} \Biggl[\frac{1}{(2\pi)^2}
  \Biggl(\frac{\sqrt{\Delta_\gamma(t_1,t_2)}}{\sigma_\gamma(t_1,t_2)+2i\epsilon(t_1-t_2)
    + \epsilon^2} + 
  V^{(n)}_\gamma(t_1,t_2)\ln\bigl(\sigma_\gamma(t_1,t_2)+2i\epsilon(t_1 - t_2) 
  + \epsilon^2\bigr)\Biggr) \\
  + H_{n,\gamma}(t_1,t_2)) \Biggr] f(t_1)h(t_2)dt_1 dt_2,
\end{multline*}
\else
\begin{multline*}
  \scr{W}_\gamma^{(0;2)}(f \otimes h) = \\ 
  \lim_{\epsilon\to0} \int_{\Bbb{R}\times\Bbb{R}} \Biggl[\frac{1}{(2\pi)^2}
  \Biggl(\frac{\sqrt{\Delta_\gamma(t_1,t_2)}}{\sigma_\gamma(t_1,t_2)+2i\epsilon(t_1-t_2)
    + \epsilon^2} + \\
  V^{(n)}_\gamma(t_1,t_2)\ln\bigl(\sigma_\gamma(t_1,t_2)+2i\epsilon(t_1 - t_2) 
  + \epsilon^2\bigr)\Biggr) \\
  + H_{n,\gamma}(t_1,t_2)) \Biggr] f(t_1)h(t_2)dt_1 dt_2,
\end{multline*}
\fi
where $\sigma_\gamma$, $\Delta_\gamma$, $V^{(n)}_\gamma$ and $H_{n,\gamma}$ are given in terms of
the corresponding functions from Appendix \ref{sec:hadamard},
i.e. $\sigma_\gamma(t_1,t_2) = \sigma\bigl(\gamma(t_1),\gamma(t_2)\bigr)$, $\Delta_\gamma(t_1,t_2) =
\Delta\bigl(\gamma(t_1),\gamma(t_2)\bigr)$, $V^{(n)}_\gamma(t_1,t_2) =
V^{(n)}\bigl(\gamma(t_1),\gamma(t_2)\bigr)$ and $H_{n,\gamma}(t_1,t_2) =
H_{(n,\gamma)}\bigl(\gamma(t_1),\gamma(t_2)\bigr)$. If $\gamma$ is a proper time
parameterized geodesic the expression is simplified by
$\sigma_\gamma(t_1,t_2) = (t_1-t_2)^2$. However even in this case
$\scr{W}_\gamma^{(0;2)}$ (and therefore the fields $\Phi_\gamma^0$) contains
geometric information of space time $(M,g)$, because the functions
$\Delta_\gamma$ and $V^{(n)}_\gamma$ depend on derivatives of $g$ of higher order
along $\gamma$ (cf. the discussion of the fundamental solutions in Appendix
\ref{sec:retadvKG}). Another problem arises from the fact that it is
not clear in general whether $\scr{P}^l(\gamma)$ is really a symplectic
subspace of $\scr{P}$. To demonstrate this consider again $l=0$. In
this case the restriction of $G$ to $\scr{P}^0(\gamma) \times \scr{P}^0(\gamma)$ is
non-degenerate iff the restriction of the function $GT$ to $\Ran \gamma$
does not vanish for $T \not= 0 \in \scr{P}$. It seems to be plausible
that the latter condition is true at least if we are considering only
a very small neighborhood of an event $p \in M$ (such that we are in
some sense ``close'' to Minkowski space). Unfortunately it is
difficult to derive a precise proof from this intuition. If $G$ is
really degenerate on $\scr{P}^l(\gamma)$ the C*-closure of the *-algebra
generated by $W(T)$, $T \in \goth{D}(\gamma,\Bbb{R})$, is not unique and we
need additional information about the embedding of $\scr{A}^l(\gamma)$ into
$\CCR(\scr{P},G)$.  

\section{A simple example in Minkowski space}
\label{sec:minkowski-space}

To make the discussion of the last section more transparent, we will
discuss in the following a simple example in Minkowski\footnote{Please
  note that most of the material presented in this section is quite
  well known \cite{borchers64}. In our context it mainly serves as an
  explicit example for the abstract concepts introduced in the
  previous sections.}
space $(\Bbb{R}^4,g)$. Hence let us introduce some notations
first. The one particle Hilbert space $\scr{K}$ is given by
$\Lin^2(\Bbb{R}^3,d\xi)$ and the commutator function $G$ is $G(f\otimes h) = \Im
\langle \OPS f, \OPS h\rangle$ where%
\footnote{Since we are on Minkowski space it is reasonable to consider
  in this section Schwartz functions and tempered distributions.}   
$\OPS : \scr{S}(\Bbb{R}^4,\Bbb{R}) \to \scr{K}$ is the real linear  
function  
\begin{equation}  
\label{eq:Ef-def}  
  (\OPS f)(\xi) = \frac{1}{\sqrt{(2\pi)^3}} \int_{\Bbb{R}^4} \frac{e^{i 
      (\lambda(\xi) t - \xi \cdot x)}}{\sqrt{\lambda(\xi)}}f(t,x) dt dx. 
\end{equation}
with $\lambda(\xi) = :=\sqrt{|\xi|^2 + m^2}$. The generating functional of the
Minkowski vacuum is given as well in terms of $\OPS$ and has the form
\begin{displaymath}
\scr{P} \ni [f] \mapsto \omega\bigl(W([f])\bigr) = \exp\left( - \frac{1}{4} \| K(f)
  \|^2 \right) = \left\langle\Omega, \exp \bigl( i \Phi_S(Kf) \bigr) \Omega \right\rangle,
\end{displaymath}
where $\Phi_S(Kf)$ denotes the Segal operator on the Fock space
$\scr{F}_+(\scr{K})$. Therefore we get $\Phi(f) = \Phi_S(Kf)$ for the free
field. Alternatively we can write
\begin{displaymath}
  \Phi(f) = \int_{\Bbb{R} \times \Bbb{R}^3} \Phi(t,x) f(t,x) dt dx,
\end{displaymath}
with the \emph{quadratic form}
\begin{displaymath}
  \Phi(t,x) = \frac{1}{\sqrt{(2\pi)^3}} \int_{\Bbb{R}^3} \left( A(\xi) e^{i (\lambda(\xi)
      t + \xi \cdot x)} + A^*(\xi) e^{-i (\lambda(\xi) t - \xi \cdot x)} \right)
  \frac{d\xi}{\sqrt{2 \lambda(\xi)}}. 
  \end{displaymath}
$A^*(\xi)$, $A(\xi)$ denote here the well known creation and annihilation
``operators''%
\footnote{$A^*(\xi)$ is defined only as a quadratic form on $\scr{H} =
  \scr{F}_S(\scr{K})$.}
at $\xi \in \Bbb{R}^3$. 

Let us consider now an inertial observer $\gamma_i : \Bbb{R} \ni t \mapsto (t,0) \in
\Bbb{R}^4$. A distribution $T$ of $\goth{D}^l(\gamma_i)$ can be expressed by
\begin{displaymath}
  T(f) = \sum_{|\alpha| \leq l} \int_\Bbb{R} a_\alpha(t)  \frac{\partial^{|\alpha|}f(t,0)}{\partial x^\alpha} dt   
\end{displaymath}
(cf. Proposition \ref{prop:trans-jet}). Therefore the map $\OPS$ can
be extended to $\goth{D}^l(\gamma_i)$ by
\begin{equation} \label{eq:33}
  \OPS T(\xi) = \sum_{|\alpha|\leq l} (-i)^{|\alpha|}
  \frac{\widehat{a_\alpha}\bigl(\lambda(\xi)\bigr)}{\sqrt{\lambda(\xi)}} \xi^\alpha.  
\end{equation}
This follows easily by calculating the Fourier transform of $T$ and
restricting $\hat T$ to the mass shell $\{ (\lambda(\xi),\xi) \in \Bbb{R}^4 \, | \,
\xi \in \Bbb{R}^3 \}$. This implies for the field operators $\Phi(T) =
\Phi_S(KT)$ and therefore:
\begin{align*}
  \Phi(T) &= \sum_{|\alpha| \leq l} a_\alpha(t) \frac{1}{\sqrt{(2\pi)^3}} \int_{\Bbb{R}^3}
  (-i)^{|\alpha|} \xi^\alpha \left( A(\xi) e^{i \lambda(\xi) t} + A^*(\xi) e^{-i \lambda(\xi) t}
  \right) \frac{d\xi}{\sqrt{2 \lambda(\xi)}} \\
  &= \sum_{|\alpha| \leq l} a_\alpha(t) \frac{\partial^{|\alpha|} \Phi(t,0)}{\partial x^\alpha}.
\end{align*}

The local von Neumann algebras $\scr{R}^l(\gamma_i)$ are given by
\begin{displaymath}
  \scr{R}^l(\gamma_i) = \{ \exp\bigl(i \Phi_S(f) \bigr) \, | \, f \in \scr{K}^l(\gamma_i)
  \}''   
\end{displaymath}
with 
\begin{equation} \label{eq:50}
  \scr{K}^l(\gamma_i) = \{ KT \, | \, T \in \goth{D}^l(\gamma_i,\Bbb{R}) \} \subset \scr{K}.  
\end{equation}
 These subspaces are best studied in polar coordinates
$\Bbb{R}^+ \times S^2 \ni (r,\kappa) \mapsto r\kappa \in \Bbb{R}^3 \setminus \{0\}$. If we identify
$\Lin(\Bbb{R}^3,d^3\xi)$ and $\Lin(\Bbb{R}^+,r^2 dr) \otimes \Lin\bigl(S^2,
d\Omega(\kappa)\bigr)$ (where $d\Omega$ denotes the standard surface element of
$S^2$) with respect to the unitary transformation
$\Lin(\Bbb{R}^3,d^3\xi) \ni f \mapsto \tilde f \in \Lin(\Bbb{R}^+,r^2 dr) \otimes
\Lin(S^2, d\Omega)$ given by $\tilde f(r,\kappa) = f(r\kappa)$ we get the following
lemma 

\begin{lem} \label{lem:1}
  The closure of the real linear subspace $\scr{K}^l(\gamma_i)$ defined in
  Equation \eqref{eq:50} is complex linear and has the form
  \begin{displaymath}
    \overline{\scr{K}^l(\gamma_i)} = \Lin(\Bbb{R}^+, r^2 dr) \otimes \scr{Y}_l,
  \end{displaymath}
  where $\scr{Y}_l \subset \Lin\bigl(S^2,d\Omega(\kappa)\bigr)$ is the (finite
  dimensional) subspace generated by spherical harmonics up to the
  order $l$. 
\end{lem}

\begin{pf}
  Let us consider first the space
  \begin{equation} \label{eq:51}
    \tilde \scr{Z}_j = \{ Z_j(f) \, | \, f \in \scr{S}(\Bbb{R},\Bbb{R})
    \} \subset \Lin^2(\Bbb{R}^+,r^2 dr)
  \end{equation}
  with 
  \begin{displaymath}
    \Bbb{R}^+ \ni r \mapsto (\tilde Z_jf)(r) = \frac{\hat f \bigl( \lambda(r)
      \bigr)}{\sqrt{\lambda(r)}} r^j \in \Bbb{C},
  \end{displaymath}
  where we have used $\lambda(r) = \sqrt{r^2 + m^2}$ in slight abuse of
  notation. $\scr{Z}_j$ is a complex linear subspace of
  $\Lin(\Bbb{R}^+,r^2 dr)$, although only real valued Schwartz
  functions are used in Equation \eqref{eq:51}. To see this, note that
  $\overline{f} = f \iff \overline{\hat f(\lambda)} = \hat f(-\lambda) \ \forall \lambda \in
    \Bbb{R}$ holds for \emph{any} Schwartz function $f \in
  \scr{S}(\Bbb{R},\Bbb{C})$ and that only the restriction
  of $\hat f$ to $\Bbb{R}^+_m := (m,\infty)$ enters in $Z_jf$. Therefore if
  $f \in \scr{S}(\Bbb{R},\Bbb{R})$ we can replace $i \hat f$ by an $h \in
  \scr{S}(\Bbb{R},\Bbb{C})$ such that $h(\lambda) = i \hat f(\lambda)$ and $h(-\lambda)
  = \overline{\hat h(\lambda)} = -i f(-\lambda)$ holds for all $\lambda \in
  \Bbb{R}^+_m$. On the interval $[-m,m]$ we can choose $h$ freely as
  long as it remains smooth and the condition $h(-\lambda) = \overline{\hat
    h(\lambda)}$ is satisfied. Hence we get $Z_j \check h = i \hat f$ which
  shows that $\scr{Z}_j$ is complex linear.

  Consider now the unitary operator $U : \Lin^2(\Bbb{R}^+,r^2dr) \ni h \mapsto
  Uh \in \Lin^2(\Bbb{R}^+_m,d\lambda)$ with
  \begin{displaymath}
    (Uh)(\lambda) = \sqrt{\lambda r(\lambda)} h\bigl(r(\lambda)\bigr) \ \mbox{and} \ r(\lambda) =
    \sqrt{\lambda^2-m^2}. 
  \end{displaymath}
  Combining it with $\tilde Z_j$ we get the map $\scr{S}(\Bbb{R}) \ni f
  \mapsto Z_j f \in \Lin^2(\Bbb{R}^+_m, d\lambda)$ with $(Z_jf)(\lambda) = r(\lambda)^{j+1/2}
  \hat f(\lambda)$. Hence we have to show that $\{ Z_j f \, | \, f \in
  \scr{S}(\Bbb{R}) \}$ is dense in $\Lin^2(\Bbb{R}^+_m,d\lambda)$. To this
  end note first that $\{ h \restr \Bbb{R}^+_m \, | \, h \in \scr{S}(\Bbb{R})
  \}$ is dense in $\Lin^2(\Bbb{R}^+_m,d\lambda)$, because $\scr{S}(\Bbb{R})$
  is dense in $\Lin^2(\Bbb{R},d\lambda)$. Therefore it is sufficient to
  approximate an arbitrary $h \restr \Bbb{R}^+_m$, $h \in
  \scr{S}(\Bbb{R})$ by a sequence $\Bbb{N} \ni k \mapsto Z_j f_k$ with $f_k \in
  \scr{S}(\Bbb{R})$. With a sequence $k \mapsto b_k$ of smooth, positive
  functions satisfying $b_k(\lambda) = 0$ for $m \leq \lambda \leq m+1/k$ and $b_k(\lambda) =
  1$ for $\lambda > m+2/k$ we can choose $f_k = (b_k r^{-j-1/2} h)^{\lor} \in 
  \scr{S}(\Bbb{R})$. Hence we get 
  \begin{displaymath}
    \| h - Z_j f_k \|^2 = \int_m^\infty | h(\lambda) - b_k(\lambda) h(\lambda) |^2 d\lambda \leq \frac{2}{k}
    \sup_{\lambda \in \Bbb{R}} |h(\lambda)|^2.
  \end{displaymath}
  Since $h \in \scr{S}(\Bbb{R})$ it is bounded, which implies $Z_j f_k \to
  h$, or in other words: $\tilde \scr{Z}_j$ is dense in
  $\Lin^2(\Bbb{R}^+,r^2 dr)$.

  The next step concerns the space 
  \begin{equation}
    \scr{Z}_j = \{ Z_j(f) \, | \, f \in \scr{D}(\Bbb{R},\Bbb{R})
    \} \subset \tilde \scr{Z}_j.
  \end{equation}
  To  prove that it is dense in $\tilde \scr{Z}_j$, and therefore
  dense in $\Lin^2(\Bbb{R}^+,r^2 dr)$ as well, we choose a polynomial
  $p(\lambda)$ such that $r(\lambda)^{j+1/2} \leq p(\lambda)$ holds for all $\lambda \in
  \Bbb{R}^+_m$. With $f \in \scr{S}(\Bbb{R},\Bbb{R})$ and $h \in
  \scr{D}(\Bbb{R},\Bbb{R})$ we get
  \begin{align}
      \| Z_j f - Z_j h\| &= \int_m^\infty | \hat f(\lambda) - \hat h(\lambda) |^2
      r(\lambda)^{2j+1} d\lambda \notag \\
      &\leq \int_\Bbb{R} \big| p(\lambda) \bigl(\hat f(\lambda) - \hat h(\lambda)\bigr) \big|^2
      \notag \\
      &\leq \int_\Bbb{R} | p(D) f(x) - p(D) h(x) |^2 dx, \label{eq:52}
  \end{align}
  where $p(D)$ is the partial differential operator with $p$ as its
  symbol. Since $\scr{D}(\Bbb{R},\Bbb{R})$ is dense in
  $\scr{S}(\Bbb{R},\Bbb{R})$ with respect to each Sobolev norm we can
  choose for all $f \in \scr{S}(\Bbb{R},\Bbb{R})$ and all $\epsilon > 0$ a $h \in
  \scr{D}(\Bbb{R},\Bbb{R})$ such that the integral in Equation
  \eqref{eq:52} is smaller than $\epsilon$. Hence  $\scr{Z}_j$ is dense in
  $\Lin^2(\Bbb{R},r^2 dr)$. 

  Now consider $\scr{K}^l(\gamma_i) = \bigoplus_{j=0}^l \scr{Z}_j \otimes \tilde
  \scr{Y}_j$, where $\tilde\scr{Y}_j \subset \Lin^2\bigl(S^2, d\Omega(\kappa)\bigr)$
  denotes the space generated by restrictions of $j^{\rm th}$ order
  homogeneous polynomials to $S^2 \subset \Bbb{R}^3$. Up to now we have
  shown that 
  \begin{displaymath}
    \overline{\scr{K}^l(\gamma_i)} = \bigoplus_{j=0}^l \Lin^2(\Bbb{R}^+,r^2 dr) \otimes
    \overline{\tilde \scr{Y}_j} = \Lin^2(\Bbb{R}^+,r^2 dr) \otimes
    \left(\bigoplus_{j=0}^l \overline{\tilde \scr{Y}_j}\right).
  \end{displaymath}
  However the space $\bigoplus_{j=0}^l \tilde \scr{Y}_j$ is generated by
  restrictions of (not necessarily homogeneous) $j^{\rm th}$ order
  polynomials to $S^2$, which is known to coincide with $\scr{Y}_l$
  (see \cite[23.38.3.4]{deng7}).
\end{pf}  

This result immediately implies the following statements concerning
the algebras $\scr{R}^l(\gamma_i)$.

\begin{thm}
  Consider the free scalar field in Minkowski space $(\Bbb{R}^4,g)$
  and the inertial observer with worldline $\gamma_i: \Bbb{R} \ni t \mapsto (t,0) \in
  \Bbb{R}^4$. 
  \begin{enumerate}
  \item 
    The family of von Neumann algebras $\scr{R}^l(\gamma_i)$, $l \in \Bbb{N}_0$
    is strictly increasing.
  \item 
    The inductive limit $\scr{R}^\infty(\gamma_i) = \left(\bigcup_{l \in \Bbb{N}_0}
      \scr{R}^l(\gamma_i)\right)''$ coincides with $\scr{B}(\scr{H})$.
  \end{enumerate}
\end{thm}

\begin{pf}
  This is an immediate consequence of Lemma \ref{lem:1} and the fact
  that the map $\scr{K} \ni f \mapsto \exp\bigl( i \Phi_S(f) \bigr)$ is strongly
  continuous. 
\end{pf}

Another consequence of Lemma \ref{lem:1} concerns the subspace
$\scr{P}^l(\gamma_i)$ of $\scr{P}$ (cf. Section \ref{sec:frskf-gen}). It is
easy to show that $\scr{P}^l(\gamma_i)$ is a symplectic subspace of
$\scr{P}$.  

\begin{prop}
  The set $\scr{P}^l(\gamma_i) = \{ [T] \in \scr{P}^l(\gamma_i) \, |  \, T \in
  \goth{D}^l(\gamma_i) \} \subset \scr{P}$ is a symplectic subspace of $\scr{P}$.   
\end{prop}

\begin{pf}
  According to Lemma \ref{lem:1} the range of the map
  \if1\aivsize
  $\OPS :  \scr{P}^l(\gamma_i) \to \Lin^2(\Bbb{R}^+,r^2 dr) \otimes \scr{Y}_l$
  \else
  \begin{displaymath}
    \OPS :  \scr{P}^l(\gamma_i) \to \Lin^2(\Bbb{R}^+,r^2 dr) \otimes \scr{Y}_l
  \end{displaymath}
  \fi
  is dense. This implies that each $T \in \scr{P}^l(\gamma_i)$ and each $\epsilon >
  0$ admit a $S \in \scr{P}^l(\gamma_i)$ with $\| i \OPS(T) - \OPS(S) \| \leq
  \epsilon$. On the other hand we know that $\OPS$ is symplectic; hence: 
  \begin{displaymath}
    | G(T,S) - \Im \langle \OPS(T), i \OPS(T) \rangle | = | \Im \langle \OPS(T), \OPS(S) - i
    \OPS(T) \rangle | \leq \epsilon \| \OPS(T) \|. 
  \end{displaymath}
  Since $T \not= 0$ implies $\Im \langle\OPS(T), i \OPS(T) \rangle = \| \OPS(T) \|^2 >
  0$ we can choose $\epsilon >0$ and $S$ in such a way that $|G(T,S)| > 0$
  holds, which was to show.
\end{pf}

The last proposition shows that the C*-algebras $\scr{A}_l(\gamma_i)$ are
defined as CCR-algebras of the symplectic spaces $\scr{P}^l(\gamma_i)$. Hence
they do not depend on the embedding $\scr{A}_l(\gamma_i) \subset \CCR(\scr{P},G)$
as it was discussed in the last Section.

\section{Time translations}
\label{sec:time-translations}

At the end of this paper we want to apply the quasi free description
described in Section \ref{sec:frskf-gen} to generalize space-time
translations of Minkowski space theories. To this end recall first
that in the special relativistic case just discussed there is a
representation $\Bbb{R}^4 \ni v \mapsto \alpha_v$ of $\Bbb{R}^4$ by *-automorphisms
of the CCR-algebra $\CCR(\scr{P},G)$, where the $\alpha_v$ are given in
terms of Bogolubov transformations by $\alpha_v\bigl(W(f)\bigr) =
W\bigl(f(\,\cdot\,-v)\bigr)$. If $g(v,v) = -1$ and $\gamma_i$ is the worldline of
the inertial observer with four velocity $v$, it is easy to see that
$t \mapsto \alpha_{tv}$ provides a one-parameter group of automorphisms of
$\scr{A}_l(\gamma_i)$, which can be interpreted physically as follows:
Consider a subcurve $\mu = \gamma_i\restr (a,b)$ and a self adjoint element $A
\in \scr{A}_l(\mu)$ describing an observable measurable by the observer
$\gamma_i$ in the time interval $(a,b)$. During the measurement $\gamma_i$ and his
measuring equipment is at rest in the global inertial frame described
by the flow $\Bbb{R} \times \Bbb{R}^4 \ni (t,x) \mapsto x + tv \in \Bbb{R}^4$, or 
equivalently by the vector field $Z \equiv v$. The observable $\alpha_{tv}(A) \in
\scr{A}_l(\mu+t)$ with $\mu+t = \gamma\restr (a+t,b+t)$ represents \emph{the
  same measurement}, i.e. it is carried out by the same observer $\gamma$,
with the same equipment, in the same inertial frame $Z$ -- but
\emph{$t$ time units later}.

On a space-time without non-trivial symmetries this construction does
not work, although the physical interpretation just given would still
make sense. The following Theorem of Demoen et. al. \cite{demoen77}
can be used to get (under some conditions) at least a family of
completely positive maps between algebras $\scr{A}_l(\gamma)$.  

\begin{thm} \label{thm:3}
  Consider a real symplectic space $(\scr{S},\goth{s})$, the
  corresponding CCR-algebra $\CCR(\scr{S},\goth{s})$ and a, not
  necessarily symplectic, linear map $L: \scr{S} \to \scr{S}$.
  \begin{enumerate}
  \item 
    $\goth{s}_L(x,y) = \goth{s}(x,y) - \goth{s}(Lx,Ly)$ defines an (in
    general degenerate) antisymmetric bilinear form on $\scr{S}$. The
    corresponding CCR-algebra (equipped with the minimal C*-norm;
    cf. \cite{manuceau73}) will be denoted by
    $\CCR(\scr{S},\goth{s}_L)$.
  \item 
    For each state $\rho$ on $\CCR(\scr{S},\goth{s}_L)$ there is a
    completely positive map $\alpha_L: \CCR(\scr{S},\goth{s}) \to
    \CCR(\scr{S},\goth{s})$ with $\alpha_L\bigl(W(x)\bigr) = \rho\bigl(W(x)\bigr)
    W(Lx)$ for all $x \in \scr{S}$.
  \end{enumerate}
\end{thm}

To get a useful generalization of the time translations $\alpha_{ta}$ in
terms of this theorem we have to specify how distributions $T \in
\goth{D}^l(\gamma)$ should be pushed forward along $\gamma$. Hence consider 
again a generic (globally hyperbolic) space time $(M,g)$ 
and an arbitrary \emph{reference frame}, i.e. a timelike vector field 
$Z$ normalized to $-1$ ($G(Z,Z) = -1$), with flow $(t,p) \mapsto F_t(p)$ and
assume that $\gamma$ is one of its integral curve. For each subcurve $\mu \subset
\gamma$ and each admissible\footnote{This means that $(t,\mu(s))$ should be
  for all $s \in \Dom(\mu)$ in the domain of the flow $F$. In the
  following we will assume this implicitly if nothing else is stated.}
$t \in \Bbb{R}$ with $\mu + t := F_t \circ \mu \subset \gamma$ the $l$-jet extension
$J^lF_t$ of the flow leads to a bundle map $(J^lF_{-t})^*: \JetC(\mu) \to
\JetC(\mu + t)$ (see Appendix \ref{sec:jetbun} for details). Hence if we
define the domain (which depends in contrast to the map $\Xi^l_{\gamma,Z}$
below \emph{not} on $Z$)
\begin{displaymath}
  \Dom(\Xi^l_\gamma) = \{ (t,T) \in \goth{D}^l(\gamma,\Bbb{R}) \, | \, F_t(\supp
  T) \subset \Ran \gamma \}
\end{displaymath}
we get the following one-parameter family of linear maps
\begin{equation} \label{eq:1}
  \Dom(\Xi^l_\gamma) \ni (t,T_\psi) \mapsto \Xi^l_{\gamma Zt} \cdot T_\psi := T_{(J^lF_{-t})^*\psi} \in
  \goth{D}(\gamma,\Bbb{R}).   
\end{equation}
Here $\psi \in \scr{D}\bigl(\Dom \gamma,\JetC(\mu)\bigr)$ is a section which
satisfies the condition from Corollary \ref{kor:1}\footnote{This is
  necessary to get a well defined map, cf. the discussion at the end
  of Section \ref{sec:distri-curves}.} and $T_\psi \in
\goth{D}^l(\gamma,\Bbb{R})$ denotes the distribution which is defined by  
$\psi$ according to Theorem \ref{thm:jet2distri}.

Note however that $\Xi^l_{\gamma Zt} T_\psi$ does not coincide with
the usual push forward $F_{t*}T_\psi$ of $T_\psi$ with $F_t$. This can be
seen most easily if we recall the fact that a regular distribution is
basically not a smooth function but a smooth $n$-form $\lambda$ (with $n =
\dim M$). If $\lambda_g$ denotes the volume element associated to the metric
and $f \in \scr{E}(M)$, the usual push-forward of $f \lambda_g$ is given by 
$(F_{t*} f) (F_{t*} \lambda_g)$, while the natural generalization of the
construction given in Equation (\ref{eq:1}) to regular distributions
would leave $\lambda_g$ fixed, i.e. $f \lambda_g \mapsto = (F_{t*}f) \lambda_g$. In
other words the definition of $\Xi^l_{\gamma Zt}$ depends on a
distinguished volume form (namely $\lambda_g$) while $F_{t*}$ is an
invariant construction. However since $\lambda_g$ is a natural object on a
Lorentzian manifold, this is not a significant drawback. On the other
hand we have the advantage that $\Xi^l_{\gamma Zt} T$ for $T \in \goth{D}^l(\gamma)$
depends only on $j^l F_t \restr \gamma$, while $F_{t*} T$ would
depend on $j^{l+1}F_t \restr \gamma$ (since the determinant of the
differential of $F_t$ comes in).

This last remark leads us directly to another point which is worth to
note: The map $T \mapsto \Xi^l_{\gamma Zt} T$ depends as just stated only on
$l$-jets of $F_t$ along $\gamma$, or in other words only the $l$-jet of $Z$
along $\gamma$ is relevant for the construction. This implies in particular
that in the case $l=0$ there is a unique map $\Dom(\Xi^0_\gamma) \ni (t,T) \mapsto
\Xi^0_{\gamma t} \in \goth{D}(\gamma,\Bbb{R})$ with $\Xi^0_{\gamma Zt} = \Xi^0_{\gamma t}$ for all $Z$
admitting $\gamma$ as an integral curve. Hence in the case $l=0$ the
construction do not depend on the reference frame. If $l=1$ holds this
not true, however there is a natural choice: $\Xi^1_{\gamma Zt}$ depends in
this case on one-jets of $F_t$, i.e. the tangent maps $T_{\gamma(s)}
F_t$. Hence it is natural to assume that $T_{\gamma(s)} F_t : T_{\gamma(s)} M \to
T_{\gamma(s+t)}M$ coincides with parallel transport along $\gamma$. Now we need
the following lemma to proceed. 

\begin{lem} \label{lem:2}
  Consider an analytic space-time $(M,g)$ and a smooth (i.e. not
  necessarily analytic) timelike, normalized vector field $Z$ with
  flow $(t,p) \mapsto F_t(p)$. For each point $p$ of $(M,g)$ there is an $\epsilon
  > 0$ such that the following statement holds: For each proper
  subcurve $\mu$ of $(-\epsilon,\epsilon) \ni s \mapsto \gamma(s) := F_s(p) \in M$, the kernel of the
  map  
  \begin{equation} \label{eq:47}
    \goth{D}^l(\mu,\Bbb{R}) \ni T \mapsto GT \in \scr{E}(M).    
  \end{equation}
  is given by (setting $\goth{D}^j(\mu,\Bbb{R}) = \{0\}$ for $j < 0$)
  \begin{equation} \label{eq:55}
    \kappa(\mu,l) = \{ (\Box - m^2) T \, | \, T \in \goth{D}^{l-2}(\mu,\Bbb{R}) \} .    
  \end{equation}
\end{lem}

\begin{pf}
   From $GT = 0$ we get $G^+T = G^-T$ and due to the support properties
  of $G^\pm$ we see that $\supp G^+T \subset I(\mu)$ holds where $I(\mu)$ denotes
  the double cone generated by $\mu$, i.e. $I(\mu) =
  I^+\bigl(F_{-\epsilon}(p)\bigr) \cap I^-\bigl(F_\epsilon(p)\bigr)$. According to
  Theorem \ref{thm:2} we can assume that the restriction of $G^+T =
  G^-T$ to the submanifold  
  \begin{displaymath}
    \Sigma(0,s,\epsilon) := \bigl[H^\pm\bigl(\gamma(s)\bigr) \setminus \{ \gamma(s)\} \bigr] \cap \bigl[ 
    I^+\bigl(\gamma(-\epsilon)\bigr) \cup I^-\bigl( \gamma(\epsilon) \bigr) \bigr]. 
  \end{displaymath}
  defined in equation \eqref{eq:46} is analytic. On the other hand we
  have $G^+T \restr \bigl( \Sigma(0,s,\epsilon) \setminus I(\mu) \bigr) \equiv 0$, hence $G^+T
  \restr \Sigma(0,s,\epsilon) \equiv 0$ due to analyticity and the fact that $\Sigma(0,s,\epsilon)
  \setminus I(\mu)$ is not empty (since $-\epsilon < a < b < \epsilon$). This implies that the
  support of $G^+T$ is contained in $\Ran \mu$. Since $(\Box -m^2) G^+T =
  T$ and $T$ is of order $l$ we get $G^+T \in \goth{D}^{l-2}(\mu,\Bbb{R})$
  (see Definition \ref{def:1}). In other words the kernel of the map
  from Equation \eqref{eq:47} is given by Equation (\ref{eq:55}) as
  stated. 
\end{pf}

To define cp-maps on the algebras $\scr{A}^l(\gamma)$ by the method given
in Theorem \ref{thm:3} we need a vector field $Z$ such that the map
$\Xi^l_{\gamma Zt}: \goth{D}^l(\mu,\Bbb{R}) \to \goth{D}^l(\mu+t,\Bbb{R})$ defined in
Equation (\ref{eq:1}) maps the kernel $\kappa(\mu,l)$ to $\kappa(\mu+t,l)$. To this
end let us give the following definition:

\begin{defi}
  Consider a timelike vector field $Z$ with flow $F_t$ and an integral
  curve $\gamma$ of it. $Z$ is called an \emph{$l^{\rm th}$-order
    infinitesimal symmetry along $\gamma$} if $j^l_{\gamma(s)}(F_{t*}g) =
  j^l_{\gamma(s)}g$ holds for all $t, s-t \in \Dom(\gamma)$. The curve $\gamma$ is
  called in this case \emph{$l^{\rm th}$-order symmetric with respect
    to $Z$}. 
\end{defi}

So, roughly speaking, a vector field is $l^{\rm th}$-order infinitesimal
symmetric along some curve $\gamma$ if the restriction of the $l$-jet
extension of the metric to the range of $\gamma$ is invariant under the
flow of $Z$. It is easy to see that each time-like curve is $0^{\rm
  th}$-order symmetric with respect to an appropriate vector field
which can be constructed locally as follows: Consider the coordinate
map 
\begin{displaymath} 
  \tilde U \ni (t,\mathbf{x}) \mapsto \exp_{\gamma(t)}\left(\sum_{i=1}^3 x^i
    e_i(t)\right) \in U \subset M
\end{displaymath}
defined in Corollary \ref{kor:1} and Equation (\ref{eq:13}).
The basis vector fields $(\partial_\nu)_{\nu=0,\ldots,3}$ associated to this chart
coincide obviously along $\gamma$ with the $e_\nu$, i.e. $\partial_\nu\bigl(\gamma(t)\bigr)
= e_\nu(t)$. Hence the flow $F_t$ of $\partial_0$ has the desired property
$(F_{t*}g)_{\gamma(s)} = g_{\gamma(s)}$. Using the fact that first order
derivatives of the metric (in an appropriate coordinate system)
vanishes along $\gamma$ if it is a geodesic, we can show in the same way
that each geodesic is first order symmetric. Therefore we have just
shown the following proposition

\begin{prop} \label{prop:3}
  Each smooth timelike curve is $0^{\rm th}$ order symmetric and each
  timelike geodesic is first order symmetric.
\end{prop}

With similar arguments we can check now that the kernel $\kappa(\mu,l)$
defined in Equation (\ref{eq:55}) is really invariant under the flow
of infinitesimal symmetric vector fields.

\begin{lem} \label{lem:3}
  Consider again an analytic space-time, a smooth timelike curve $\gamma$
  and a subcurve $\mu \subset \gamma$ such that Lemma \ref{lem:2} holds. For each
  vector field $Z$, infinitesimally symmetric of order $l-1$ along $\gamma$
  and each admissible $t$ we have $\Xi^l_{\gamma Zt} \cdot \kappa(\mu,l) = \kappa(\mu+t,l)$. For
  $l=0,1$ the statement is satisfied for any vector field with $\gamma$ as
  integral curve. 
\end{lem}

\begin{pf}
  If $l=0,1$ is satisfied,  $\kappa(\mu,l)$ is trivial which implies
  immediately the assertion. Hence assume $l\geq2$. Each element $T$ of
  $\kappa(\mu,l)$ has according to Lemma \ref{lem:2} the form 
  \if\aivsize
  \begin{displaymath}
    T(f) = T_\psi \bigl((\Box - m^2) f\bigr)  = \int_\Bbb{R} \psi_{\gamma(t)} \cdot j^{l-2}_{\gamma(t)}
    \bigl( (\Box - m^2)f \bigr) dt = \int_\Bbb{R} \bigl( P_\psi \cdot (\Box - m^2)
    f \bigr)\bigl(\gamma(t)\bigr) dt,
  \end{displaymath}
  \else
  \begin{align*}
    T(f) = T_\psi \bigl((\Box - m^2) f\bigr)  &= \int_\Bbb{R} \psi_{\gamma(t)} \cdot j^{l-2}_{\gamma(t)}
    \bigl( (\Box - m^2)f \bigr) dt \\ &= \int_\Bbb{R} \bigl( P_\psi \cdot (\Box - m^2)
    f \bigr)\bigl(\gamma(t)\bigr) dt,
  \end{align*}
  \fi
  where $\psi \in \scr{D}\bigl(\Ran \gamma, \JetRa{l-2}(\gamma)\bigr)$, $T_\psi$ denotes
  the corresponding distribution in $\goth{D}^{l-2}(\gamma,\Bbb{R})$
  (cf. Theorem \ref{thm:jet2distri}) and $P_\psi$ is the differential
  given by a smooth extension of $\psi$ to a small neighbourhood of $\Ran
  \gamma$ (cf. Proposition \ref{prop:jetbun-1} and Equation
  (\ref{eq:56})). Proposition \ref{prop:2} and Equation (\ref{eq:1})
  imply that 
  \if1\aivsize
   \begin{equation} \label{eq:57} 
    \bigl(\Xi^l_{\gamma Zt} T\bigr)(f) = \int_\Bbb{R} \bigl( F_t^* P_\psi (\Box - m^2) F_{t*} f
    \bigr) \bigl(\gamma(t)\bigr) dt = \int_\Bbb{R} \bigl( F_t^* P_\psi F_{t*}
    F_t^* (\Box - m^2) F_{t*} f \bigr) \bigl(\gamma(t)\bigr) dt
  \end{equation}
  \else
  \begin{align} 
    \bigl(\Xi^l_{\gamma Zt} T\bigr)(f) &= \int_\Bbb{R} \bigl( F_t^* P_\psi (\Box - m^2) F_{t*} f
    \bigr) \bigl(\gamma(t)\bigr) dt \notag \\ 
    &= \int_\Bbb{R} \bigl( F_t^* P_\psi F_{t*} F_t^* (\Box - m^2) F_{t*} f
    \bigr) \bigl(\gamma(t)\bigr) dt \label{eq:57} 
  \end{align}
  \fi
  holds. The Klein-Gordon operator depends on the metric $g$ and its
  first derivatives. Since $P_\psi$ is a differential operator of order
  $l-2$ this implies that the expression in Equation (\ref{eq:57})
  depends on differentials of $g$ up to order $l-1$. However we only
  need to know its value along $\gamma$ and $F_t$ leaves by assumption the
  $l-1$ jet of $g$ along $\gamma$ invariant. Therefore we get
  \begin{displaymath}
    \bigl(\Xi^l_{\gamma Zt} T\bigr)(f) = \int_\Bbb{R} \bigl( F_t^* P_\psi F_{t*}
    (\Box - m^2) f \bigr) \bigl(\gamma(t)\bigr) dt = T_{(J^{l-2}F_t) \circ \psi \circ
      F_{-t}} \bigl( (\Box-m^2)f \bigr),
  \end{displaymath}
  and therefore $\Xi^l_{\gamma Zt} T \in \kappa(\mu + t,l)$ which was to show.
\end{pf}

Now we are ready to prove the following theorem, which is the main
result of this section.

\begin{thm} \label{thm:1}
  Consider an analytic space-time $(M,g)$ and a smooth (i.e. not
  necessarily analytic) curve $\gamma$. Each $s \in \Dom \gamma$ admits a subcurve
  $\mu \subset \gamma$ with $s \in \Dom \mu$ such that the following statements hold:
  \begin{enumerate}
  \item \label{item:3}
    If $l \geq 2$ there exists for each $(l-1)^{\rm th}$ order infinitesimal
    symmetry along $\gamma$ and each admissible $t \in \Bbb{R}$ a functional
    $\overline{\alpha}^l_{\gamma Zt} : \goth{D}^l(\gamma,\Bbb{R}) \to \Bbb{C}$ and a
    cp-map $\alpha^l_{\gamma Zt} : \scr{A}^l(\mu) \to \scr{A}^l(\mu+t)$ with
    $\alpha^l_{\gamma Zt} \cdot W(T) = \overline{\alpha}^l_{\gamma Zt}(T)
    W\bigl(\Xi^l_{\gamma Zt} T\bigr)$. 
  \item \label{item:4}
    If $l=0,1$ the same is true for any vector field with $\gamma$ as an
    integral curve.
  \item \label{item:5}
    If $l=0$ the construction is independent of $Z$. Hence we will
    write $\alpha^0_{\gamma t}$ respectively $\overline{\alpha}^0_{\gamma t}$ instead of
    $\alpha^0_{\gamma Zt}$ and $\overline{\alpha}^0_{\gamma Zt}$.
  \end{enumerate}
\end{thm}

\begin{pf}
  According to Lemma \ref{lem:3} the linear map $\scr{P}^l(\mu) \ni
  [T] \mapsto [\Xi^l_{\gamma Zt} T] \in \scr{P}^l(\mu + t) \subset \scr{P}$ is well
  defined. If $\scr{P}^l(\mu)$ is a \emph{symplectic} subspace of
  $\scr{P}$ Theorem \ref{thm:3} implies immediately the
  assertion. However we can not assume in general that $G$ is
  non-degenerate on $\scr{P}^l(\mu)$ (cf. the corresponding discussion
  in Section \ref{sec:frskf-gen}). Hence we need an additional
  argument: By Zorns lemma we can find an algebraic complement $\tilde
  \scr{P}$ of $\scr{P}^l(\mu)$ in $\scr{P}$ and we can extend the linear
  map just defined by  $\scr{P}^l(\mu) \oplus \tilde \scr{P} \ni (T,S) \mapsto F_{t*}T \in
  \scr{P}$ to $\scr{P}$. Now Theorem \ref{thm:3} applies, which proves
  item \ref{item:3} and \ref{item:4}. Item \ref{item:5} is then an
  immediate concequence of the fact that $\Xi^0_{\gamma t}$ is independent of
  $Z$ as well (cf. the discussion after before Lemma \ref{lem:2}).
\end{pf}

\section{Conclusions}
\label{sec:concl}

Let us discuss now some applications of our results to the description
of observer dependent effects in quantum field theory. Maybe the most
important one is particle creation due to acceleration, or more
generally: observer dependent particle concepts. To get an idea how to
proceed in this direction, restrict, for simplicity, the analysis to
the case $l=0$ and consider again the free field on Minkowski space
$(\Bbb{R}^4,g)$ together with the inertial observer $\gamma_i$ from the
last section. The subgroup    
$\Bbb{R}^4 \ni               
(x^0,\mathbf{x}) \mapsto F_t(x^0,\mathbf{x})   
:= (x^0+t,\mathbf{x}) \in \Bbb{R}^4$             
of the translation group is represented in a canonical way by     
Bogolubov automorphisms $\alpha_t$ of $\CCR(\scr{P},G)$. The   
C*-algebras $\scr{A}^0(\gamma)$ are on the other hand invariant under the  
$\alpha_t$ and $\alpha_t \restr \scr{A}^0(\gamma)$ coincides with 
$\alpha^0_{\gamma_it}$ if we choose $\overline{\alpha}^0_{\gamma_it} \equiv 1$ 
(cf. Theorem \ref{thm:1}).  The free field vacuum $\omega_0$ is (obviously)
a ground state with respect to the group $\alpha_t$. Hence the restriction
of $\omega_0$ to $\scr{A}^0(\gamma_i)$ is a ground state with respect to the one
parameter group $\alpha^0_{\gamma_it}$. In a  similar way $\omega_0$ becomes a
KMS-state with respect to $\alpha^0_{\gamma_at}$ if $\gamma_a$ is a uniformly
accelerating observer; this is a consequence of the well known result
of Bisognano and Wichmann \cite{bisognano75}. 

This observation suggests that a reasonable generalization of notions
like ground state, KMS-state and particle should be based on
properties of the cp-maps $\alpha^0_{\gamma t}$. The most naive approach is 
to consider for an arbitrary observer $\gamma$ and a state $\omega$ in a general
space-time $(M,g)$ the functions
\begin{displaymath}
\Bbb{R} \ni \mapsto \omega\bigl(A\alpha^0_{\gamma t}(B)\bigr) \in \Bbb{C}, \quad A,B \in
\scr{A}^0(\gamma) 
\end{displaymath}
and to generalize the original definitions of ground and KMS states in
the most direct way, e.g. to look at analyticity properties of these
functions. However this is most probable to naive. It is in particular
not very likely that an observer with non-constant acceleration really
sees a thermal particle spectrum (with constant temperature) during a
finite measuring interval. Maybe more realistic is to assume that the
properties of ground and KMS-states can be retained only in an
approximation, e.g. some kind of Taylor developments around each
instant of proper time of the observer. 

In this context it is interesting that we can use the idea of the
proof of Theorem \ref{thm:1} to identify $\alpha^0_{\gamma t}$ with
$\alpha^0_{\gamma_it} (=\alpha_t)$ at least in a certain local sense. More precisely
consider the cp maps (we ignore temporarily the restriction to  
small pieces of the curves, wich is necessary in the proof of Theorem
\ref{thm:1})
\begin{displaymath}
  \scr{A}_0(\gamma) \ni W(T) \mapsto \beta^0_{\gamma\gamma_i}\bigl(W(T)\bigr) =
  \overline{\beta}^0_{\gamma\gamma_i}(T) W(T) \in \scr{A}^0(\gamma_i) 
\end{displaymath}
and
\begin{displaymath}
  \scr{A}^0(\gamma_i) \ni W(T) \mapsto \beta^0_{\gamma_i\gamma}\bigl(W(T)\bigr) =
  \overline{\beta}^0_{\gamma_i\gamma}(T) W(T) \in \scr{A}^0(\gamma), 
\end{displaymath}
defined according to Theorem \ref{thm:3} by an obvious identification
of distributions from $\goth{D}^0(\gamma)$ and $\goth{D}^0(\gamma_i)$, and appropriate
functionals $\overline{\beta}^0_{\gamma_i\gamma}$, $\overline{\beta}^0_{\gamma\gamma_i}$. Hence if
we set 
\begin{displaymath}
  \overline{\beta}^0_{\gamma_i\gamma}(T+t) \overline{\beta}^0_{\gamma\gamma_i}(T) = \overline{\alpha^0_{\gamma t}},
\end{displaymath}
$\scr{A}^0(\gamma) \ni A \mapsto \beta^0_{\gamma_i\gamma} \circ \alpha_t \circ \beta^0_{\gamma\gamma_i} \in \scr{A}(\gamma)$
is a cp-map of the form discussed in Theorem \ref{thm:1}. This implies
that we can investigate properties of the state $\omega$ on $\scr{A}^0(\gamma)$
by considering $\omega \circ \beta^0_{\gamma_i\gamma}$ on $\scr{A}^0(\gamma_i)$, which is a great 
advantage because the $\alpha_t$ form in contrast to the $\alpha^0_{\gamma t}$ an
automorphism group which is easier to study. In addition we have the
possibility to compare $\omega$ directly with the Minkowski vacuum
$\omega_0$. This requires of course a detailed study of $\overline{\alpha}^0_{\gamma
  t}$ and $\overline{\beta}^0_{\gamma_i\gamma}$. However this task is simplified by
the fact that both functionals are, according to Theorem
\ref{thm:3}, generating functionals of states on appropriate CCR
algebras. 

A different complex of questions to which these ideas can be applied
as well concerns averaged energy conditions, which are considered
recently in a number of papers (see e.g. \cite{fewster00a,verch00} and
the references therein). The problem dicussed there arises from the
fact that the expectation value of the energy momentum tensor of a
quantum field does not satisfy the usual energy conditions, which
breaks many thoerems in general relativity (e.g.singularity theorems)
. In some situations however positivity of energy is not required
at each event of space-time, but only averaged along timelike or null
curves. At least in the timelike case the quantum fields $\Phi^l\gamma$ and
the corresponding algebras $\scr{A}^l(\gamma)$ provide an optimal framework
to study this kind of problem. This is in particular the case for the
scheme outlined in the last paragraph: The identification of each
$\scr{A}^l(\gamma)$ with the Minkowski space algebra $\scr{A}(\gamma_i)$ given
by the cp-maps $\beta_{\gamma\gamma_i}$ and $\beta_{\gamma_i\gamma}$ might lead to an easy
generalization of results wich are currently only available in the
flat case. 

Another interesting application concerns the question whether an
observer $\gamma$ can determine its acceleration by quantum
measurements. More precisely: is it possible for $\gamma$ (and how) to
decide (at least) whether he is in free fall or not by measuring
observables from $\scr{A}^l(\gamma)$. In Minkowski space we can pose the
related but much simpler question, how to distinguish between an
inertial observer $\gamma_i$ and a uniformly accelerated one
$\gamma_a$. According to our discussion at the beginning of this section,
one possible answer is to count particles in the Minkowski vacuum. The
temperature of the observed spectrum (which should be thermal of
course) is a direct measure for the acceleration. In the general case
however this scheme is not applicable, because there is no
distinguished global state which we can use as a reference. Hence the
observer $\gamma$ (possibly together with ``helpers'' traveling on
neighboring wordlines) has to perform a more complex protocol 
consisting of several steps, e.g.: First prepare some states locally
by distinguished procedures and then count particles. In any case
however the knowledge of approximate ground and KMS states, as
outlined above, seems to be necessary.

A closely related, but slightly different question is: Is it possible
to determine the space-time uniquely (up to isomorphisms) by a quantum
field theory? At least a partial answer is available can be found in
\cite{KEYL94,KEYL98a} and \cite{wollenberg98} (see also
\cite{buchholz98} for a completely different approach)
In these papers it is shown (using commutation relations in
\cite{KEYL94,KEYL98a} and algebras of curves in \cite{wollenberg98})
that the conformal structure of space-time can be fixed uniquely by a
net of local C*-algebras under quite general conditions. Hence the
discussion of the last paragraph, the knowledge of all geodesics in
space-time, provides exactly the missing information. However the
result of Theorem \ref{thm:1} might lead to a simpler approach:
According Proposition \ref{prop:3} each timelike geodesic is first
order symmetric. The converse however is not true, because each
integral curve of a timelike Killing vector field is $l^{\rm th}$ order
symmetric for any $l \in \Bbb{N}$. Hence the question is, whether
conformally equivalent metrics can be distinguished by their sets of
first order symmetric curves. If the answer is true, we can ask
whether first symmetric curves can be characterized by the existence
(and possibly particular properties) of the first order
time-translation maps $\alpha^1_{\gamma t}: \scr{A}^1(\mu) \to \scr{A}^1(\mu + t)$. 

\section*{Acknowledgments}

I like to thank A.~S.~Holevo for pointing out to me reference
\cite{demoen77}. 

\begin{appendix}

\section{Wave front sets of distributions}
\label{sec:wfs}

 In this appendix we will summarize some material about wave front
 sets of distributions used throughout the paper. A detailed
 presentation can be found in Chapter VIII of H{\"o}rmanders book
 \cite{HOER1}. Let us start with a distribution $T$ on
$\Bbb{R}^n$. A pair $(x_0,\xi_0) \in \Bbb{R}^n \times (\Bbb{R}^n \setminus \{0\})$ is
called a \emph{regular directed point} of $T$, if there is a
neighbourhood $U$ of $x_0$ and a conic neighbourhood $V$ of $\xi_0$ such
that for all $f \in \cni{U}{\Bbb{R}}$ and each $N \in \Bbb{N}$ there is a
constant $C_{f,N}$ with\footnote{Note that the Fourier transform of a
  compactly supported distribution is allways smooth.}
\begin{displaymath}
  |\langle T, e^{-i \langle \, \cdot \,,\xi\rangle} f \rangle| \leq \frac{C_{f,N}}{(1+|\xi|)^N} \quad \forall \xi 
  \in V.
\end{displaymath}

\begin{defi} \label{def:2}
  The \emph{wave-front set} of $\WF(T)$ is the set of all $(x_0,\xi_0) \in
\Bbb{R}^n \times (\Bbb{R}^n \setminus \{0\})$ which are not regular directed.
\end{defi}

$\WF(T)$ is closely related to the \emph{singular support} $\singsupp
T$ of $T$. More precisely an element $x_0 \in \Bbb{R}^n$ is in
$\singsupp(T)$ iff there is a $\xi_0 \in \Bbb{R}^n \setminus \{0\}$ with $(x_0,\xi_0)
\in \WF(T)$. In contrast to $\singsupp(T)$ however the wave front set
does not only tell us where the singularity of $T$ is located but also
by which high frequency oscillations it is caused. 

Consider now a smooth manifold $M$. Distributions on $M$ can be
defined, as in the flat case, as continuous linear functionals on the
space $\scr{D}(M)$ of smooth, complex valued and compactly supported
functions on $M$, equipped with the usual topology. However in
contrast to euclidean space there is no \emph{natural} embedding of
$\scr{D}(M)$ into its topological dual $\scr{D}'(M)$. There are
basically two ways to solve this problem: First we can introduce in
addition the space $\scr{D}(M,\Lambda^nT^*M)$ of smooth, compactly
supported, complex densities on $M$ and its dual space $\scr{D}'(M,
\Lambda^nT^*M)$. Now we can embed $\scr{D}(M,\Lambda^nT^*M)$ naturally into
$\scr{D}'(M)$ by mapping a density $\theta$ to the distribution $\theta(f) :=
\int_M f(x) \theta(x)$. In a similar way we can embed $\scr{D}(M)$ into
$\scr{D}'(M,\Lambda^nT^*M)$. Alternatively we can select 
 a distinguished volume $\lambda$ form (this is possible only if $M$ is
 orientable of course) and to identify $\scr{D}(M)$ with 
$\scr{D}(M,\Lambda^nT^*M)$ by the isomorphism $\scr{D}(M) \ni f \mapsto f\lambda \in
\scr{D}(M,\Lambda^nT^*M)$. Since we are considering exclusively orientable
Lorentzian manifolds in this paper a distinguished volume form is
naturally given. Therefore we will use in most cases the second
possibility and interpret $\scr{D}(M)$ as a subspace of
$\scr{D}'(M)$. However we should keep in mind that such an
identification depends basically on the volume form $\lambda$.

Let us come back to wave front sets now. If $(M_u,u)$ is a coordinate
chart of $M$ we can consider the \emph{push forward} forward $\langle u_* T,
f\rangle = \langle T, (f \circ u)\rangle$ of $T \in \scr{D}'(M)$ and get a distribution $u_*T$ 
on $u(M_u) \subset \Bbb{R}^n$, which has the wave front set
$\WF(u_*T_u)$. Using different coordinate systems, it turns out that
the elements of $\WF(u_*T_u)$ transforms like co-vectors
(cf. \cite[Theorem 8.2.4]{HOER1}). Hence we can define $\WF(T)$ as
follows: 

\begin{defi}
  The wave front set $\WF(T)$ of a distribution $T$ on the manifold
  $M$ is the unique subset of the cotangent bundle $T^*M \setminus \{0\}$
  satisfying   
\begin{displaymath}  
  \WF(T) \restr M_u = (T^*u)^{-1} \WF(u_*T_u), \quad \forall \
  \mbox{coordinate charts} \ (M_u,u). 
\end{displaymath}  
Here $(T^*M_u,T^*u)$ denotes the coordinate chart of $T^*M$ induced  
by $(M_u,u)$. 
\end{defi}

The notion of wave front set allows us to extend a number of
operations to distributions. Of special importance for us are
\emph{products of distributions} and \emph {restrictions to 
  submanifolds}. The basic idea to do that is to extend those
operations in a continuous way from spaces of smooth functions to
distribution spaces. This concept requires however a special form of
convergence because the weak topology is not appropriate. Let us
discuss the euclidean case first, since the generalization to
manifolds is straight forward. 

\begin{defi} \label{def:scrd-gamma-u}
  Consider an open subset $U \subset \Bbb{R}^n$ and a closed cone $\Gamma \subset U \times
  \Bbb{R}^n \setminus \{0\}$ and define
  \begin{equation} \label{eq:17}
    \scr{D}_\Gamma'(U) = \{ T \in \scr{D}'(U) \, | \, \WF(T) \subset \Gamma\}. 
  \end{equation}
  We say that a sequence $\Bbb{N} \ni j \mapsto T_j \in \scr{D}'_\Gamma(U)$ is
  converging in $\scr{D}'_\Gamma(U)$ to $T \in \scr{D}'_\Gamma(U)$ if
  \begin{enumerate}
  \item 
    $j \mapsto T_j$ converges weakly to $T$,
  \item \label{item:9}
    for each $(x_0,\xi_0) \in (U \times \Bbb{R}^n \setminus \{0\}) \setminus \Gamma$ there is a
    function $f \in \scr{D}(U)$ with $f(x_0) \not= 0$ and a conic
    neighbourhood $V$ of $\xi_0$ such that
    \begin{equation} \label{eq:16}
      \lim_{j\to\infty} \sup_{\xi \in V} |\xi|^N | \widehat{fT_j}(\xi) -
      \widehat{fT}(\xi) | \to 0 \ \forall N \in \Bbb{N}_0
    \end{equation}
  \end{enumerate}
\end{defi}

Note that $T \in \scr{D}'_\Gamma(U)$ is equivalent to the following: For each
$(x_0,\xi_0) \in (U \times \Bbb{R}^n \setminus \{0\}) \setminus \Gamma$ there are $f, V$ as above with
\begin{displaymath}
  \sup_{\xi \in V} |\xi|^N |\widehat{fT}(\xi)| < \infty \ \forall N \in \Bbb{N}_0.
\end{displaymath}
This implies especially that each constant sequence is
converging. Note in addition that due to weak convergence
$\widehat{fT_j}$ converges uniformly on each compact set. This implies 
that (\ref{eq:16}) is equivalent to
\begin{equation} \label{eq:20}
  \sup_{j \in \Bbb{N}} \sup_{\xi \in V} |\xi|^N |\widehat{fT_j}(\xi)| < \infty \ \forall N
  \in \Bbb{N}_0. 
\end{equation}

To generalize this definition to distributions on a manifold $M$ we
can use again coordinate representations. If in particular $\WF(T) \subset
\Gamma$ holds for $T \in \scr{D}'(M)$ and a closed cone $\Gamma \subset T^*M \setminus \{0\}$ we
have
\begin{displaymath} 
  \WF(u_*T) \subset T^*u( T^*M_u \cap \Gamma) =: T^*u \cdot \Gamma,
\end{displaymath}
where $(M_u,u)$ is a coordinate chart of $M$ and $(T^*M_u,T^*u)$ the
corresponding coordinate system of $T^*M$. Hence we can define:

\begin{defi} \label{def:scrd-gamma-m}
  Consider in analogy to Equation (\ref{eq:17}) the set
  \begin{displaymath}
    \scr{D}_\Gamma'(M) = \{ T \in \scr{D}'(M) \, | \, \WF(T) \subset \Gamma\},
  \end{displaymath}
  where $M$ is a smooth manifold and $\Gamma \subset T^*M \setminus \{0\}$ a closed cone. A 
  sequence $\Bbb{N} \ni j \mapsto T_j \in \scr{D}_\Gamma'(M)$ converges in
  $\scr{D}_\Gamma'(M)$ to $T \in \scr{D}_\Gamma'(M)$ if $j \mapsto u_*T_j$ converges for 
  each coordinate chart $(M_u,u)$ in $\scr{D}_{T^*u
    \Gamma}\bigl(u(M_u)\bigr)$ to $u_*T$. 
\end{defi}

Now we are ready to discuss restrictions of distributions $T \in
\scr{D}'(M)$ to submanifolds $\Sigma \subset M$. Note in this context that we
need volume forms on $M$ \emph{and} on $\Sigma$ to embed $\scr{D}(M)$ and
$\scr{D}(\Sigma)$ into $\scr{D}'(M)$ respectively
$\scr{D}'(\Sigma)$. However if $(M,g)$ is a Lorentzian manifold and
$\Sigma$ is non-null (i.e. the pull back of $g$ to $\Sigma$ is
nondegenerate) this is no problem, because there 
are canonical choices on $M$ and on $\Sigma$ (the volume forms induced
by the corresponding metrics).

\begin{thm} \label{thm:distri-restr}
  Let $M$ be a manifold and $\Sigma$ a submanifold with normal bundle
  \begin{equation} \label{eq:22}
    N(\Sigma) := \{ \lambda \in T^*M\restr S \, | \, \lambda(v) = 0 \ \forall v \in T_{\pi(\lambda)}\Sigma \}.  
  \end{equation}
  For every distribution $T \in \scr{D}'(M)$ with $\WF(T) \cap N(\Sigma) = \emptyset$
  The restriction $T\restr \Sigma$ can be defined for every distribution $T
  \in \scr{D}'(M)$ with $\WF(T) \cap N(\Sigma) = \emptyset$ in exactly one way such that 
  \begin{enumerate}
  \item 
    it coincides with usual restrictions for $T \in \scr{D}(M)$,
  \item 
    the wave front set of the restriction satisfies
    \if1\aivsize
    \begin{displaymath}
      \WF(T\restr \Sigma) \subset i_\Sigma^*\WF(T) = \{ \theta \in T^*\Sigma \, | \, \exists \lambda \in \WF(T)
      \ \mbox{with} \ \theta(v) = \lambda(Ti_\Sigma v) \ \forall v \in T_{\pi(\theta)}\Sigma \} 
    \end{displaymath}
    \else
    \begin{multline*}
      \WF(T\restr \Sigma) \subset i_\Sigma^*\WF(T) = \\ \{ \theta \in T^*\Sigma \, | \, \exists \lambda \in
      \WF(T) \ \mbox{with} \ \theta(v) = \lambda(Ti_\Sigma v) \ \forall v \in T_{\pi(\theta)}\Sigma \} 
    \end{multline*}
    \fi
  \item 
    and for each closed cone $\Gamma \subset T^*M \setminus \{0\}$ the map $\scr{D}_\Gamma(M) \ni
    T \mapsto T\restr \Sigma \in \scr{D}_{i_\Sigma^* \Gamma}(\Sigma)$ is sequential continuous,
    i.e. this means that for each sequence $\Bbb{N} \ni j \mapsto T_j$
    converging in $\scr{D}_\Gamma(M)$ to $T$ the sequence $j\mapsto T_j\restr \Sigma$
    converges in $\scr{D}_{i_\Sigma^*\Gamma}$ to $T\restr \Sigma$.
  \end{enumerate}
\end{thm}

The proof of this theorem can be found in \cite{HOER1} (see Theorem
8.2.4, cf. also Corollary 8.2.7). Using restrictions of distributions
we are now able to define products as well, because a product is
simply the restriction of the tensor product to the diagonal. Hence we
need the following result about the wave front set of tensor products
(cf. Theorem 8.2.9 of \cite{HOER1}). 
 
\begin{prop} \label{prop:wf-tensprod}
  For $T,S \in \scr{D}'(M)$ we have
  \if1\aivsize
  \begin{displaymath}
    \WF(T \otimes S) \subset \bigl( \WF(T) \times \WF(S)\bigr) \cup \bigl( [\supp(T) \times
    \{0\}] \times \WF(S)\bigr) \cup \bigl(\WF(T) \times [\supp(S) \times \{0\}]\bigr)
  \end{displaymath}
  \else
   \begin{multline*}
    \WF(T \otimes S) \subset \bigl( \WF(T) \times \WF(S)\bigr) \cup \\ 
    \bigl( [\supp(T) \times \{0\}] \times \WF(S)\bigr) \cup \bigl(\WF(T) \times [\supp(S)
    \times \{0\}]\bigr) 
  \end{multline*}
  \fi
  for the wave front set of the tensor product.
\end{prop}

\begin{prop} \label{prop:wf-tensprod-2}
  Consider now two sequences $\Bbb{N} \ni j \mapsto T_j \in \scr{D}_\Gamma'(M)$ and 
  $\Bbb{N} \ni k \mapsto S_k \in \scr{D}_\Theta'(M)$ converging in $\scr{D}_\Gamma'(M)$
  respectively in $\scr{D}_\Theta'(M)$ to $S$ and $T$. The sequence $k \mapsto
  T_k \otimes S_k$ of tensor products converges to $T \otimes S$ in $\scr{D}_{\Gamma \odot
    \Theta}(M)$ where $\Gamma \odot \Theta$ denotes the cone
  \begin{equation} \label{eq:18}
    \Gamma \odot \Theta := (\Gamma \times \Theta) \cup \bigl([M \times \{0\}] \times \Theta\bigr) \cup \bigl(\Gamma \times [M
    \times \{0\}]\bigr).
  \end{equation}
\end{prop}

\begin{pf}
  Since this statement is not explicitly proved in \cite{HOER1}, we
  will give a sketch of a prove here. First of all let us consider
  only the corresponding Euclidean problem, i.e. $M$ is an open subset 
  of $\Bbb{R}^n$ and $\Gamma, \Theta$ are closed cones in $M \times \Bbb{R}^n \setminus
  \{0\}$. According to Definition \ref{def:scrd-gamma-m} it is easy to
  derive the more general statement from this special case.

  We have to check now whether the sequence $j \mapsto T_j \otimes S_j$ satisfies
  the conditions from Definition \ref{def:scrd-gamma-u}. Weak
  convergence to $T \otimes S$ is obvious, which implies that only item
  \ref{item:9} has to be shown. Hence we have to show that for
  $(x_1,x_2;\xi_1,\xi_2) \in \bigl(M^2 \times (\Bbb{R}^{2n} \setminus \{0\})\bigr)  \setminus (\Gamma \odot
  \Theta)$ a function $f \in \scr{D}(M^2)$ with $f(x_1,x_2) \not= 0$ and a
  conic neighbourhood $V \subset \Bbb{R}^{2n}$ of $(\xi_1,\xi_2)$ exist such
  that Equation (\ref{eq:20}) holds. The definition of $\Gamma \odot \Theta$ in
  Equation (\ref{eq:18}) implies that the only nontrivial case is
  $(x_1,\xi_1) \in \Gamma$ and $(x_2,\xi_2) \notin \Theta$ (or vice versa).

  Consider first that $(x_1,\xi_1) \in \Gamma$ holds. Hence we can not assume
  that item \ref{item:9} of Definition \ref{def:scrd-gamma-u} is
  satisied for $j \to T_j$ and $(x_1,\xi_1)$. However $f_1T_j$ and $f_1T$
  are compactly supported distributions for each $f_1 \in
  \scr{D}(M)$, i.e. $f_1T_j, f_1T \in \scr{E}'(M)$. Hence by the
  Paley-Wiener theorem we get constants $C_j$, $N_j$ such that 
  \begin{equation} \label{eq:21}
    | \widehat{f_1T_j}(\eta_1) | \leq C_j (1+|\eta_1|)^{N_j} \ \forall \eta_1 \in \Bbb{R}^n
  \end{equation}
  holds. In addition we have, due to weak convergence, $f_1 T_j(h) \to
  f_1 T(h)$ for $j\to\infty$ and for all $h \in \scr{E}(M)$. Therefore the set
  $\{|f_1T_j(h)| \, | \, j \in \Bbb{N} \}$ is bounded for all $h \in
  \scr{E}(M)$. But $\scr{E}(M)$ is a Fr{\`e}chet space and the uniform
  boundedness principle applies. This means that 
  \begin{displaymath}
    |f_1T_j(h)| \leq C \max_{|\alpha| \leq N} \sup_{x \in K} |D^\alpha h|
  \end{displaymath}
  holds for a compact subset $K \subset \Bbb{R}^n$ and some constants $C, N$
  \emph{independent of $j$}. Applying this inequality to $h(x) =
  \exp(ix\cdot\eta)$ we see that Equation (\ref{eq:21}) holds as well for
  constants $C, N$ independant of $j$.

  Since $(x_2,\xi_2) \notin \Theta$ we can assume that there is a function $f_2 \in
  \scr{D}(M)$ with $f_2(x_2)$ and a conic neighbourhood $V \subset
  \Bbb{R}^n$ of $\xi_2$ such that
  \begin{displaymath}
    \sup_{j \in \Bbb{N}} \sup_{\substack{\eta_2 \in V_2\\|\eta_2|=1}} \sup_{\lambda \in
      \Bbb{R}^+} \lambda^{\tilde{N}} |\widehat{fS_j}(\lambda\eta_2)| < \infty \ \forall
    \tilde{N} \in \Bbb{N}_0 
  \end{displaymath}
  is satisfied. Together with Equation (\ref{eq:21}) and an
  appropriately chosen $V_1$ we get
  \if1\aivsize
  \begin{multline*}
    \sup_{j \in \Bbb{N}} \ \sup_{\substack{(\eta_1,\eta_2) \in V_1\times
        V_2\\|(\eta_1,\eta_2)|=1}} \ \sup_{\lambda \in \Bbb{R}^+} \lambda^{\tilde{N}}
    |\widehat{f_1T_j(\lambda\eta_1)}||\widehat{f_2S_j}(\lambda\eta_2)| \leq 
    \sup_{j \in \Bbb{N}} \sup_{\substack{\eta_2 \in V_2\\|\eta_2|=1}} \sup_{\lambda \in
      \Bbb{R}^+} \lambda^{\tilde{N}} (1 + \lambda^N) |\widehat{f_2S_j}(\lambda\eta_2)| < \infty
  \end{multline*}
  \else
  \begin{multline*}
    \sup_{j \in \Bbb{N}} \ \sup_{\substack{(\eta_1,\eta_2) \in V_1\times
        V_2\\|(\eta_1,\eta_2)|=1}} \ \sup_{\lambda \in \Bbb{R}^+} \lambda^{\tilde{N}}
    |\widehat{f_1T_j(\lambda\eta_1)}||\widehat{f_2S_j}(\lambda\eta_2)| \leq \\
    \sup_{j \in \Bbb{N}} \sup_{\substack{\eta_2 \in V_2\\|\eta_2|=1}} \sup_{\lambda \in
      \Bbb{R}^+} \lambda^{\tilde{N}} (1 + \lambda^N) |\widehat{f_2S_j}(\lambda\eta_2)| < \infty
  \end{multline*}
  \fi
  and this completes the proof.
\end{pf}

Combining the last two propositions with Theorem
\ref{thm:distri-restr} we get the following statement about products
of distributions (cf. Theorem 8.2.10 of \cite{HOER1} and Theorem
2.5.10 of \cite{hoermander71}):

\begin{thm} \label{thm:distri-prod}
  Consider two distributions $T,S \in \scr{D}'(M)$ with the following
  property 
  \if1\aivsize
  \begin{displaymath}
    (x,0) \notin \WF(T) \oplus \WF(S) := \{ (x, k_1 + k_2) \in T^*M \, | \, (x,k_1) \in 
    \WF(T) \ \mbox{and} \ (x,k_2) \in \WF(S) \}.
  \end{displaymath}
  \else
  \begin{multline*}
    (x,0) \notin \WF(T) \oplus \WF(S) := \\ \{ (x, k_1 + k_2) \in T^*M \, | \, (x,k_1) \in 
    \WF(T) \ \mbox{and} \ (x,k_2) \in \WF(S) \}.
  \end{multline*}
  \fi
  Then the product $TS \in \scr{D}'(M)$ can be defined in a unique way
  such that the following condition holds:
  \begin{enumerate}
  \item 
    If $T,S$ are regular, $TS$ coincides with the usual product of
    functions.
  \item 
    For each pair of closed cones $\Gamma, \Theta \subset T^*M \setminus \{0\}$ with $\Gamma \oplus \Theta \subset
    T^*(M \times M) \setminus \{0\}$ the map $\scr{D}_\Gamma'(M) \times \scr{D}_\Theta'(M) \ni (T,S) \mapsto 
    TS \in \scr{D}'(M)$ is sequentially continuous, i.e. if $j \mapsto T_j$
    converges in $\scr{D}_\Gamma'(M)$ to $T$ and $j \mapsto S_j$ converges in
    $\scr{D}_\Theta'(M)$ the sequence of tensor products converges weakly
    in $\scr{D}'(M)$.
  \end{enumerate}
  The wave front set of $TS$ is given by 
  \begin{displaymath}
    \WF(TS) \subset \WF(T) \cup  \WF(S) \cup \bigl( \WF(T) \oplus \WF(S) \bigr).
  \end{displaymath}
\end{thm}

In Section \ref{sec:qf:qf-on-wl} we are mainly interested in limits of
the form $\lim_{k\to\infty} T(S_k)$ where $S_k \to S$ in $\scr{D}_\Theta'(M)$ for
compactly supported $S$ and an appropriate cone $\Theta$. 
\emph{Heuristically} this means that we want to calculate
\begin{equation}\label{eq:9}
  T(S) = \int_M T(x) S(x) \lambda_g = \int_M T(x) S(x) \mathbf{1}(x) \lambda_g =
  (TS)(\mathbf{1}), 
\end{equation}
where $\lambda_g$ denote the volume element on $M$ induced by $g$ and
$\mathbf{1}: M \to \Bbb{R}$ is the function with $\mathbf{1}(x) = 1$, $\forall
x \in M$. Note that the last equality in Equation (\ref{eq:9}) follows
from the fact that $TS$ is compactly supported if $S$ is, and that
$TS$ can, in this case, be extended to a linear functional on
$\scr{E}(M)$. If the sequence $\Bbb{N} \ni k \mapsto S_k \in \scr{D}(M)$
converges as described to $S$ we get (again heuristically)
\begin{multline*}
  (TS)(\mathbf{1}) = \int_M T(x) S(x) \mathbf{1}(x) \lambda_g = \lim_{k\to\infty} 
  \int_M T(x) S_k(x) \mathbf{1}(x) \lambda_g(dx) = \\ 
  \lim_{k \to \infty} \int_M T(x) S_k(x) \lambda_g(dx) = \lim_{k\to\infty} T(S_k).
\end{multline*}
This indicates that $\lim_{k\to\infty}T(S_k)$ exists if the product $TS$
exists and is in this case equal to $TS(\mathbf{1})$ (and therefore
independent from the the sequence $S_k$). The following theorem shows
that this conjecture is true.

\begin{thm} \label{thm:distri-extend}
  Consider two distributions $T \in \scr{D}'(M)$, $S \in \scr{E}'(M)$
  (i.e. $S$ is compactly supported) with the same property as in
  Theorem \ref{thm:distri-prod} ($(x,0) \notin \WF(S) \oplus \WF(T)$) and a
  sequence $\Bbb{N} \ni k \mapsto S_k \in \scr{D}(M)$ converging in
  $\scr{D}_\Theta'(M)$ to $S$, where $\Theta$ satisfies $\WF(T) \oplus \Theta \subset T^*M \setminus \{0\}$. 
  \begin{enumerate}
  \item 
    The product $TS$ exists and is compactly supported. This implies
    $TS \in \scr{E}'(M)$.
  \item 
    The limit $T(S) := \lim_{k\to\infty} T(S_k)$ exists and is equal to
    $TS(\mathbf{1})$, which implies in particular that $T(S)$ does not 
    depend on the sequence $S_k$.
  \end{enumerate} 
\end{thm}

\begin{pf}
  The first statement is an immediate concequence of Theorem
  \ref{thm:distri-prod} and the fact that $S$ is compactly
  supported. To prove the second one consider the sequence $j \mapsto (T,
  S_j) \in \scr{D}_{\WF(T)}'(M) \times \scr{D}_\Theta'(M)$ which converges
  obviously to $(T,S) \in \scr{D}_{\WF(T)}'(M) \times \scr{D}_\Theta'(M)$. Theorem 
  \ref{thm:distri-prod} implies that the sequence of products $j \mapsto
  TS_j$ converges weakly to $TS$. Since the $S_j$ are compactly
  supported and converge to a compactly supported distribution $S$
  there is a compact subset $K$ of $M$ with $\supp S_j \subset K$ for all $j
  \in \Bbb{N}$ and $\supp S \subset K$. Weak convergence of the sequence $j \mapsto
  TS_j$ implies in addition $S_j T(f) \to ST(f)$ for a smooth compactly
  supported function  $f$ with $f(x) = 1$ for all $x \in K$. Hence we
  get $T(S_j) = S_j T(\mathbf{1}) \to ST(\mathbf{1})$ for $j \to \infty$ as
  stated in the theorem.
\end{pf}

For the rest of this appendix we will provide some methods, which are
needed in the paper to calculate some wave front sets. The first one
concerns distributions of the type $\scr{K} f$ where $\scr{K}:
\scr{D}(M) \to \scr{D}'(N)$ is a continuous linear map (and $M,N$ are
manifolds). According to \cite[Theorem 8.2.12]{HOER1} we have

\begin{thm} \label{thm:wf-kern-distri}
  Let $M,N$ be manifolds and $K \in \scr{D}'(N \times N)$. If the
  corresponding linear transformation from $\scr{D}(M)$ to
  $\scr{D}'(N)$ is denoted by $\scr{K}$ we have:
  \begin{displaymath}
    \WF(\scr{K}f) \subset \{ (x,\xi) \, | \, (x,y,\xi,0) \in \WF(K) \ \mbox{for $y
      \in \supp f$} \}.
  \end{displaymath}
\end{thm}

In addition we need the following result about sums of distributions
and their wave front sets (see \cite[Ch. VII]{HOER1}).

\begin{prop} \label{prop:1}
  Consider the sum $T+S$ of two distributions $T,S \in \scr{D}'(M)$. Its
  wave front set can be estimated from above by $\WF(T+S) \subset
  \WF(T)+\WF(S)$. 
\end{prop}

And finally two results which are needed to calculate the wave front
set of the distributions $T_\psi$ defined im section \ref{sec:qf:qf-on-wl}
(cf. Section VIII.2 of \cite{HOER1} for proofs).

\begin{prop} \label{prop:wf-delta}
  The wave front set of a distribution $S$ given by 
  \begin{displaymath}
    S(f) = \int_\Sigma f(x) \lambda(x) 
  \end{displaymath}
  with a volume element $\lambda$ on a submanifold $\Sigma \subset M$ is contained in
  the normal bundle $N(\Sigma)$ of $\Sigma$ (see Equation (\ref{eq:22})). 
\end{prop}

\begin{prop} \label{prop:wf-diffop}
  For each distribution $T$ on $M$ and each differential operator $P$
  we have $\WF(PT) \subset \WF(T)$.
\end{prop}

\section{Proof of the extension theorem}
\label{sec:pf-ext-thm}

Let us discuss now the proof of Theorem \ref{prop:extend-qf} which we
have postponed in section \ref{sec:QFsingTF} to this appendix. Hence
consider a quantum field $f \mapsto \Phi(f)$ wich is extendible in the sense of 
Definition \ref{def:extend-qf} to a distribution space $\goth{D}$. We
have to show first that for each $T \in \goth{D}$ the operator $\Phi(T)$
exists as described in Definition \ref{eq:qf-extend-0}. The following
lemma gives a necessary condition.

\begin{lem} \label{lem:qf-extend-1}
  For each $T \in \goth{D}$ there is a closed cone $\Gamma \subset T^*M \setminus \{0\}$
  which contains the wave front sets of all distributions $f \mapsto \langle u,
  \Phi(f) v\rangle$, $u,v \in D_0$ and satisfies in addition the relation $\WF(T) 
  \oplus \Gamma \subset T^*M \setminus \{0\}$
\end{lem}

\begin{pf}
   We can choose $u,v \in D_0$ such that $\Phi(F) \Omega = u$ and $\Phi(G)\Omega = v$
   holds with $F= f_1 \otimes \cdots \otimes f_n \in \scr{D}(M^n)$ and $G = g_1 \otimes \cdots
  \otimes g_m \in \scr{D}(M^m)$. Hence we have $f \mapsto \langle u, \Phi(f) v\rangle =
  \scr{W}^{(n+m+1)}(F^* \otimes f \otimes G)$. This implies together with Theorem
  \ref{thm:wf-kern-distri} that the wave front set of the distribtuion
  $\langle u, \Phi(\,\cdot\,)v\rangle$ is contained in the set 
  \if1\aivsize
  \begin{multline*}
    \Gamma := \{ (x,\xi) \, | \, (x_1,\ldots,x_n,x,x_{n+1},\ldots,x_{n+m};0,\ldots,0,\xi,0,\ldots,0) \in
    \WF(\scr{W}^{(n+m+1)}) \\ \mbox{for $(x_1,\ldots,x_{n+m}) \in M^{n+m}$}\},
  \end{multline*}
  \else
  \begin{multline*}
    \Gamma := \{ (x,\xi) \, | \, (x_1,\ldots,x_n,x,x_{n+1},\ldots,x_{n+m};0,\ldots,0,\xi,0,\ldots,0)
    \in \\
    \WF(\scr{W}^{(n+m+1)}) \mbox{for $(x_1,\ldots,x_{n+m}) \in M^{n+m}$}\},
  \end{multline*}
  \fi
  where $\xi$ is located in the tuple $(0,\ldots,\xi,\ldots,0)$ on the $n+1$
  position. On the other hand we have according to Equation
  (\ref{eq:24}) 
  \if1\aivsize
  \begin{displaymath}
    \Gamma(f_1 \otimes \cdots \otimes T \otimes \cdots \otimes g_m) = (M^n \times \{0\}) \times \WF(T) \times (M^m
    \times  \{0\}) \subset T^*M^{n+m+1}.
  \end{displaymath}
  \else
  \begin{multline*}
    \Gamma(f_1 \otimes \cdots \otimes T \otimes \cdots \otimes g_m) = \\ 
    (M^n \times \{0\}) \times \WF(T) \times (M^m \times  \{0\}) \subset T^*M^{n+m+1}.
  \end{multline*}
  \fi
  Hence we get by Definition \ref{def:extend-qf} $\WF(T) \oplus \Gamma \subset T^*M \setminus
  \{0\}$ as stated.
\end{pf}

This lemma implies by Theorem \ref{thm:distri-extend} that $\Phi(T)$
exists for each $T \in \goth{D}$ as a quadratic form. To prove that
$\Phi(T)$ exists even as an operator we will use the following result.

\begin{lem} \label{kor:distri-extend}
  Consider two manifolds $M, N$, distributions $S \in \scr{D}'(M \times
  N)$, $T = T^{(1)} \otimes T^{(2)} \in \scr{E}'(M \times N)$ such that $\WF(S) \oplus
  (\WF(T_1) \odot \WF(T_2)) \subset T^*(M \times N) \setminus \{0\}$ holds, where $\WF(T_1) \odot
  \WF(T_2)$ is defined as in Equation (\ref{eq:18}) and two sequences
  $\Bbb{N} \ni k \mapsto T^{(1)}_k \in \scr{D}(M)$, $\Bbb{N} \ni l \mapsto T^{(2)}_l \in
  \scr{D}(N)$ converging in $\scr{D}_{\WF(T^{(1)})}(M)$ respectively
  $\scr{D}_{\WF(T^{(2)})}(N)$ to $T^{(1)}$ respectively
  $T^{(2)}$. Then the \emph{double sequence} $\Bbb{N}^2 \ni (k,l) \mapsto
  S(T^{(1)}_k \otimes T^{(2)}_l) \in \Bbb{C}$ converges \emph{uniformly in
    $k,l$} to $TS(\mathbf{1})$, i.e. for each $\epsilon > 0$ there is a $N_\epsilon
  \in \Bbb{N}$ such that 
 \begin{displaymath}
    | S(T^{(1)}_k \otimes T^{(2)}_l) - (TS)(\mathbf{1}) | < \epsilon \ \forall k, l > N_\epsilon. 
  \end{displaymath}
\end{lem}

\begin{pf}
  Only the statement concerning the uniformity of the convergence of
  $(k,l) \mapsto S(T^{(1)}_k \otimes T^{(2)}_l)$ is not an immediate concequence
  of Theorem \ref{thm:distri-extend}. Hence assume the statement is
  false, i.e. there is an $\epsilon > 0$ such that for each $N \in \Bbb{N}$
  there are $(k_N,l_N) \in \Bbb{N}^2$ with $k_N >  
  N$ and $l_N > N$ and
  \begin{equation} \label{eq:23}
    | S(T^{(1)}_{k_N} \otimes T^{(2)}_{l_N}) - TS(\mathbf{1}) | \geq \epsilon \ \forall N \in
      \Bbb{N}. 
  \end{equation}
  However the sequences $N \mapsto T^{(1)}_{k_N}$ and $N \mapsto T^{(2)}_{l_N}$
  are subsequences of $k \mapsto T^{(1)}_k$ and $l \mapsto T^{(2)}_l$ which
  implies that they converge in $\scr{D}_{\WF(T^{(1)})}'(M)$ respectively  
  $\scr{D}_{\WF(T^{(2)})}'(M)$ to $T^{(1)}$ and $T^{(2)}$. Hence Equation
  (\ref{eq:23}) contradicts Theorem \ref{thm:distri-extend}.
\end{pf}

Now we are able to prove the existence of the operators $\Phi(T)$ as
stated in item \ref{item:10} of Theorem \ref{prop:scrW-extend}. 

\begin{prop} \label{prop:extend-thm-2}
  If the quantum field $f \mapsto \Phi(f)$ is extendible to the distribution
  space $\goth{D} \subset \scr{E}'(M)$ the operators $\Phi(T)$ exist for each
  $T \in \goth{D}$ in the sense of Definition \ref{eq:qf-extend-0}.
\end{prop}

\begin{pf} 
  Consider $T \in \goth{D}$ and a sequence  $\Bbb{N} \ni k \mapsto T_k \in
  \scr{D}(M)$ which converges in $\scr{D}_{\WF(T)}'(M)$ to $T$. We
  have seen in lemma \ref{lem:qf-extend-1} that $k \mapsto \langle u, \Phi(T_k) v \rangle$
  converges  as well and depends only on $T$, not on the sequence $k \mapsto 
  T_k$. Now we have to show convergence of the sequence $\Bbb{N} \ni k \mapsto
  \Phi(T_k)u \in \scr{H}$ for all $u \in D_0$. To this end let us assume
  first that $T$ and the $T_k$ are real valued. Then we have 
  \if1\aivsize
   \begin{displaymath}
    \| \Phi(T_k)u - \Phi(T_l)u\|^2=\|\Phi(T_k)\|^2+\|\Phi(T_l)\|^2 - \langle \Phi(T_l)u, \Phi(T_k)u\rangle
    - \langle\Phi(T_k)u,\Phi(T_k)u\rangle 
  \end{displaymath}
  \else
  \begin{multline*}
    \| \Phi(T_k)u - \Phi(T_l)u\|^2=\\ \|\Phi(T_k)\|^2+\|\Phi(T_l)\|^2 - \langle \Phi(T_l)u, \Phi(T_k)u\rangle
    - \langle\Phi(T_k)u,\Phi(T_k)u\rangle 
  \end{multline*}
  \fi
  and therefore
  \if1\aivsize
  \begin{multline}
    \label{eq:3}
  \| \Phi(T_k)u - \Phi(T_l)u\|^2 = \scr{W}^{(2n+2)}(\mathbf{f}^* \otimes T_l \otimes T_l \otimes \mathbf{f}) 
  + \scr{W}^{(2n+2)}(\mathbf{f}^* \otimes T_k \otimes T_k \otimes \mathbf{f}) \\
  - \scr{W}^{(2n+2)}(\mathbf{f}^* \otimes T_k \otimes T_l \otimes \mathbf{f}) 
  - \scr{W}^{(2n+2)}(\mathbf{f}^* \otimes T_l \otimes T_k \otimes \mathbf{f}),
  \end{multline}
  \else
  \begin{multline}
    \label{eq:3}
  \| \Phi(T_k)u - \Phi(T_l)u\|^2 = \\ \scr{W}^{(2n+2)}(\mathbf{f}^* \otimes T_l \otimes
  T_l \otimes \mathbf{f}) + \scr{W}^{(2n+2)}(\mathbf{f}^* \otimes T_k \otimes T_k \otimes
  \mathbf{f}) \\ 
  - \scr{W}^{(2n+2)}(\mathbf{f}^* \otimes T_k \otimes T_l \otimes \mathbf{f}) 
  - \scr{W}^{(2n+2)}(\mathbf{f}^* \otimes T_l \otimes T_k \otimes \mathbf{f}),
  \end{multline}
  \fi
  where we have choosen, as in the proof of \ref{lem:qf-extend-1} $u,v
  \in D_0$ such that $\Phi(\mathbf{f}) \Omega = u$ and $\Phi(G)\Omega = v$ holds with $\mathbf{f}= f_1 \otimes \cdots
  \otimes f_n \in \scr{D}(M^n)$ and $G = g_1 \otimes \cdots \otimes g_m \in \scr{D}(M^m)$. 
  By assumption we have $\WF(\mathbf{f}^* \otimes T \otimes T \otimes \mathbf{f}) \subset \Theta$ and $\Theta \oplus
  \WF(\scr{W}^{(2n+2)}) \subset T^*(M^{2n+2}) \setminus \{0\}$ with $\Theta = (M^n \times \{0\}) \times
  (\WF(T) \odot \WF(T)) \times (M^n \times \{0\})$  (cf. Definition \ref{def:extend-qf}). Hence
  convergence of $k \mapsto T_k$ implies together with
  \ref{kor:distri-extend}
  that there is a $W \in \Bbb{C}$ such that $\forall \epsilon 
  > 0 \ \exists N \in \Bbb{N}$ with
  \begin{displaymath}
    \left| \scr{W}^{(2n+2)}(\mathbf{f}^* T_k \otimes T_l \otimes \mathbf{f}) - W \right| < \epsilon \quad \forall
    k,l > N. 
  \end{displaymath}
  In other words each summand on the left hand side of equation
  (\ref{eq:3}) differs from $W$ only by $\epsilon/4$ provided $k,l$ are big
  enough; i.e we have
  \begin{gather*}
    \left| \scr{W}^{(2n+2)}(\mathbf{f}^* \otimes T_l \otimes T_l \otimes \mathbf{f}) - W \right| +
    \left|\scr{W}^{(2n+2)}(\mathbf{f}^* \otimes T_k \otimes T_k \otimes \mathbf{f}) - W \right| \leq
    \frac{\epsilon}{2} \\
    \left| \scr{W}^{(2n+2)}(\mathbf{f}^* \otimes T_k \otimes T_l \otimes \mathbf{f}) - W \right| +
    \left|\scr{W}^{(2n+2)}(\mathbf{f}^* \otimes T_l \otimes T_k \otimes \mathbf{f}) - W \right| \leq
    \frac{\epsilon}{2}.
  \end{gather*}
  This implies $\| \Phi(T_k)u - \Phi(T_l)u\|^2 \leq \epsilon/2$ for $k,l > N$, i.e. $k \mapsto
  \Phi(T_k)u$ is a Cauchy sequence, converging to an element $\Phi(T)u$ of
  $\scr{H}$.

  This defines an operator $\Phi(T)$ with domain $D \subset \scr{H}$ 
  because independence of the limit $\lim_{k\to\infty}\Phi(T_k)u$ from the
  sequence $k \mapsto T_k$ follows from the corresponding property of $\lim_{k\to\infty}
  \langle v, \Phi(T_k) u \rangle$ for all $v \in D_0$ (cf. Definition \ref{eq:qf-extend-0}). 

  If $T$ is complex valued we can apply the arguments just discussed
  to the real and imaginary part separately, and we get again an
  operator $\Phi(T)$. This completes the proof.
\end{pf}

The next step in the proof concerns the existence of an invariant
domain $D$. If $D$ exists, its elements are given by linear
combinations of expressions of the form $\Phi(T^{(1)}) \cdots \Phi(T^{(n)}) \Omega$
with $\mathbf{T} = T^{(1)} \otimes \cdots \otimes T^{(n)} \in \goth{D}^{\otimes n}$. Hence it
is natural to consider limits of the form $\lim_{k\to\infty}
\Phi(\mathbf{T}_k)\Omega$ where $\Bbb{N} \ni k \mapsto \mathbf{T}_k = T^{(1)}_k \otimes \cdots \otimes
T^{(n)}_k \in \scr{D}(M^n)$ converges in $\scr{D}_{\Gamma(\mathbf{T})}(M^n)$ to
$\mathbf{T}$. The next lemma shows that they allways exist.

\begin{lem} \label{lem:qf-extend-3}
  Consider a sequence $\Bbb{N} \ni k \mapsto \mathbf{T}_k = T^{(1)}_k \otimes \cdots \otimes
  T^{(n)}_k \in \scr{D}(M^n)$ converging in $\scr{D}_{\Gamma(\mathbf{T})}(M^n)$
  to $\mathbf{T} = T^{(1)} \otimes \cdots \otimes T^{(n)} \in \goth{D}^{\otimes n}$.
  \begin{enumerate}
  \item \label{item:1}
    The limit $\lim_{k \to \infty} \Phi(\mathbf{T}_k)\Omega =: u(\mathbf{T}) \in
    \scr{H}$ exists and depends only on $\mathbf{T}$.
  \item \label{item:2}
    If $u(\mathbf{T}) = 0$ holds, we get $u(S \otimes \mathbf{T}) =
    0$ for all $S \in \goth{D}$.
  \end{enumerate}
\end{lem}

\begin{pf}
  \ref{item:1}. We use the same idea as in the first part of
  Proposition \ref{prop:extend-thm-2}. Hence consider $l,k \in
  \Bbb{N}^n$ and   
  \if1\aivsize
  \begin{displaymath}
    \| \Phi(\mathbf{T}_k) \Omega -
    \Phi(\mathbf{T}_l)\Omega\|^2=\|\Phi(\mathbf{T}_k)\|^2+\|\Phi(\mathbf{T}_l)\|^2 - \langle
    \Phi(\mathbf{T}_l) \Omega, \Phi(\mathbf{T}_k)\Omega\rangle - \langle\Phi(\mathbf{T}_k)
    \Omega,\Phi(\mathbf{T}_l) \Omega\rangle  
  \end{displaymath}
  \else
  \begin{multline*}
    \| \Phi(\mathbf{T}_k) \Omega -
    \Phi(\mathbf{T}_l)\Omega\|^2=\\ \|\Phi(\mathbf{T}_k)\|^2+\|\Phi(\mathbf{T}_l)\|^2 - \langle
    \Phi(\mathbf{T}_l) \Omega, \Phi(\mathbf{T}_k)\Omega\rangle - \langle\Phi(\mathbf{T}_k)
    \Omega,\Phi(\mathbf{T}_l) \Omega\rangle  
  \end{multline*}
  \fi
  which leads to
  \if1\aivsize
  \begin{multline}  \label{eq:5}
    \| \Phi(\mathbf{T}_k) \Omega - \Phi(\mathbf{T}_l) \Omega\| =
    \scr{W}^{(2n)}(\mathbf{T}_k^* \otimes \mathbf{T}_k) +
    \scr{W}^{2n}(\mathbf{T}_l^* \otimes \mathbf{T}_l) -
    \scr{W}^{(2n)}(\mathbf{T}_l^* \otimes \mathbf{T}_k) - \\
    \scr{W}^{(2n)}(\mathbf{T}_k^* \otimes \mathbf{T}l). 
  \end{multline}
  \else
  \begin{multline}  \label{eq:5}
    \| \Phi(\mathbf{T}_k) \Omega - \Phi(\mathbf{T}_l) \Omega\| = 
    \scr{W}^{(2n)}(\mathbf{T}_k^* \otimes \mathbf{T}_k) +
    \scr{W}^{2n}(\mathbf{T}_l^* \otimes \mathbf{T}_l) - \\
    \scr{W}^{(2n)}(\mathbf{T}_l^* \otimes \mathbf{T}_k) - 
    \scr{W}^{(2n)}(\mathbf{T}_k^* \otimes \mathbf{T}l). 
  \end{multline}
  \fi
  Extendibility of the quantum field implies together with lemma
  \ref{kor:distri-extend} that each of the terms
  $\scr{W}^{(2n)}(\,\cdot\,)$ converges to the same value $W$. Hence we have 
    \begin{gather*}
    \left| \scr{W}^{(2n)}(\mathbf{T}_l^* \otimes \mathbf{T}_l) - W \right| +
    \left|\scr{W}^{(2n)}(\mathbf{T}_k^* \otimes \mathbf{T}_k) - W \right| \leq
    \frac{\epsilon}{2} \\
    \left| \scr{W}^{(2n)}(\mathbf{T}_k^* \otimes \mathbf{T}_l) - W \right| +
    \left|\scr{W}^{(2n)}(\mathbf{T}_l^* \otimes \mathbf{T}_k)  - W \right| \leq
    \frac{\epsilon}{2},
  \end{gather*}
  provided $l,k$ are big enough. This implies together with Equation
  (\ref{eq:5}) convergence of $\Phi(\mathbf{T}_l)\Omega$ to an element
  $u(\mathbf{T}) \in \scr{H}$. The fact that $u(\mathbf{T})$ depends
  only on $\mathbf{T}$ and not on the sequence $k \mapsto \mathbf{T}_k$
  follows with the same argument as in Proposition
  \ref{prop:extend-thm-2}.

  \ref{item:2}. 
  By assumption we have
    $\Phi(\mathbf{T}_l) \Omega \to 0, \quad l \to \infty$ 
  for each sequence $\Bbb{N} \ni l \mapsto \mathbf{T}_l \in \scr{D}(M^n)$ 
  converging to $\mathbf{T}$ as described above. This implies 
  \begin{equation} \label{eq:4}
    \scr{W}^{n+m+1}(\mathbf{f}^*\otimes S_k \otimes \mathbf{T}_l) = \langle v, \Phi(S_k)
    u(\mathbf{T}_l) \rangle \to 0 \quad l \to \infty 
  \end{equation}
  for each $v = u(\mathbf{f})$ with $\mathbf{f} \in \scr{D}(M^m)$, each
  sequence $\Bbb{N} \ni k \mapsto S_k \in \scr{D}(M)$ converging in
  $\scr{D}_{\WF(S)}(M)$ to $S$ and each fixed $l \in \Bbb{N}$. On the
  other hand we can consider the multisequence $(k,l) \to S_k \otimes
  \mathbf{T}_l$ which converges in $\scr{D}_{\Gamma(S \otimes
    \mathbf{T})}(M^{n+1})$ to $S \otimes \mathbf{T}$. This implies that $(k,l) \to \Phi(S_k)
  \Phi(\mathbf{T}_k) \Omega$ converges uniformly in $(k,l)$ to $u(S \otimes
  \mathbf{T})$. Hence to prove that $u(S \otimes \mathbf{T}) = 0$ it is
  sufficient to show that for each $v \in D$ there is a subsequence
  $\Bbb{N} \ni j \mapsto \Phi(S_{k_j}) \Phi(\mathbf{T}_{l_j}) \Omega$ of $(k,l) \to \Phi(S_k) 
  \Phi(\mathbf{T}_l) \Omega$ such that
  \begin{displaymath}
    \langle v, \Phi(S_{k_j}) \Phi(\mathbf{T}_{l_j}) \rangle \to 0 \quad k\to \infty
  \end{displaymath}
  holds. However this is a direct consequence of Equation
  (\ref{eq:4}). Therefore we get $\Phi(S) u(S \otimes \mathbf{T}) = 0$ as
  stated.
\end{pf}

Now we are ready to prove the existence of an invariant dense domain
$D \subset \scr{H}$ for all $\Phi(T)$ (see item \ref{item:11} of Theorem
\ref{prop:extend-qf}). 

\begin{prop} \label{prop:qf-extend-4}
  The operators $\Phi(T)$ can be extended to the invariant domain
  \begin{displaymath}
    D := \Lh \{ u(\mathbf{T})  \, | \, T \in \goth{D}^{\otimes n}, \ n \in \Bbb{N} \},
  \end{displaymath}
  which is a subset of the domain $D(\overline{\Phi(T)})$ of the closure
  of $\Phi(T)$.
\end{prop}

\begin{pf}
  Obviously $D_0 \subset D$ hence $D$ is dense in $\scr{H}$. In addition we
  have $\Phi(T) u(\mathbf{f})$ $= u(T \otimes \mathbf{f})$ for all $\mathbf{f}
  \in \scr{D}(M^n)$ and $T \in \goth{D}$. Hence we can define an extension
  of $\Phi(T)$ to the domain $D$ by $\Phi(T) u(\mathbf{T}) := u(T \otimes
  \mathbf{T})$ provided the left hand site depends only on
  $u(\mathbf{T}) \in \goth{D}^n$ and not on $\mathbf{T}$, but this is a
  consequence of item \ref{item:2} of Lemma \ref{lem:qf-extend-3}.

  To complete the proof of item \ref{item:11} we have to show that all
  $\Phi(T)$ are closed and $D$ is a subspace of the domain of the closure
  of $\Phi(T)$. To prove closedness consider $\Phi(\bar T) u$ for $u \in
  D_0$. Obviously we have $\Phi(\bar T)u = \lim_{l\to\infty} \Phi(\bar{T}_l)$ where
  $T_l$ converges weakly to $T$. But $\Phi(\bar{T}_l) \subset \Phi(T_l)^*$ and
  this implies $\Phi(\bar T) \subset \Phi(T)^*$. Hence the domain of $\Phi(T)^*$ is
  dense, which implies closability of $\Phi(T)$. Since the construction of
  the operator $\Phi(T) : D \to \scr{H}$ given above implies immediately
  that its graph is contained in the closure of the graph of $\Phi(T):
  D_0 \to \scr{H}$, which in turn coincides with the graph of the
  closure of $\Phi(T): D_0 \to \scr{H}$, we get $D \subset D(\overline{\Phi(T)})$
  and this completes the proof of the second statement.
\end{pf}

The last statement of Theorem \ref{prop:extend-qf} (item
\ref{item:12}) is now a simple consequence of Proposition
\ref{prop:qf-extend-4} and Lemma \ref{lem:qf-extend-3}. Hence Theorem
\ref{prop:extend-qf} is proved.

\section{Jet bundles and differential operators}
\label{sec:jetbun}

In this appendix we want to summarize some material about jet bundles used
throughout the paper. For a more detailed presentation we want to refer to the
book of Saunders \cite{saunders89}. 

Hence consider two manifolds $M,N$, a point $x \in M$, the set $\cni{x}{N}$
of smooth maps from an open neighbourhood of $x$ to $N$ and a nonnegative
integer $l$. Two such maps $f, h$ coincide in $x$ up to the $l^{\rm th}$
order, iff $f(x) = h(x)$ and their Taylor expansion (in coordinates around $x$
and $f(x)$) coincide up to the $l^{\rm th}$ order. It is easy to check that
this defines an equivalence relation. The corresponsding equivalence
class of $f$ is called the \emph{k-jet of $f$ at $x \in M$} and denoted
by $j^k_xf$; the set of all equivalence classes is denoted by
$J^k_x(M,N)$. If we take the union
\begin{displaymath}
 J^k(M,N) :=  \bigcup_{x \in M} J^k_x(M,N)
\end{displaymath}
we get a manifold which is a \emph{fiber bundle} with respect to the
\emph{source-projection}
\begin{displaymath}
  J^k(M,N) \ni j^k_xf \mapsto \pi^k(j^k_xf) = x \in M  
\end{displaymath}
and as well with respect to the \emph{target projection}
\begin{displaymath}
  J^k(M,N) \ni j^k_xf \mapsto \pi^k_0(j^k_xf) = f(x) \in N.  
\end{displaymath}
An adapted coordinate system is given in terms of a chart $(M_u,u)$ of
$M$ and $(N_v,v)$ of $N$ by 
\begin{multline} \label{eq:32}
  J^l(M_u,N_v) \ni j^l_xf \mapsto \bigl(u(x), v[f(x)], \tilde{f}'[u(x)], \ldots,
  \tilde{f}^{(l)}[u(x)]\bigr) \\ \in u(M_u) \times v(N_v) \times L(\Bbb{R}^m,\Bbb{R}^n) \times
  L^2_s(\Bbb{R}^m,\Bbb{R}^n) \times \ldots \times L^l_s(\Bbb{R}^m,\Bbb{R}^n)
\end{multline}
where $\tilde{f}=v\circ f\circ u^{-1}$ denotes the local representative of $f$,
$m=\dim(M)$, $n=\dim(N)$ and $L^k_s(\Bbb{R}^m,\Bbb{R}^n)$ is the space 
of $k$--linear, symmetric maps $(\Bbb{R}^m)^k \to \Bbb{R}^n$. Each
smooth function $f$ on $M$ defines a section, the \emph{$l$--jet
  extension} of $f$, by
\begin{displaymath}
  j^lf : M \to J^l(M,\Bbb{K}), \ x \mapsto j^lf(x) := j^l_xf.  
\end{displaymath}
However not all sections of $J^l(M,\Bbb{K})$ are of this form.

To discuss the relation of jet bundles to (linear) differential
operators consider $J^l(M,\Bbb{K})$ with $\Bbb{K} = \Bbb{R}$ or
$\Bbb{K}=\Bbb{C}$. It is a vector bundle with respect to the source
projection $\pi^k$, i.e. each fiber $J_x^k(M,\Bbb{K})$ of
$\pi^k:J^l(M,\Bbb{K}) \to M$ is a $\Bbb{K}$--vector space. A linear
differential operator $P: \ci{M}{\Bbb{K}} \to \ci{M}{\Bbb{K}}$ of order
$l$ defines a morphism $\psi_P$ of vector bundles (i.e. $\psi_P$ is linear
on each fiber) by
\begin{displaymath}
 \psi_P: J^l(M,\Bbb{K}) \to M \times \Bbb{K}, j^l_pf \mapsto \psi_P(j^l_xf):=
 \bigl(p,(Pf)(x)\bigr).   
\end{displaymath}
If on the other hand a vector bundle morphism
$\psi:J^l(M,\Bbb{K}) \to M \times \Bbb{K}$ is given, we get immediately a
differential operator 
\begin{equation} \label{eq:56}
  P_\psi: \ci{M}{\Bbb{K}} \to \ci{M}{\Bbb{K}}, f \mapsto P_\psi f := \psi \circ j^lf  
\end{equation}
In other words differential operators $P$ and morphisms $\psi$ are in a
one to one correspondence. Since the morphism $\psi$ defines a linear map
$\psi_x$ for each $x \in M$ by $\psi(j^l_xf)=\bigl(x,\psi_x(j^l_xf)\bigr)$ we can
identify $\psi$ with a \emph{section} of the vector bundle
$J^l(M,\Bbb{K})^*$ \emph{dual}\footnote{Note that stars as sub- and
  superscripts appear in the context of vectorbundles customarily in
  two different meanings: They denote dual maps and dual spaces on the
  one hand and pull backs and push forwards on the other.} to
$J^l(M,\Bbb{K})$, i.e. $J^l(M,\Bbb{K}) = \bigcup_{x\in
  M}J^l_x(M,\Bbb{K})^*$. Hence we have proved the following 

\begin{prop} \label{prop:jetbun-1}
  The space of smooth sections $\Gamma\bigl(J^l(M,\Bbb{K})^*\bigr)$
  coincides  with the space of linear, $l^{\rm th}$ order,
  $\Bbb{K}$--valued differential operators on $M$. 
\end{prop}

Let us consider now a (smooth) diffeomorphism $F$ of $M$. It defines
the \emph{pull back} $F^*f = f \circ F$ and the \emph{push forward} $F_*f
= f\circ F^{-1}$ of the smooth function $f: M \to \Bbb{K}$. A simple
calculation shows that $J^lF(j^l_xf) := j^l_x(F_*f)$ is well defined,
i.e. does not depend on the representative $f \in j^l_xf$. Hence we get
an automorphism $J^lF: J^l(M,\Bbb{K}) \to J^l(M,\Bbb{K})$ of the vector
bundle $J^l(M,\Bbb{K})$ which covers $F: M \to M$, i.e. the following
diagram commutes
\begin{displaymath}
  \begin{CD}
    J^l(M,\Bbb{K}) @>J^lF>> J^l(M,\Bbb{K})\\
    @V\pi^l_0VV @V\pi^l_0VV \\
    M @>F>> M.
  \end{CD}
\end{displaymath}
If we consider now for each $x \in M$ the dual $J^l_xF^*:
J_{F(x)}(M,\Bbb{K})^* \to J_x(M,\Bbb{K})^*$ of the linear
isomorphism $J^l_x(M,\Bbb{K}) \ni j_x^lf \mapsto J^l_xF(j_x^lf) \in
J_{F(x)}^l(M,\Bbb{K})$, we get the vector bundle isomorphism $J^lF^* : 
J^l(M,\Bbb{K})^* \to J^l(M,\Bbb{K})^*$ dual to $J^lF$, which given by
$J^lF^*\restr J^l_{F(x)}(M,\Bbb{K})^* = J^l_xF^*$. In contrast to
$J^lF$ it covers $F^{-1}$ rather than $F$: 
\begin{displaymath}
  \begin{CD}
    J^l(M,\Bbb{K})^* @>J^lF^*>> J^l(M,\Bbb{K})^*\\
    @VVV @VVV \\
    M @>F^{-1}>> M.
  \end{CD}
\end{displaymath}
If $P$ is a differential operator of order $l$, we 
can apply $J^lF^*$ to the section $\psi_p \in
\Gamma\bigl(J^l(M,\Bbb{K})^*\bigr)$ associated to $P$ by Proposition
\ref{prop:jetbun-1}. We get obviously another section of
$J^l(M,\Bbb{K})^*$ and therefore a new differential operator, which is
in the following way associated to $P$:

\begin{prop} \label{prop:2}
  If $Pf = \psi_P \circ j^lf$ holds for a section $\psi_P$ of $J^l(M,\Bbb{K})$
  and a linear differential operator of $l^{th}$ order, the \emph{pull 
    back} $F^*P$ of $P$ by a diffeomorphism $F: M \to M$, given by
  $(F^*P)f := F^*(P(F_*f))$ satisfies $(F^*P)f = (J^lF^* \circ \psi_P \circ F) \cdot j^lf$.
\end{prop}

\section{Fundamental solutions of the Klein-Gordon equation}
\label{sec:retadvKG}

In this appendix we want to discuss solutions of the inhomogeneous
Klein-Gordon equation with distributions $T \in \goth{D}^l(\gamma)$ as source 
term. As a first step let us recall some well known properties of the
retarded and the advanced fundamental solutions (see \cite{DIMOCK80}
and the references therein for details).

\begin{thm} \label{thm:retadv}
  Consider the Klein-Gordon operator $\Box_g - m^2$ on a globally
  hyperbolic space-time $(M,g)$. There are distributions $G^\pm \in
  \scr{D}'(M \times M)$, which are uniquely determined by the following
  porperties ($f \in \scr{D}(M)$):  
  \begin{enumerate}
  \item 
    The distributions $h \mapsto G^\pm(f,h)$ are regular. The corresponding
    continuous maps $\scr{D}(M) \to \scr{E}(M)$ will be denoted by $G^\pm$
    as well. 
  \item 
    The function $G^\pm f$ solves the inhomogeneous equation $(\Box_g -
    m^2)h = f$, i.e. we have $G^\pm (\Box_g - m^2) f = (\Box_g - m^2) G^\pm f =
    f$.
  \item 
    The support of $G^\pm f$ is contained in the causal future/past of
    $\supp f$, i.e. $\supp G^\pm f \subset J^\pm(\supp f)$.
  \item \label{item:13}
    Hence the support of the distribution $f \mapsto G^\mp_p(f) := (G^\pm f)(p)$
    ($p \in M$) satisfies $\supp G^\pm_p \subset J^\pm(p)$.
  \end{enumerate}
\end{thm}

The $G^\pm$ can be extended in exactly one way to continuous maps $G^\pm : 
\scr{E}'(M) \to \scr{D}'(M)$: For $T \in \scr{E}'(T)$ the distribution
$G^\pm T$ is given by $(G^\pm T)(f) = T(G^\mp f)$, where $f \in \scr{D}(M)$. As
in the regular case $G^\pm T$ solves the inhomogeneous Klein-Gordon
equation with $T$ as source term. More precisely: $G^\pm T$ is the unique 
weak solution of $(\Box_g - m^2) G^\pm T = T$ with support in $J^\pm(\supp
T)$.

One mayor problem in the theory of (linear) partial differential
equations is to determine the singularities of the solution if the
singularities of the initial data and/or the inhomogeneity is
given. In the hyperbolic case a complete answer is given by the
\emph{propagation of singularities} theorem of Duistermaat and
H{\o}rmander, which we will state here in the special Klein-Gordon case
(see \cite{duistermaat72} for a proof and for the general
statement). To this end it is necessary to introduce some terminology
first: $L \subset T^*M$ denotes the set of nonvanishing null covectors,
i.e. 
\begin{displaymath}
  L = \{ \theta \in T^*M \setminus \{0\} \, | \, g(\theta,\theta) = 0 \} 
\end{displaymath}
and $\sim \, \subset L \times L$ is the \emph{bicarachteristic relation} associated to
$\Box_g - m^2$, i.e.$(p_1,\theta_1) \sim (p_2,\theta_2)$ iff 1. $p_1$ and $p_2$ can be
connected by a null geodesic $\nu$ and 2. the vectors $\theta^\#_j \in T_{p_j}^*M$ 
with $g(\theta^\#_j, \,\cdot\,) = \theta_j$ are tangent to $\nu$.

\begin{thm} \label{thm:prop-sing-thm}
  Consider a weak solution $S \in \scr{D}'(M)$ of the inhomogeneous
  Klein-Gordon equation $(\Box_g -m^2)S = T$. Then we have 
  \begin{enumerate}
  \item 
    $\WF(S) \setminus \WF(T) \subset L$ and
  \item 
    $(p_1,\theta_1) \in \WF(S) \setminus \WF(T)$ and $(p_1,\theta_1) \sim (p_2,\theta_2)$ imply
    $(p_2,\theta_2) \in \WF(S) \setminus \WF(T)$ (i.e. $\WF(S) \setminus \WF(T)$ is invariant
    under the bicharacteristic flow).
  \end{enumerate}
\end{thm}

Let us consider now a smooth timelike curve $\gamma$ and distributions $T \in
\goth{D}^l(\gamma)$ ($l \in \Bbb{N}_0$). The propagation of singularities
theorem allows us immediately to locate the singularities of $G^\pm T$.

\begin{kor} \label{kor:kg-inhom-1}
  If $T \in \goth{D}^l(\gamma)$ holds, we get $\WF(G^\pm T) \subset \WF(T)$. Hence the
  singular support of $G^\pm T$ is contained in the range of $\gamma$.
\end{kor}

\begin{pf}
  We have to show that $\WF(G^\pm T) \setminus \WF(T)$ is empty. Hence assume
  $(p_1,\theta_1) \in \WF(G^\pm T) \setminus \WF(T)$. It defines a unique, maximally
  extended null geodesics $\nu$ by $\nu(0) = p_1$ and $g(\nu'(0),\,\cdot  \,)
  = \theta_1$. Since $\WF(T) \cap L = \emptyset$ holds the bicaracteristic strip given 
  by $\nu$, i.e. $t \mapsto \bigl(\nu(t),g(\nu'(t),\,\cdot\,)\bigr)$, lies entirely in
  $\WF(G^\pm T) \setminus \WF(T)$. Hence by Theorem \ref{thm:prop-sing-thm} we get
  $\bigl(\nu(t),g(\nu'(t),\,\cdot\,)\bigr) \in \WF(G^\pm T) \setminus \WF(T)$ for all $t$
  in the domain of $\nu$.

  Due to global hyperbolicity there is in addition a Cauchy 
  surface $C$ in the past (in the $G^+$ case) or future (for $G^-$)
  of $\supp T$. Since $C$ is a Cauchy surface, $\nu$ hits it in a point
  $\nu(t_2) = p_2$. But $p_2 \not\in \supp G^\pm T$ holds, due to the condition
  $\supp (G^\pm T) \subset J^\pm(\supp T)$. This implies obviously $(p_2,
  g\bigl(\nu'(t_2),\,\cdot\,)\bigr) \not\in  \WF(G^\pm T) \setminus \WF(T)$, in
  contradiction to the result of the last paragraph. Hence $\WF(G^\pm T)
  \setminus \WF(T) = \emptyset$ as stated. 
\end{pf}

Hence we have seen that the weak solutions $G^\pm T$ are regular on $M \setminus
\Ran \gamma$. In Section \ref{sec:frskf-gen} however this information is
not quite sufficient, because analyticity properties of $G^\pm T$ are
used in the proof of Theorem \ref{thm:1}. To proceed in this
direction it is necessary to determine the distributions $G^\pm_p$
defined in item \ref{item:13} of Theorem \ref{thm:retadv} more
explicitly. Let us introduce some notations first: If $\scr{O} \subset M$ is 
an open convex set the \emph{square of the geodesic distance} $\Gamma(p,q)$ 
between two points $p,q \in \scr{O}$ is a well defined smooth function
$\scr{O} \times \scr{O} \ni (p,q) \mapsto \Gamma(p,q) \in \Bbb{R}$. For each fixed $p \in
\scr{O}$ the function $\Gamma_p(\,\cdot\,) := \Gamma(p,\,\cdot\,)$ gives rise to a pair
of distributions $\delta_\pm(\Gamma_p)$ defined formally as the composition of
$\Gamma_p$ with the delta distribution. More precisely for a test function
$f$ with support in $\scr{O}$ we define
\begin{equation} \label{eq:39}
  \langle\delta_\pm(\Gamma_p),f\rangle := \frac{1}{2} \int_{\Bbb{R}^3} \frac{
    (u_{p*}f)(\pm|\mathbf{x}|,\mathbf{x}) \sqrt{|\det \bigl[(u_{p*}g)_{\mu
        \nu}(\pm|\mathbf{x}|,\mathbf{x})\bigr] |}}{|x|} dx, 
\end{equation}
where $\scr{O} \ni q \mapsto u_p(q) = (x^0,\mathbf{x})$ is a normal coordinate 
system around $p$  with the $x^0$ axis future directed, and 
$u_{p*} f, (u_{p*}g)_{\mu \nu}$ are the local
representatives\footnote{I.e. $u_{p*}f$, $u_{p*}g$ denote the push
  forwards of $f$ and $g$ by the coordinate map $u$.} of $f$
respectively $g$ in this chart. Finally we need the so called
\emph{van Vleck -- Morette determinant} $\Delta : \scr{O} \times \scr{O} \to
\Bbb{R}$ which is defined in  an arbitrary coordinate system $u :
\scr{O} \to u(\scr{O}) \subset \Bbb{R}^4$  by   
\if1\aivsize
\begin{equation} \label{eq:38}
  [(u \times u)_* \Delta](x,y) := \frac{1}{16} \frac{1}{\sqrt{|\det \bigl[
      (u_*g)_{\mu \nu}(x)\bigr] \det \bigl[ (u_* g)_{\mu \nu}(y)|\bigr]}}
  \left| \det \frac{\partial^2 [(u \times u)_*\Gamma] (x,y)}{\partial x^\mu \partial y^\nu}  
  \right|, 
\end{equation}
\else
\begin{multline} \label{eq:38}
  [(u \times u)_* \Delta](x,y) := \\ \frac{1}{16} \frac{1}{\sqrt{|\det \bigl[
      (u_*g)_{\mu \nu}(x)\bigr] \det \bigl[ (u_* g)_{\mu \nu}(y)|\bigr]}}
  \left| \det \frac{\partial^2 [(u \times u)_*\Gamma] (x,y)}{\partial x^\mu \partial y^\nu}  
  \right|, 
\end{multline}
\fi
where $(u \times u)_* \Delta$, $(u \times u)_* \Gamma$ and $(u_*g)_{\mu \nu}$ denote the local 
representatives of $\Delta$, $\Gamma$ and $g$ in the chosen chart. 

\begin{thm} \label{thm:anal-fund-sol} 
  Consider an analytic space-time and an open convex set $\scr{O} \subset M$
  which is at the same time globally hyperbolic.
  \begin{enumerate}
  \item 
    The square of the geodesic distance $\Gamma$ and the van Vleck --
    Morette determinant $\Delta$ (\ref{eq:38})  are analytic functions on
    $\scr{O} \times \scr{O}$.
  \item 
    For each $p \in \scr{O}$ we have
    \begin{equation} \label{eq:41}
      G^\pm_p \restr \scr{O} = \frac{1}{2\pi} \sqrt{\Delta(p,\,\cdot\,)} \delta_\pm(\Gamma_p) +
      \chi_{J^\pm(p,\scr{O})}  V(p,\,\cdot\,), 
    \end{equation}
    where $\delta_\pm(\Gamma_p)$ is the distribution from Equation (\ref{eq:39}),
    $V: \scr{O} \times \scr{O} \to \Bbb{R}$ is a smooth function and 
    $\chi_{J^\pm(p,\scr{O})}$ denotes the characteristic function of
    $J^\pm(p,\scr{O}) := J^\pm(p) \cap \scr{O}$. 
  \item 
    Each $p \in \scr{O}$ has a neighbourhood $\scr{Q}_p \subset \scr{O}$ such 
    $V(p,\,\cdot\,)$ is given on $\scr{Q}_p$ by an absolutely convergent
    power series 
    \begin{equation} \label{eq:40}
      V(p,q) = \sum_{j=0}^\infty V_j(p,q) \frac{\Gamma^j(p,q)}{j!}.
    \end{equation}
    Hence $V(p,\,\cdot\,)$ is analytic on $\scr{Q}_p$.
  \item 
    The coeficient functions $V_j$ are defined by the \emph{Hadamard
      recursion relations}: 
    \if1\aivsize
    \begin{equation} \label{eq:43}
      V_j(p,q) = -\frac{1}{4} \Delta(p,q) \int_0^1\frac{\Bbb{1} \otimes (\Box_g -
        m^2)V_{j-1}\bigl(p,\nu(s)\bigr)}{\Delta\bigl(p,\nu(s)\bigr)} s^j ds 
      \quad \mbox{for $j \in \Bbb{N}_0$} \ \mbox{and} \ V_{-1} = \Delta. 
    \end{equation}
    \else
    \begin{equation} \label{eq:43}
      V_j(p,q) = -\frac{1}{4} \Delta(p,q) \int_0^1\frac{\Bbb{1} \otimes (\Box_g -
        m^2)V_{j-1}\bigl(p,\nu(s)\bigr)}{\Delta\bigl(p,\nu(s)\bigr)} s^j ds 
    \end{equation}
    for $j \in \Bbb{N}_0$ and $V_{-1} = \Delta$.
    \fi
    The integrations are carried out along the unique geodesic $\nu:
    [0,1] \to \scr{O}$ with $\nu(0) = p$ and $\nu(1) = q$.
  \item 
    The neighborhood $\scr{Q}_p$ depends continuously on $p$, 
    i.e. $\scr{Q}_p$ can be chosen in such a way that $V(p',\,\cdot\,)$ is 
    analytic on $\scr{Q}_p$ for all $p' \in \scr{Q}_p$. 
  \end{enumerate}
\end{thm}

This is a well known result which follows directly from Hadamards work 
on second order, hyperbolic, partial differential equations
\cite{hadamard32}; see also \cite[Chapter 4.3]{FRIEDL} and
\cite{dewitt60}. In the non-analytic case the series (\ref{eq:40})
does not converge and leads therefore only to an asymptotic expansion
of $V$; see \cite{FRIEDL} and the references therein for
details. However we do not want to proceed in this direction, because
our main interest concerns the following extension of Corollary
\ref{kor:kg-inhom-1}.

\begin{thm} \label{thm:2}
  Consider an analytic manifold $(M,g)$ and a smooth (but not 
  necessarily analytic) timelike curve $\gamma : (a,b) \to M$. Each $t \in
  (a,b)$ has a neighborhood $\scr{I}_{t,\epsilon} := (t-\epsilon,t+\epsilon) \subset (a,b)$
  such that for each $T \in \goth{D}\bigl(\gamma \restr (t-\epsilon,t+\epsilon)\bigr)$
  the function $G^\pm T \setminus \Ran \gamma$ is analytic on
  \begin{equation}\label{eq:46}
    \Sigma(t,\epsilon) := \bigl[H^\pm\bigl(\gamma(s)\bigr) \setminus \{ \gamma(s)\} \bigr] \cap \bigl[ 
    I^+\bigl(\gamma(t-\epsilon)\bigr) \cup I^-\bigl( \gamma(t+\epsilon) \bigr) \bigr]. 
  \end{equation}
  Here $H^\pm(p) =  J^\pm(p) \setminus I^\pm(p)$ denotes the future/past horismo of
  $p \in M$ (cf. \cite[Chapter 14]{ON}), i.e. the manifold generated
  by future/past pointing null geodesics starting in $p$. 
\end{thm}

\begin{pf}
  We will look only at the $G^+$ case, since $G^-T$ can be treated 
  similarly. Due to Corollary \ref{kor:kg-inhom-1} the distribution
  $G^+T$  is regular on $M \setminus \Ran \gamma$. This means it can be identified
  with a smooth function which we can restrict to the submanifolds
  $\Sigma(t,\epsilon)$. To investigate the analyticity behaviour of this 
  restriction, let us consider now an open convex normal neighbourhood
  of $\gamma(t)$ and choose $\epsilon > 0$ in such way that the double cone 
  \begin{displaymath}
    \scr{O} := I^+\bigl(\gamma(t-\epsilon)\bigr) \cup I^-\bigl(\gamma(t+\epsilon) \bigr)
  \end{displaymath}
  is contained in it. Using Equation (\ref{eq:41}) we can decompose 
  the restriction of $G^+T$ to $\scr{O}$ into the sum $(G^+T)(f) =
  (G^+_1T)(f) + (G^+_2T)(f)$, where $f$ is a smooth test function with
  support in $\scr{O}$ and $G^+_j(f)$, $j=1,2$ are the functions 
  \begin{displaymath}
    G^+_1(f)(p) = \Delta(p,\,\cdot\,) \delta_+(\Gamma_p)(f) \ \mbox{and} \ G^+_2(f)(p) =
    \int_{J^+(p,\scr{O})} V(p,q) f(q) \lambda(q).
  \end{displaymath}

  Before we start to calculate $G^+_1(f)$ note that we can assume
  without loss generality that $T \in \goth{D}^0(\gamma)$ holds
  with
  \begin{displaymath}
    T(f) := \int_{-\epsilon}^\epsilon h(s) f\bigl(\gamma(s+t)\bigr) ds,   
  \end{displaymath}
  where $h : (-\epsilon,\epsilon) \to \Bbb{R}$ is a compactly supported smooth
  function, because the higher order case differs only by additional
  derivatives in the $p$ variable, which do not affect analyticity.
  Consider now the coordinate system $\scr{O} \setminus \Ran \gamma \ni p \mapsto
  u(p) \in \tilde \scr{O}$ which is defined as follows: 1. Identify the
  tangent spaces $T_{\gamma(s)}M$ for all $s \in (t-\epsilon,t+\epsilon)$ with the
  $\Bbb{R}^4$ such that parallel transport becomes constant and the
  $x^0$-axis is future pointing. 2. Set $u^{-1}(x^0,\mathbf{x}) =
  \exp_{\gamma(t+x^0)}(|\mathbf{x}|, \mathbf{x})$ for all
  $(x^0,\mathbf{x})$ with $x^0 \in (-\epsilon,\epsilon)$ and with
  $\exp_{\gamma(t+x^0)}(|\mathbf{x}|,\mathbf{x}) \in \Sigma(t+x^0,\epsilon)$. 
  The restriction of this coordinate map to the future horismo is
  analytic (since it coincides with the exponential map 
  there) although the curve is only assumed to be smooth. According to
  Equation (\ref{eq:39}) $T\bigl(G^-_1(f)\bigr)$ can be expressed in
  this chart by: 
  \begin{displaymath}
    T\bigl(G^-_1(f)\bigr) = \int_{-\epsilon}^\epsilon h(s) \int_{\Bbb{R}^3}
    \frac{(u_*f)(s,\mathbf{x}) [(\Id \times u)_*
      \Delta]\bigl(\gamma(t+s);s,\mathbf{x}) k(s,\mathbf{x})}{|\mathbf{x}|}
    d\mathbf{x} ds,    
  \end{displaymath}
  where $u_* f$ and $[(\Id \times u)_*\Delta]\bigl(\gamma(t+s,\,\cdot\,)$ are the local
  representatives of $f \in \scr{D}(\scr{O})$ and of\footnote{$(\Id \times
    u)_*\Delta$ denotes the push-forward of $\Delta$ by the diffeomorphism
    $\scr{O} \times \scr{O} \ni (p,q) \mapsto (\Id \times u)(p,q) = \bigl(p,u(q)\bigr) \in 
    \scr{O} \times u(\scr{O})$.} $\Delta\bigl(\gamma(t+s),\,\cdot\,)$ in the coordinate 
  system $u$. Furthermore $k(s,\mathbf{x})$ is defined by 
  \if0\aivsize
  \clearpage
  \fi
  \begin{displaymath}
    k(s,\mathbf{x}) := \sqrt{| \det [(u_{\gamma(t+s)*}g)_{\mu
        \nu}(|\mathbf{x}|,\mathbf{x})]|}, 
  \end{displaymath}
  where $(u_{\gamma(t+s)*}g)_{\mu \nu}$ denotes the representative of $g$ in 
  normal coordinates \if0\aivsize \linebreak[4] \fi $u_{\gamma(t+s)} :
  \scr{O} \to u_{\gamma(t+s)}(\scr{O}) \subset \Bbb{R}^4$ around $\gamma(t+s)$;
  cf. Equation (\ref{eq:39}). Hence the Distribution $G^+_1T$
  coincides on $\scr{O} \setminus \Ran \gamma$ with the smooth function
  \begin{displaymath}
    (G^+_1T)\bigl(u^{-1}(s,\mathbf{x})\bigr) =  h(s)
    \frac{(u_*f)(s,\mathbf{x}) [(\Id \times u)_*
      \Delta]\bigl(\gamma(t+s);s,\mathbf{x}) k(s,\mathbf{x})}{|\mathbf{x}|
      \sqrt{|\det \bigl[(u_*g)_{\mu \nu}(s,\mathbf{x})\bigr]}} d\mathbf{x}
    ds. 
  \end{displaymath}
  Since the functions $\scr{O} \times \scr{O} \ni (p,q) \mapsto u_p(q)$ and
  $\Sigma(t+s,\epsilon) \ni q \mapsto u(q)$ are analytic, $G^+_1T \restr \Sigma(t+s,\epsilon)$ is
  analytic as well. 

  Consider now the second term. According to Proposition
  \ref{prop:gothDl} $T G^+_2(f)$ is given by
  \begin{displaymath}
    \int_{t-\epsilon}^{t+\epsilon} \int_{J^+(\gamma(s),\scr{O})} h(s) V(\gamma(s),q) f(q) \lambda(q) ds,
  \end{displaymath}
  where $\lambda(q)$ denotes the volume form defined by the metric. Hence 
  the distribution $G^+_2 T$ can be identified with the $L^1_{\rm loc}$
  function 
  \begin{displaymath} 
    (G_2^+ T)(q) = \int_{s(q)}^{t+\epsilon} h(s) V(\gamma(s),q) ds,
  \end{displaymath}
  where $s(q)$ is defined by $q \in \Sigma\bigl(s(q),\epsilon\bigr)$. This function
  is smooth on $\scr{O} \setminus \Ran T$ since the distributions $G^+T$  
  and $G^+_1T$ are regular and $G_2^+T = G^+T - G_1^+T$. By Theorem
  \ref{thm:anal-fund-sol} we can choose $\epsilon$ in such a way that
  $V(\gamma(s),\,\cdot\,)$ is analytic on $I^+\bigl(\gamma(t-\epsilon)\bigr) \cup I^-\bigl( 
  \gamma(t+\epsilon) \bigr)$ for each $s \in (t-\epsilon,t+\epsilon)$. Hence $G_2 T \restr
  \Sigma(t+s,\epsilon)$ is analytic too, and this completes the proof.
\end{pf}

\section{The global Hadamard condition}
\label{sec:hadamard}
 
In this appendix we will give the original definition of global
Hadamard states which we have postponed in Sec. \ref{sec:frskf-gen}
because it is somewhat involved. 

Consider first a smooth Cauchy surface $\Sigma$ of space-time
$(M,g)$. A \emph{causal normal neighbourhood} of $\Sigma$ is an open
neighbourhood $N$ of $\Sigma$ such that $\Sigma$ is a Cauchy surface
for $\Sigma$ and such that for all $x_1,x_2\in N$ with $x_1 \in
J^+(x_2)$ there is a convex normal neighbourhood which contains
$J^-(x_1) \cap J^+(x_2)$. This implies that the square of the geodesic 
distance is a well defined smooth function on $N\times N$. The
existence of $N$ is guaranteed by Lemma 2.2 of \cite{KAY+WALD91}.

In addition we need two open regions $\scr{O}, \scr{O}' \subset
M\times M$ with the following properties: $\overline{\scr{O}'} \subset
\scr{O}$, $\scr{O}' \subset N\times N$ and $\scr{O}$ is a
neighbourhood of the set of causally related points $(x_1,x_2) \in
M\times M$ such that $J^+(x_1) \cap J^-(x_2)$ and $J^+(x_2) \cap
J^-(x_1)$ are contained in a convex normal neighbourhood. On $\scr{O}$ 
the square of the geodesic distance $\Gamma(x_1,x_2)$ is again well
defined and smooth such that we get a function $V^{(n)} \in
\cni{\scr{O}}{\Bbb{C}}$ 
\begin{displaymath}
  V^{(n)}(x_1,x_2) := \sum_{j=0}^nV_j(x_1,x_2)\Gamma^j
\end{displaymath}
for each $n\in\Bbb{N}$, where the $V_j$ are the functions from Equation
(\ref{eq:43}). 

Finally we need a smooth, global time function $T: \scr{M} \to
\Bbb{R}$, a smooth function $\chi\in\ci{N\times N}{\Bbb{R}}$ with 
\begin{displaymath}
  \chi(x_1,x_2) = 
  \begin{cases}
    0 & \mbox{for} \  (x_1,x_2) \not\in \scr{O}\\
    1 & \mbox{for} \  (x_1,x_2) \in \scr{O}',
  \end{cases}
\end{displaymath}
and the van Vleck-Morette determinant $\Delta: N \times N \to \Bbb{R}$
(\ref{eq:38}). This leads for each $n \in \Bbb{N}$ and
$\epsilon>0$ to a complex valued function $G^{T,n}_\epsilon$:
\if1\aivsize
\begin{displaymath}
  G^{T,n}_\epsilon(x_1,x_2) := \frac{1}{(2\pi)^2} \left(
    \frac{\sqrt{\Delta(x_1,x_2)}}{\Gamma+2i\epsilon\bigl(T(x_1) - 
        T(x_2) \bigr) + \epsilon^2} + 
      v^{(n)}(x_1,x_2)\ln\biggl(\Gamma+2i\epsilon\bigl(T(x_1) - T(x_2)
        \bigr) + \epsilon^2\biggr)\right)
\end{displaymath}
\else
\begin{multline*}
  G^{T,n}_\epsilon(x_1,x_2) := \frac{1}{(2\pi)^2} \Biggl(
    \frac{\sqrt{\Delta(x_1,x_2)}}{\Gamma+2i\epsilon\bigl(T(x_1) - 
        T(x_2) \bigr) + \epsilon^2} \\ + 
      v^{(n)}(x_1,x_2)\ln\biggl(\Gamma+2i\epsilon\bigl(T(x_1) - T(x_2)
        \bigr) + \epsilon^2\biggr)\Biggr)
\end{multline*}
\fi
where the branch-cut of the logarithm is taken to be on the negative
real axis. Now we are ready to give the following definition

\begin{defi}
  A global Hadamard state is a quasi-free regular state $\omega$ on the
  CCR-algebra $\CCR(\scr{S},E)$ whose two-point function
  $\scr{W}^{(2)}$ has the following structure: There exists a sequence
  of functions $H^{(n)} \in C^n(N\times N,\Bbb{R})$ such that for all $f_1,
  f_2 \in \cni{N}{\Bbb{C}}$ and for all $n \in \Bbb{N}$ we have
  \begin{displaymath}
    \scr{W}^{(2)}(f_1\otimes f_2) = \lim_{\epsilon\to 0}\int_{N\times N} 
    \scr{W}^{T,n}_\epsilon(x_1,x_2)f_1(x_1)f_2(x_2) \lambda_g(x_1)\land 
    \lambda_g(x_2)
  \end{displaymath}
  with
  \begin{displaymath}
    \scr{W}^{T,n}_\epsilon(x_1,x_2) :=
    \chi(x_1,x_2)G^{T,n}_\epsilon(x_1,x_2)+H^{(n)}(x_1,x_2). 
  \end{displaymath}
  This definition does not depend on the choice of $\Sigma, N, \chi$ and
  $T$ (see \cite{KAY+WALD91}).
\end{defi}

\end{appendix}

\section*{Index of notations}

\if1\aivsize
\begin{longtable}{p{3.2cm}p{12cm}}
\else
\begin{longtable}{p{3.2cm}p{8.2cm}}
\fi
\endhead
\endfoot
$\scr{E}(M)$ & complex valued, smooth functions on the manifold $M$
\\[1ex]
$\scr{E}(M,\Bbb{R})$ & real valued smooth functions on the manifold
$M$ \\[1ex]
$\scr{E}(M,E)$ & smooth sections of vector bundle $E \to M$\\[1ex]
$\scr{E}'(M)$ & complex valued, compactly supported distributions on
the manifold $M$\\[1ex] 
$\scr{E}'(M, \Bbb{R})$ & real valued, compactly supported
distributions on the manifold $M$\\[1ex] 
$\scr{E}'(M,E)$ & compactly supported, distributional sections of the
vectorbundle $E \to M$\\[1ex]
$\scr{D}(M)$ & complex valued, compactly supported, smooth  functions
on the manifold $M$\\[1ex] 
$\scr{D}(M, \Bbb{R})$ & real valued, compactly supported, smooth
functions on the manifold $M$\\[1ex]
$\scr{D}(M,E)$ & compactly supported, smooth sections of the vector
bundle $E \to M$\\[1ex]
 $\scr{D}'(M)$ & complex valued distributions on the manifold
 $M$\\[1ex]  
$\scr{D}'(M, \Bbb{R})$ & real valued distributions on the manifold
$M$\\[1ex]  
$\scr{D}'(M,E)$ & distributional sections of the vectorbundle $E \to
M$\\[1ex] 
$\Lin(D_0,\scr{H})$ & (unbounded) operators on the Hilbert space $\scr{H}$
with domain $D_0 \subset \scr{H}$.\\[1ex]
$f \mapsto \Phi(f)$ & hermitian quantum field (Section \ref{sec:qf})\\[1ex]
$D_0$ & common, dense domain of the quantum field $\Phi$ (Section
\ref{sec:qf})\\[1ex] 
$\Omega$ & vacuum vector of the quantum field $\Phi$\\[1ex]
$\goth{A}$ & Borchers-Uhlmann algebra (Section \ref{sec:qf})\\[1ex]
$\scr{W}$, $\scr{W}^{(n)}$ & Wightman functional and $n$-point
function (Section \ref{sec:qf})\\[1ex]
$\scr{R}(\scr{O})$ & local von Neumann algebra associated to
space-time region $\scr{O} \subset M$ (Section \ref{sec:qf})\\[1ex]
$\scr{B}(M)$ & open, relatively compact subsets of the manifold $M$
(Section \ref{sec:qf})\\[1ex]
$\WF(T)$ & wave front set of distribution $T$ (Definition
\ref{def:2})\\[1ex]
$\scr{D}'_\Gamma(M)$ & ($M$ manifold, $\Gamma$ closed cone) Definition
\ref{def:scrd-gamma-m}\\[1ex]  
$\Gamma \oplus \Sigma$ & ($\Gamma, \Sigma \subset T^*M$ closed cones) Equation (\ref{eq:19}), Theorem
\ref{thm:distri-prod}\\[1ex] 
$\goth{D} \subset \scr{D}(M)$ & space of test distributions (Section
\ref{sec:QFsingTF})\\[1ex]
$\Phi(T)$, $T \in \goth{D}$ & extended quantum field (Definitions
\ref{eq:qf-extend-0}, \ref{def:extend-qf} and Theorem
\ref{prop:extend-qf})\\[1ex]
$D \subset \scr{H}$ & common, dense domain of the extended quantum field
(Theorem \ref{prop:extend-qf})\\[1ex]
$\goth{A}(\goth{D})$ & Equation (\ref{eq:bu-goth-d})\\[1ex]
$\goth{W}$, $\goth{W}^{(n)}$ & extended Wightman functional and
$n$-point function (Proposition \ref{prop:scrW-extend})\\[1ex]
$\Gamma \odot \Sigma$ & ($\Gamma, \Sigma \subset T^*M$ closed cones) Equation (\ref{eq:18})\\[1ex]
$\Gamma(\mathbf{T})$ & ($\mathbf{T} \in \scr{D}'(M^n)$) Definition
\ref{def:extend-qf}\\[1ex] 
$\goth{D}(\gamma)$, $\goth{D}^\infty(\gamma), \goth{D}^l(\gamma)$ & ($\gamma$ smooth curve)
Definition \ref{def:1}\\[1ex]
$J^l(M,C) \to M$ & $l$-jet bundle over manifold $M$ (Appendix
\ref{sec:jetbun})\\[1ex]
$\JetC(\gamma)$ & ($\gamma$ smooth curve) Theorem \ref{thm:jet2distri}\\[1ex]
$T_\psi \in \goth{D}^l(\gamma)$ & Theorem \ref{thm:jet2distri}\\[1ex]
$\goth{D}(M)$ & Equation (\ref{eq:gothD-def})\\[1ex]
$E \boxtimes F$ & exterior tensor product of vector bundles $E, F$
(Section \ref{sec:qf:qf-on-wl})\\[1ex]
$\goth{A}^l(\gamma)$ & ($l \in \Bbb{N}$, $\gamma$ smooth, timelike curve) Equation
(\ref{eq:59})\\[1ex] 
$\scr{W}^{(l;n)}_\gamma$ & ($l,n \in \Bbb{N}$, $\gamma$ smooth, timelike curve) Equation
(\ref{eq:scrWsubgammal-def})\\[1ex] 
$\scr{H}^l(\gamma) \subset \scr{H}$ & Theorem \ref{thm:Phik-def}\\[1ex]
$\Phi^l(\gamma)$ &  Theorem \ref{thm:Phik-def}\\[1ex]
$\scr{R}^l(\gamma)$ & ($l \in \Bbb{N}$, $\gamma$ smooth, timelike curve) Equation
(\ref{eq:scrRsubk-gamma-def})\\[1ex]
$\scr{R}(\gamma)$ & ($\gamma$ smooth, timelike curve) Equation
(\ref{eq:scrR-gamma-def})\\[1ex] 
$\eta^l_\gamma$ & ($l \in \Bbb{N}$, $\gamma$ smooth, timelike curve) Equation
(\ref{eq:2})\\[1ex]
$\scr{M}^l_\gamma(\mu)$ & ($l \in \Bbb{N}$, $\gamma$ smooth, timelike curve, $\mu \subset
\gamma$) Equation (\ref{eq:36})\\[1ex]
$G$ & retarded minus advanced solution (Section
\ref{sec:frskf-gen})\\[1ex]
$(\scr{P},G)$ & symplectic space defined by $G$ (Section
\ref{sec:frskf-gen})\\[1ex]
$\CCR(\scr{P},G)$ & CCR-algebra associated to symplectic space
$(\scr{P},G)$ (Section \ref{sec:frskf-gen})\\[1ex] 
$\iota_\Sigma$ & ($\Sigma$ Cauchy surface) Equation (\ref{eq:42})\\[1ex]
$K$ & one particle structure (Section \ref{sec:frskf-gen})\\[1ex]
$\scr{A}(\scr{O})$ & local C*-algebra (Equation (\ref{eq:60})\\[1ex]
$\scr{A}^l(\gamma)$ & ($l \in \Bbb{N}$, $\gamma$ smooth, timelike curve) Equation
(\ref{eq:61})\\[1ex] 
$\scr{P}^l(\gamma) \subset \scr{P}$ & ($l \in \Bbb{N}$, $\gamma$ smooth, timelike curve)
Section \ref{sec:frskf-gen}\\[1ex]
$\gamma_i$ & inertial observer in Minkowski space (Section
\ref{sec:minkowski-space})\\[1ex]
$\Xi^l_{\gamma Zt}$ & Equation (\ref{eq:1}) \\[1ex]
$\alpha^l{\gamma Zt}$ &  Theorem \ref{thm:1} \\[1ex]
$\overline{\alpha}^l_{\gamma Zt}$  & Theorem \ref{thm:1} \\[1ex]
$\alpha^0{\gamma t}$ &  Theorem \ref{thm:1} \\[1ex]
$\overline{\alpha}^0_{\gamma t}$  & Theorem \ref{thm:1} \\[1ex]

\end{longtable}



\end{document}